\documentclass[preprint,12pt]{elsarticle}

\usepackage[english]{babel}
\usepackage[T1]{fontenc} 
\usepackage{times}

\usepackage{amsthm} 
\usepackage{amsmath} 
\usepackage{amssymb} 
\usepackage[normalem]{ulem}
\usepackage[usenames,dvipsnames]{xcolor}
\usepackage{graphicx}
\usepackage{overpic}
\usepackage{xifthen}
\usepackage{verbatim}
\usepackage{booktabs}
\usepackage{float}

\usepackage{paralist} 

\newtheorem{theorem}{Theorem}
\newtheorem{lemma}[theorem]{Lemma}

\usepackage{tikz}

\usepackage{mathtools} 

\usepackage{url}
\usepackage{breakurl}
\usepackage{hyperref}
\usepackage{natbib}
\hypersetup{
  colorlinks,
  citecolor=Blue,
  linkcolor=Blue,
  urlcolor=Blue,
  breaklinks=true
}

\newcommand{\pyfunc}[1]{\texttt{#1}}

\usepackage{listings}

\newcommand{\pyline}[1]{\lstinline[language=python,basicstyle=\ttfamily\small\mdseries,stringstyle=\color{strings}]{#1}}

\newcommand{\pycode}[3]%
{
\definecolor{number}{gray}{0.6}
\definecolor{keywords}{rgb}{1.0,0.3,0.3}
\definecolor{comments}{rgb}{0.1,0.65,0.1}
\definecolor{strings}{rgb}{0.3,0.0,1.0}
\lstset{language=python,
        morekeywords={switch,case},
        sensitive=true,
        showspaces=false, 
        basicstyle=\ttfamily\small\mdseries,
      	keywordstyle=\bfseries\color{keywords},
        commentstyle=\color{comments},
        stringstyle=\color{strings},
        numbers=left,
        numberstyle=\scriptsize\color{number},
        stepnumber=1,
        breaklines=true,
        frame=none,
        showstringspaces=false,
        tabsize=4,
        xleftmargin=\bigskipamount,
        xrightmargin=\bigskipamount,
        aboveskip=\bigskipamount,
        belowskip=0pt
}
\lstinputlisting[caption=#2, label=#3, frame=tb]{#1}
}
\newcommand{\pyfrag}[5]%
{
\definecolor{number}{gray}{0.6}
\definecolor{keywords}{rgb}{1.0,0.3,0.3}
\definecolor{comments}{rgb}{0.1,0.65,0.1}
\definecolor{strings}{rgb}{0.3,0.0,1.0}
\lstset{language=python,
        morekeywords={switch,case},
        sensitive=true,
        showspaces=false, 
        basicstyle=\ttfamily\small\mdseries,
        keywordstyle=\bfseries\color{keywords},
        commentstyle=\color{comments},
        stringstyle=\color{strings},
        numbers=left,
        numberstyle=\scriptsize\color{number},
        stepnumber=1,
        breaklines=true,
        frame=none,
        showstringspaces=false,
        tabsize=4,
        xleftmargin=\bigskipamount,
        xrightmargin=\bigskipamount,
        aboveskip=\bigskipamount,
        belowskip=\bigskipamount
}
\lstinputlisting[caption=#2, label=#3, frame=tb, firstnumber=#4, firstline=#4,
  lastline=#5]{#1}
}

\usepackage[normalem]{ulem} 
\usepackage{cancel}

\newcounter{bla}

\journal{Computer Physics Communications}

\definecolor{c0}{HTML}{000000} 
\definecolor{c1}{HTML}{9EC05B} 
\definecolor{c2}{HTML}{3288BD} 
\definecolor{c3}{HTML}{D53E3F} 
\definecolor{c4}{HTML}{E8B42B} 
\definecolor{c5}{HTML}{B82EE6} 
\definecolor{c9}{HTML}{FFFFFF} 

\definecolor{dg}{HTML}{006400} 

\newcommand{\OSMPS}{OSMPS}

\newcommand{\miosixtyx}{\emph{Penguin Relion 2x(Intel X5675) 12 cores 3.06GHz}} 
\newcommand{\mioseventeenx}{\emph{2x(Intel Xeon E5-2680 Dodeca-core) 24 Cores 2.50GHz}} 
\newcommand{\py}{Python}
\newcommand{\fort}{Fortran}

\newcommand{\gershgorin}{Ger{\v s}gorin}

\newcommand{\psitem}[2]{\emph{#1:~}#2\newline}
\newcommand{\mpst}[5]{
  \ifthenelse{\equal{#5}{}}{
    \mpstnoidx{#1}{#2}{#3}{#4}
  }{
    \mpstidx{#1}{#2}{#3}{#4}{#5}
}}

\newcommand{\mpstidx}[5]{
  \ifthenelse{\equal{#2}{}}{#1_{#4}^{\lbrack #5 \rbrack #3}
  }{\ifthenelse{\equal{#4}{}}{#1_{#2}^{\lbrack #5 \rbrack #3}
  }{#1_{#2,#4}^{\lbrack #5 \rbrack #3}}}}

\newcommand{\mpstnoidx}[4]{
  \ifthenelse{\equal{#2}{}}{#1_{#4}^{#3}
  }{\ifthenelse{\equal{#4}{}}{#1_{#2}^{#3}
  }{#1_{#2,#4}^{#3}}}}

\newcommand{\mpot}[6]{
  \ifthenelse{\equal{#6}{}}{\mpotnoidx{#1}{#2}{#3}{#4}{#5}
  }{\mpotidx{#1}{#2}{#3}{#4}{#5}{#6}}}

\newcommand{\mpotidx}[6]{
  \ifthenelse{\equal{#2}{}}{#1_{#5}^{\lbrack #6 \rbrack #3, #4}
  }{\ifthenelse{\equal{#5}{}}{#1_{#2}^{\lbrack #6 \rbrack #3, #4}
  }{#1_{#2,#5}^{\lbrack #6 \rbrack #3, #4}}}}

\newcommand{\mpotnoidx}[5]{
  \ifthenelse{\equal{#2}{}}{#1_{#5}^{#3, #4}
  }{\ifthenelse{\equal{#5}{}}{#1_{#2}^{#3, #4}
  }{#1_{#2,#5}^{#3, #4}}}}

\newcommand{\dt}{\mathrm{d}t}
\newcommand{\Tens}[4][]{\ifthenelse{\equal{#1}{}}
   {  
      #2_{#3}^{#4}
   }
   {  
      #2_{#3}^{\lbrack #1 \rbrack \, #4}
   }
}

\newcommand{\peloc}{\varepsilon_{\mathrm{local}}}
\newcommand{\pevar}{\varepsilon_{\mathrm{V}}}
\newcommand{\pelanc}{\varepsilon_{\mathrm{l}}}

\newcommand{\erhoi}{\epsilon_{\mathrm{local}}}
\newcommand{\erhoij}{\epsilon_{\mathrm{corr}}}
\newcommand{\eener}{\epsilon_{\mathrm{E}}}
\newcommand{\eentr}{\epsilon_{\mathrm{S}}}
\newcommand{\echi}{\epsilon_{\chi}}
\newcommand{\edt}{\epsilon_{dt}}
\newcommand{\emeth}{\epsilon_{\mathrm{method}}}
\newcommand{\etot}{\epsilon_{}}
\newcommand{\etebdt}{\epsilon_{\mathrm{TEBD2}}}
\newcommand{\etebdf}{\epsilon_{\mathrm{TEBD4}}}
\newcommand{\epsi}{\epsilon}
\newcommand{\erhom}{\tilde{\epsilon}_{-}}
\newcommand{\erhop}{\tilde{\epsilon}_{+}}
\newcommand{\erhopm}{\tilde{\epsilon}_{\pm}}
\newcommand{\eloc}{\epsilon_{\mathrm{local}}}

\newcommand{\Tcpu}{T_{\mathrm{CPU}}}
\newcommand{\Tjob}{T_{\mathrm{Job}}}

\makeatletter
\newcommand{\pushright}[1]{\ifmeasuring@#1\else\omit\hfill$\displaystyle#1$\fi\ignorespaces}
\makeatother

\newcommand{\e}{\mathrm{e}}
\renewcommand{\i}{\mathrm{i}}
\newcommand{\1}{\mathbb{I}}

\newcommand{\ket}[1]{\left|{#1}\right\rangle}

\newcommand{\bra}[1]{\left\langle{#1}\right|}

\newcommand{\braket}[2]{\left\langle #1 \middle| #2 \right\rangle}

\newcommand{\sandwich}[3]{\left\langle #1 \right| #2 \left| #3 \right\rangle}

\DeclareMathOperator{\tr}{Tr}
\newcommand{\Ptr}[2]{\mathrm{Tr_{#1}} #2}
\newcommand{\PTr}[2]{\mathrm{Tr_{#1}} \left( #2 \right)}

\definecolor{daniel}{rgb}{0,.1,1}

\definecolor{michael}{RGB}{150,00,00} 

\definecolor{lincoln}{RGB}{40,180,40}

\definecolor{ulmcolor}{rgb}{.4,0,.6}

\definecolor{ulmmethod}{rgb}{0,.3,.6}

\definecolor{dangercolor}{rgb}{0.8,0.,0.}

\newcommand{\csm}{Department of Physics, Colorado School of Mines, Golden,
  Colorado 80401, USA}
\newcommand{\jila}{JILA, NIST and University of Colorado, Boulder, 
  Colorado 80309-0440, USA}
\newcommand{\jhu}{The Johns Hopkins University Applied Physics Laboratory,
  Laurel, MD, 20723, USA}

\begin{document}

\begin{frontmatter}

\title{Open source Matrix Product States: Opening ways to simulate
  entangled many-body quantum systems in one dimension}

\author[a]{Daniel Jaschke\corref{author}}
\author[a,b]{Michael L.\ Wall\corref{thanks}}
\author[a]{Lincoln D.\ Carr}

\cortext[author] {Corresponding author.\\\textit{E-mail address:} djaschke@mines.edu}
\cortext[thanks] {Present address: \jhu}
\address[a]{\csm}
\address[b]{\jila}



\begin{abstract}

%

  Numerical simulations are a powerful tool to study quantum systems beyond     
  exactly solvable systems lacking an analytic expression. For one-dimensional
  entangled quantum systems, tensor network methods, amongst them Matrix Product
  States (MPSs), have attracted interest from different fields of quantum physics
  ranging from solid state systems to quantum simulators and quantum computing.
  Our open source MPS code provides the community with a
  toolset to analyze the statics and dynamics of one-dimensional quantum systems.
  Here, we present our open source library, Open Source Matrix Product States (OSMPS),
  of MPS  methods implemented in \py{} and \fort{}2003. The library includes tools for
  ground state calculation and excited states via the variational ansatz. We
  also support ground states for infinite systems with translational invariance.
  Dynamics are simulated with different algorithms, including three algorithms
  with support for long-range interactions. Convenient features include built-in
  support for fermionic systems and number conservation with rotational
  $\mathcal{U}(1)$ and discrete $\mathbb{Z}_2$ symmetries for finite
  systems, as well as data parallelism with MPI. We explain the principles
  and techniques used in this library along with examples of how to efficiently
  use the general interfaces to analyze the Ising and Bose-Hubbard models. This
  description includes the preparation of simulations as well as dispatching
  and post-processing of them.

\end{abstract}


\end{frontmatter}


{\bf PROGRAM SUMMARY}

\begin{small}
\noindent
\psitem{Program title}{Open Source Matrix Product States (\OSMPS{}), v2.0}
\psitem{Program summary and documentation}{\url{http://openmps.sourceforge.io/}}
\psitem{Program obtainable from}{\url{http://sourceforge.net/p/openmps}}
\psitem{Licensing provisions}{GNU GPL v3 (Minor parts follow the copyright
  of the Expokit package.)}
\psitem{Programming language}{\py{}, \fort{}2003, MPI for parallel computing}
\psitem{Compilers (\fort{})}{gfortran, ifort, g95}
\psitem{Operating system}{Linux, Mac OS X, Windows}
\psitem{Supplementary material}{We provide programs to reproduce selected
  figures in the Appendices.}
\psitem{Nature of the problem}{Solving the ground state and dynamics of
  a many-body entangled quantum system is a challenging problem; the Hilbert
  space grows exponentially with system size. Complete diagonalization of
  the Hilbert space to floating point precision is limited to less than
  forty qubits.}
\psitem{Solution method}{Matrix Product States in one spatial dimension
  overcome the exponentially growing Hilbert space by truncating the
  least important parts of it. The error can be well controlled.
  Local neighboring sites are variationally optimized in order to
  minimize the energy of the complete system. We can target the ground
  state and low lying excited states. Moreover, we offer various methods
  to solve the time evolution following the many-body Schr\"odinger equation.
  These methods include e.g. the Suzuki-Trotter decompositions using local
  propagators or the Krylov method, both approximating the propagator on the
  complete Hilbert space.}
\end{small}

\tableofcontents

\section{Introduction                                                          \label{sec:intro}}

Numerical methods have been widely used to study physical systems in quantum
mechanics that are not exactly solvable. In many-body systems we encounter with
the exponentially growing Hilbert space a challenge to develop methods which
can still simulate quantum systems on a classical computer. Starting with the
Density Matrix Renormalization Group (DMRG)~ \cite{White1992,White1993}, a wide range
of tensor network methods have been developed. Especially in one dimension,
where numerical scaling and conditioning are best, such methods offer strong
alternatives to other methods such as Quantum Monte
Carlo~\cite{Prokofev1998exact,Sandvik1991} and the Truncated Wigner
approximation~\cite{Polkovnikov2010phase,Schachenmayer2015}. Applications
include solid state systems, ultracold atoms and molecules, Rydberg atoms,
quantum information and quantum computing, and Josephson junction-based
superconducting electro-mechanical nano devices.                                
Matrix Product States (MPSs) \cite{Vidal2003,Schollwoeck2011,Orus2014,Chan2016}
define the tensor network at the foundation of the DMRG method. MPSs themselves
represent a pure quantum state constructed on local Hilbert spaces of
lattice sites or discretized systems. They handle the exponentially growing
Hilbert space by limiting the entanglement between any two parts of the system.
Numerical methods using MPSs support static results such as ground states and
time evolution of pure states. In principle, highly entangled states can be
represented as MPSs, but due to the upper bound of entanglement set as a parameter,
they are only accurate as long as the entanglement does not exceed this bound
guaranteeing feasible computation times. This point is addressed in the main
part of this paper in detail. Moreover, MPSs can exploit intrinsic
characteristics of the systems such as symmetries~\cite{Singh2010,Singh2011}.
Beyond MPSs, tensor network methods are extended for multiple
purposes such as 2D systems via Projected Entangled Pair States (PEPS)
\cite{Verstraete2008}, tree tensor networks
(TTNs) \cite{Shi2006}, open systems with quantum trajectories (QT)
\cite{Dalibard1992,Dum1992} or Matrix Product Density Operators (MPDOs)
\cite{Verstraete2004,Zwolak2004}. Although there have been many developments
in many-body quantum simulation over the last ten to twenty years
\cite{CarrQPT}, from multi-scale entanglement renormalization ansatz (MERA)
\cite{Vidal2008} to minimally entangled typical thermal states (METTS)
\cite{Stoudenmire2010} to dynamical mean-field theory \cite{Georges1996}
to time-dependent density functional theory (TDDFT) \cite{Runge1984,TDDFTNote},
MPS methods are the most often used and well-established for strongly
correlated quantum systems and appropriate for large-scale open source
development. The impact of these methods is represented by the large
number of open source packages for MPS and DMRG \cite{ALPS,BlockDMRG,Wouters2014,
DMRGPlusPlus,GarciaRipollMPS,ITensor,KoehlerMPS,MilstedMPS,MPSToolkit,
OpenTEBDPackage,SimpleDMRG,SnakeDMRG,deChiara2008,Urbanek2016,Uni10}, not
counting proprietary efforts of multiple other groups.

We present in this paper our open source Matrix Product State (\OSMPS{})
library, which is available on SourceForge \cite{OSMPSPackage}. We have over
2300 downloads since its initial release in January 2014. The library or a
derivative of the library has been used in various publications
\cite{Anisimovas2016,Bellotti2017,Dhar2016,Dolfi2014,Gardas2016,Gong2016,
Gong2016B,JaschkeLRQIC,Koller2016,Maghrebi2015,Russomanno2016,Vargas2015,
Wall2012,Wall2013,Wall2013NJP,Weimer2014}. Our implementations cover
features such as variational ground and excited state
searches and real time evolution for finite systems as well as ground states
of infinite systems \cite{McCulloch2008}. We provide built-in features such as
support for symmetries, e.g. rotational  $\mathcal{U}(1)$ symmetry used for
number conservation in the Bose-Hubbard model and discrete $\mathbb{Z}_{2}$
symmetry occurring in the quantum Ising model, and present them in the
case studies of this article. These symmetries lead to a speedup in
terms of computation time and allow us to address specific states. Our
libraries also support data parallel execution via Message Passing Interface (MPI)
to utilize modern
high performance computing resources efficiently. We illustrate the
algorithms in our library together with examples of models.
%
%

One motivation for the development of \OSMPS{} is our focus on ultracold
molecules and other quantum simulator architectures incorporating long-range
interacting synthetic quantum matter. Where
some MPS-based algorithms are limited to nearest neighbor terms in
one-dimensional systems, molecules and many other systems have long-range
interactions, e.g. due to dipolar effects. In order to treat such systems,
many of the algorithms in OSMPS, including dynamics algorithms, feature
support for long-range interactions.

This paper is intended for two audiences: First, tensor network methods and
our interfaces are introduced for researchers not familiar with such methods,
but in need of numerical simulations of correlations, entanglement, and
dynamics in many-body systems. On the other
hand, experienced researchers within the tensor networks community should
have a clear way to understand the concrete and useful details of our
implementations. The paper is organized as follows. In Sec.~\ref{sec:basics},
we introduce the general idea of tensor networks, and provide in addition
appropriate references for further reading. We continue with the example of
the Ising model in Sec.~\ref{sec:design} to demonstrate the variational ground
state search including the general setup of systems and then highlight the
other algorithms in the following sections. Section~\ref{sec:algorithms}
describes the variational search for excited states and the infinite MPS (iMPS) for the
thermodynamic limit. The time evolution methods including Krylov, Time-Evolving Block
Decimation (TEBD), Time-Dependent Variational Principle (TDVP),
and local Runge-Kutta follow in Sec.~\ref{sec:timeevo}.
We describe future developments ahead of the conclusion in
Sec.~\ref{sec:conclusion}. The appendices cover topics such as convergence
studies for the algorithms, convenient features, and technical information for
installation and the scope of the open source project.


\section{Basic concepts in tensor network techniques                           \label{sec:basics}}

In this section, we briefly review the concepts of tensor network techniques.
Readers familiar with MPS algorithms can continue on to Sec.~\ref{sec:design}
discussing the design of simulations specifically for \OSMPS{}.
The MPS algorithms rely heavily on the Schmidt decomposition of a quantum
system, which can be explained best in the case of two subsystems $1$ and
$2$ and their wave function $\ket{\psi_{1,2}}$. Each subsystem is
defined on a local Hilbert space $\mathcal{H}_{k}$ of
dimension $d_{k}$, and is spanned by an orthonormal basis of states
$\{|i_k\rangle\}$. The joint Hilbert space of subsystem $1$ and $2$ is formed
via the tensor product $\mathcal{H} = \mathcal{H}_{1,2} = \mathcal{H}_{1}
\otimes \mathcal{H}_{2}$ and has a dimension $d_{1} \times d_{2}$. The Schmidt
decomposition is then based on a set of local wave functions
$\ket{\psi^{[1]}_{\alpha}}$ and $\ket{\psi^{[2]}_{\alpha}}$
\begin{eqnarray}                                                                \label{eq:schmidt}
  | \psi_{1,2} \rangle =
  \sum_{\alpha=1}^{\chi_{\max}} \lambda_{\alpha} | \psi^{[1]}_{\alpha} \rangle
  | \psi^{[2]}_{\alpha} \rangle \, .
\end{eqnarray}
The Schmidt decomposition corresponds to a singular value decomposition (SVD)
where the singular values are the $\lambda_{\alpha}$. The number of non-zero
singular values serves as a coarse measure of entanglement between the systems
1 and 2, known as the Schmidt number or Schmidt rank. We dub the maximal number
of singular values as $\chi_{\max}$. Tracing out over either
of the subsystems, we obtain the reduced density matrix of the other subsystem.
This demonstrates that the eigenvalues of the reduced density matrices
$\rho_{1}=\mathrm{Tr}_2| \psi_{1,2} \rangle\langle  \psi_{1,2} |$ and
$\rho_{2}=\mathrm{Tr}_1| \psi_{1,2} \rangle\langle  \psi_{1,2} |$ are
$\{\lambda_{\alpha}^2\}$ \cite{NielsenChuang}.

The Schmidt decomposition is unique up to rotations within subspaces of
degenerate singular values for a chosen basis. Local unitary
transformations such as basis transformations affecting each
half of the bipartition separately are possible in general, because they do
not change the entanglement, i.e., the number and value of non-zero
singular values. MPSs generalize the Schmidt decomposition by
allowing for rotations within the Schmidt bases $|\psi^{[k]}_{\alpha}\rangle$
that keep the amount of bipartite entanglement between the two subsystems
fixed:
\begin{eqnarray}                                                                \label{eq:schmidtMPS}
  | \psi_{1,2} \rangle
  = \sum_{\alpha=1}^{\chi_{\max}} \sum_{i_{1}=1}^{d_1} \sum_{i_{2} =1}^{d_2}
  \mpst{A}{}{i_1}{\alpha}{1} \mpst{A}{\alpha}{i_2}{}{2}
  \ket{i_{1}} \ket{i_{2}} \, .
\end{eqnarray}
In addition, while the Schmidt decomposition is only defined for a bipartite
system, the MPS form can be extended to any number of degrees of freedom.

We take the example of a two-site system to illustrate the connection between
the Schmidt and MPS decompositions.
Subsystem $1$ is the site with $k=1$ and subsystem two corresponds to the site
with $k=2$. The MPS decomposition of such a system is described in
Eq.~\eqref{eq:schmidtMPS}. We rewrite any state $\ket{\psi_{\alpha}}$ in
Eq.~\eqref{eq:schmidt}
as a matrix of complex numbers $c_{i_1i_2}$ and its corresponding basis states,
that is $\sum_{i_1i_2}c_{i_1i_2}|i_1\rangle |i_2\rangle$. In
Eq.~\eqref{eq:schmidt}, each dimension $k$ is spanned by $\chi_{\max}$ orthonormal
vectors. We can rewrite the sets of these vectors as matrices (rank-two
tensors), where each column of the matrix represents a vector in
the case of site $1$, and each row is filled with one vector for site $2$.
Then, the matrix $A^{[ k ]}$
in Eq.~\eqref{eq:schmidtMPS} represents a rank-2 tensor for site $k$ and the
singular values are contained in either $A^{[ k ]}$. A single element of the
matrix $A^{[ k ]}$ can be written as $\mpst{A}{}{j_k}{\alpha}{k} =
\braket{j_k}{\psi_{\alpha}^{[k]}}$ in the two-site case above.
Thus, the indices of the tensors $A^{[k]}$ correspond to the
local Hilbert space $i_k$ and the singular values of the Schmidt decomposition
$\alpha$. Throughout the paper we note the site index of a tensor in brackets,
i.e., the $k$ in $\mpst{A}{}{}{}{k}$.

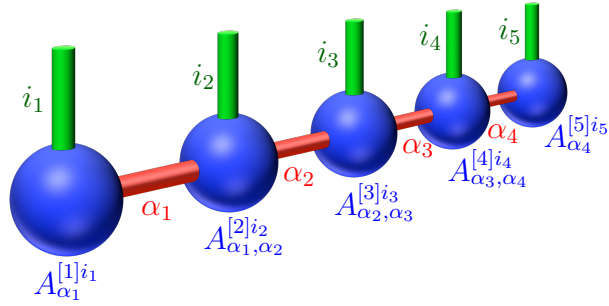
\begin{figure}[t]
  \begin{center}
    \begin{overpic}[width=0.55 \columnwidth, unit=1mm]{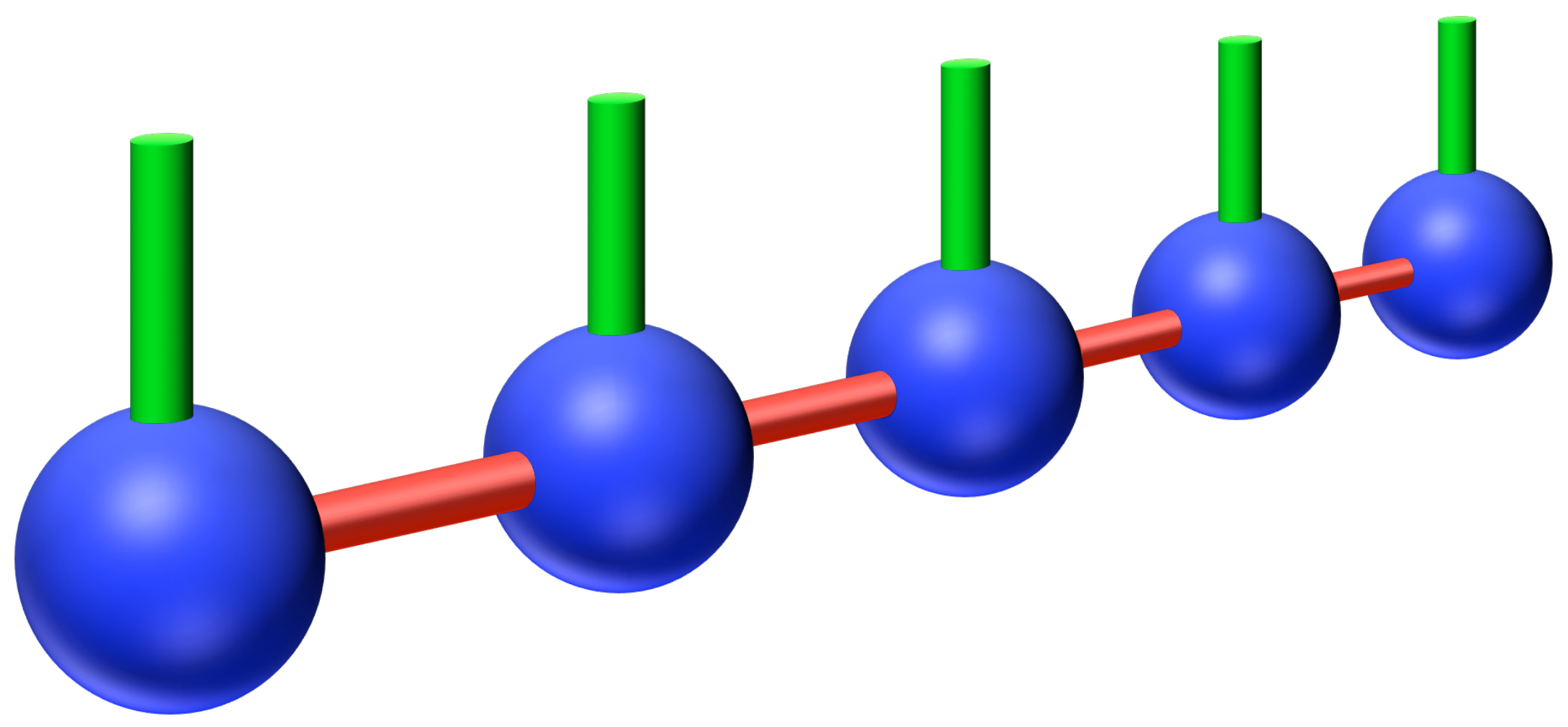}
      \put( 5, -6){\color{blue}$\mpst{A}{}{i_1}{\alpha_1}{1}$}
      \put(35,  2){\color{blue}$\mpst{A}{\alpha_1}{i_2}{\alpha_2}{2}$}
      \put(58,  8){\color{blue}$\mpst{A}{\alpha_{2}}{i_3}{\alpha_{3}}{3}$}
      \put(78, 14){\color{blue}$\mpst{A}{\alpha_{3}}{i_4}{\alpha_{4}}{4}$}
      \put(95, 20){\color{blue}$\mpst{A}{\alpha_{4}}{i_5}{}{5}$}
      \put( 3, 27){\color{dg}$i_1$}
      \put(33, 31){\color{dg}$i_2$}
      \put(55, 35){\color{dg}$i_3$}
      \put(73, 38){\color{dg}$i_4$}
      \put(87, 39){\color{dg}$i_5$}
      \put(24,  8){\color{red}$\alpha_1$}
      \put(49, 14){\color{red}$\alpha_2$}
      \put(70, 19){\color{red}$\alpha_3$}
      \put(85, 21.5){\color{red}$\alpha_4$}
    \end{overpic}\vspace{0.1cm}
    \caption{\emph{Tensor network representation of an MPS} of five sites for
      open boundary conditions. The state is decomposed into local tensors
      representing each site. These tensors are connected to their nearest
      neighbors via the Schmidt decomposition where singular values are
      truncated to maintain feasible run times for larger systems. The indices
      for the local Hilbert space $i_{k}$ are connected for measurements,
      e.g. for the norm $\langle \psi | \psi \rangle$, with their complex
      conjugated counterpart.
                                                                                \label{fig:tensornetwork_mps}}
  \end{center}
\end{figure}

In order to generalize this decomposition for a system with $L$ sites,
successive SVDs lead to one tensor per site where the tensors are now rank-2
at the boundaries and rank-3 in the bulk of the system, as shown in the
representation as a tensor network in Fig.~\ref{fig:tensornetwork_mps}:
\begin{eqnarray}                                                                 \label{eq:mps}
  | \psi_{1, \ldots, L} \rangle =
  \sum_{\alpha_{1} \cdots \alpha_{L-1}} \sum_{i_1 \cdots i_L}
  \mpst{A}{}{i_1}{\alpha_{1}}{1}
  \mpst{A}{\alpha_1}{i_2}{\alpha_2}{2} \cdots
  \mpst{A}{\alpha_{L-2}}{i_{L-1}}{\alpha_{L-1}}{L-1}
  \mpst{A}{\alpha_{L-1}}{i_{L}}{}{L}
  \ket{i_1} \cdots \ket{i_L}
  \, .
\end{eqnarray}
If we allow the indices $\alpha_j$ to run over exponentially large values
$\sim d^{L/2}$ ($d=d_k$ the local dimension, assumed uniform for simplicity),
such a representation is exact, but manipulating this exact representation
also scales exponentially with the system size and we do not gain anything over
exact diagonalization methods.

We now introduce the key approximation in the MPS algorithm, that is the
truncation of the Hilbert space according to the singular values from the
Schmidt decomposition. The essential idea here is to replace the number of
singular values $\chi_{\max}$ with a reduced number $\chi$: for instance,
if there are singular values less than $10^{-16}$ there is no reason to count them
toward the amount of entanglement between the two subsystems. Thus $\chi$
becomes the \emph{reduced Schmidt rank}, the major convergence parameter
of the whole MPS algorithm, as we will show.
We obtain these singular values for any splitting in
two connected subsystems, and the truncation is encoded in the maximum range
of the auxiliary indices $\alpha_{i}$. Considering an approximated state
$\ket{\psi'}$ truncated to the first $\chi$ singular values at some particular
bond of the normalized state $\ket{\psi}$ with $\chi_{\max}$ singular values
at that bond, the overlap between the two states is
\begin{eqnarray}                                                                \label{eq:overlap}
  \braket{\psi'}{\psi}
 &=& \frac{\sum_{i=1}^{\chi} \lambda_{i}^{2}}
     {\sqrt{\sum_{i=1}^{\chi} \lambda_{i}^{2}}}
  = \sqrt{\sum_{i=1}^{\chi} \lambda_{i}^{2}} \, .
\end{eqnarray}
We prefer the overlap instead of the 2-norm of the overlap since it allows us
to relate our result to the quantum fidelity in our case of pure states with
real overlaps as $\mathcal{F} = \langle \psi \ket{\psi'}$.
We define the truncation error made in this step as
$\eloc = 1 - \langle \psi \ket{\psi'}$ and obtain
\begin{eqnarray}                                                                \label{eq:elocal}
  \eloc
 &=& 1 - \sqrt{\sum_{i=1}^{\chi} \lambda_{i}^{2}}
  \le \sum_{i=\chi + 1}^{\chi_{\max}} \lambda_{i}^{2} \, .
\end{eqnarray}
This upper bound is useful since it relates directly to the truncated singular
values. Such a truncation corresponds to a truncation of entanglement
through two entanglement measures: the Schmidt rank $\chi$ and the
von Neumann entropy $S$. The former is simply the number of non-zero
singular values, and is an entanglement monotone. The von Neumann
entropy, or bond entropy, $S$ is given as
\begin{eqnarray}
  S = - \sum_{i=1}^{\chi} \lambda_{i}^{2} \log(\lambda_{i}^{2}) \, .
\end{eqnarray}
%
\begin{figure}[t]
  \begin{center}
    \vspace{-0.5cm}
    \includegraphics[width=0.6 \columnwidth]{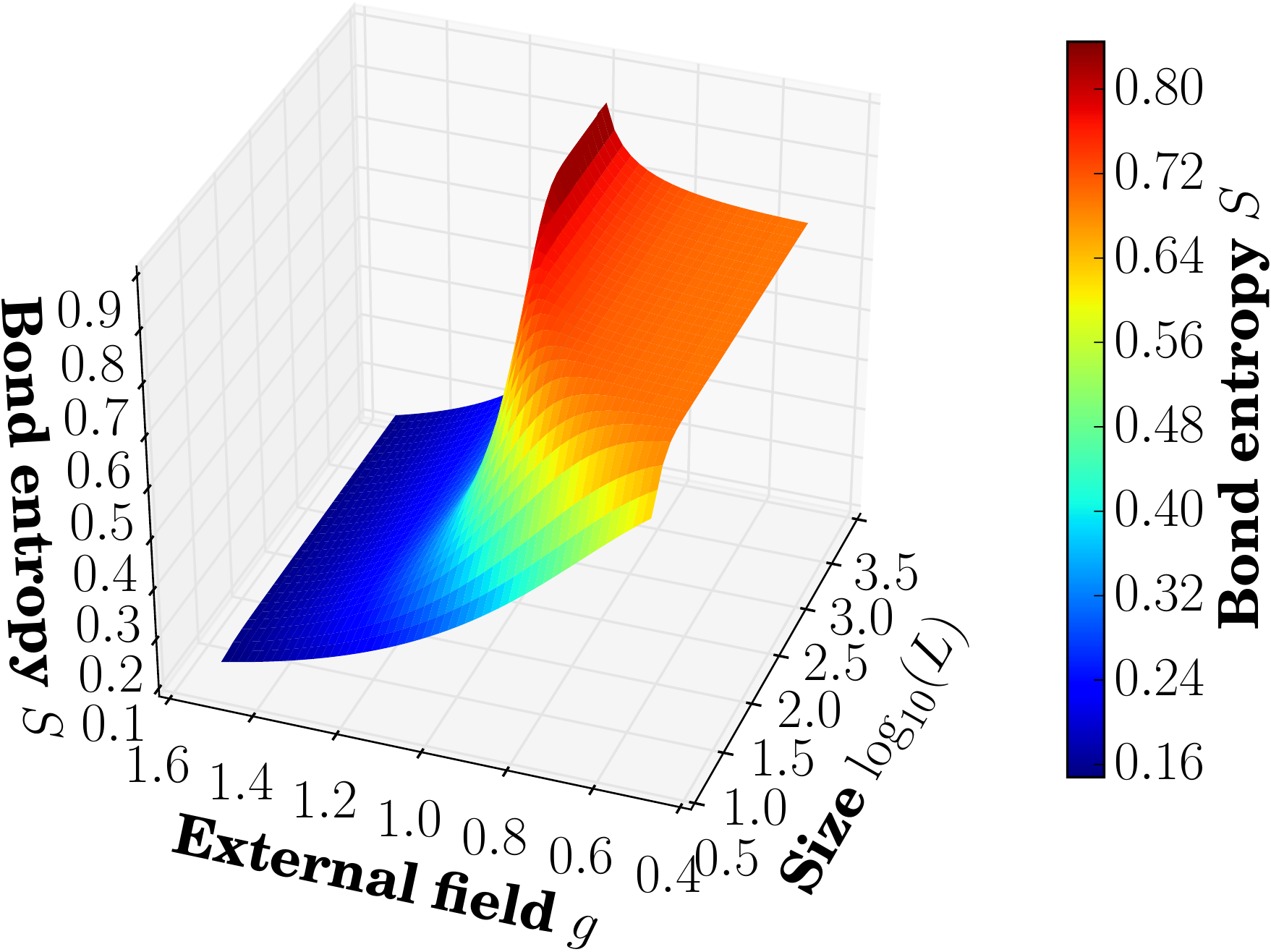}
    \vspace{-0.2cm}
    \caption{\emph{The compression of the quantum state in the MPS} acts on
      the singular values of the Schmidt decomposition in
      Eq.~\eqref{eq:schmidt}. The bond entropy or von Neumann entropy
      at the center bond for the nearest-neighbor quantum Ising model $H = - J \sum
      \sigma_{k}^{z} \sigma_{k+1}^{z} - J g \sum \sigma_{k}^{x}$, defined later
      in Eq.~\eqref{eq:Ising}, peaks around the critical point for
      increasing system sizes.
                                                                                \label{fig:01_Ising_BE3d}}
  \end{center}
\end{figure}
In Fig.~\ref{fig:01_Ising_BE3d} we show the bond entropy for the bipartition
at the center bond for the quantum Ising
model with transverse field as a function of the system size and external field.
Errors in the MPS approach originate in high entanglement; therefore, simulations
for increasing system sizes and around the quantum critical point are more
vulnerable to smaller cutoffs $\chi$. The quantum critical point for the Ising model is
$h_c = 1.0$ and becomes visible as a red ridge in Fig.~\ref{fig:01_Ising_BE3d}
for large system sizes. The ground state of the ferromagnetic phase in the
limit of zero external field, also called
the Greenberger-Horne-Zeilinger (GHZ) state, is the superposition of all spins
up and all spins down, i.e. $\left( | \uparrow \cdots \uparrow
\rangle + | \downarrow \cdots \downarrow \rangle \right) / \sqrt{2}$. We
expect an entropy of $S = - \log(0.5) \approx 0.69$, which agrees well with the
results in the Fig.~\ref{fig:01_Ising_BE3d}. For gapped 1D
systems with short-range interactions, the so-called area law for
entanglement~\cite{Eisert2010} states that the entanglement at any bipartition
is independent of the length of the subsystems (and hence of the system size $L$).
Since the bond entropy is an entanglement measure, this upper bound can be used
as the gap opens away from the critical point. At the critical point, the
entanglement grows logarithmically with the subsystem size.

We introduce a list of basic operations that can be performed on tensor
networks, and explain the orthogonality center, an isometrization or gauge,
used in the \OSMPS{} algorithms. For these linear algebra operations on
tensors, we suppress the basis kets of the quantum states for simplicity
throughout the paper. One key feature of every MPS with open boundary
conditions is that introducing an orthogonality center leads to faster local
measurements and error reduction in the truncation~\cite{Verstraete2004PBC}.
We introduce the left and right canonical form of tensors according to
\begin{eqnarray}
  \mpst{A}{\alpha_{k-1}}{i_{k}}{\alpha_{k}}{k}
  = \mpst{L}{\alpha_{k-1}}{i_{k}}{\alpha_{k}}{k}
  \; \; \mathrm{if} \; \; \begin{cases}
    \mpst{A}{}{}{}{j} = \mpst{L}{}{}{}{j} \; \forall \; j < k \\
    \mathrm{and~}
    \sum_{\alpha_{k-1}, i_{k}}
    \mpst{A}{\alpha_{k-1}}{i_{k}}{\alpha_{k}}{k}
    (\mpst{A}{\alpha_{k-1}}{i_{k}}{\alpha_{k}'}{k})^{\ast}
    = \1_{\alpha_{k}, \alpha_{k}'} \, ,
  \end{cases}                                                                   \nonumber \\
  \mpst{A}{\alpha_{k-1}}{i_{k}}{\alpha_{k}}{k}
  = \mpst{R}{\alpha_{k-1}}{i_{k}}{\alpha_{k}}{k}
  \; \; \mathrm{if} \; \; \begin{cases}
    \mpst{A}{}{}{}{j} = \mpst{R}{}{}{}{j} \; \forall \; j > k \\
    \mathrm{and~}
    \sum_{\alpha_{k}, i_{k}}
    \mpst{A}{\alpha_{k-1}}{i_{k}}{\alpha_{k}}{k}
    (\mpst{A}{\alpha_{k-1}'}{i_{k}}{\alpha_{k}}{k})^{\ast}
    = \1_{\alpha_{k-1}, \alpha_{k-1}'} \, .
  \end{cases}                                                                   \label{eq:oc}
\end{eqnarray}
The left (right) canonical forms $\mpst{L}{}{}{}{k}$ $\left(\mpst{R}{}{}{}{k}\right)$
are unitary
matrices, e.g. from the SVD obtained from the Schmidt decomposition in Eq.~\eqref{eq:schmidt}.
These conditions apply if the singular values have not been multiplied into the tensor.
We define the orthogonality center as the site which has only left
orthogonal tensors on the left side and right orthogonal tensor on the right
side. This feature becomes beneficial for measurements as the contractions in the
condition of Eq.~\eqref{eq:oc} do not have to be calculated knowing that the
result is the identity~$\1$. Stated equivalently, the tensor of the
orthogonality center $A^{[k]i_k}_{\alpha \beta}$ consists of the coefficients
of the wave function in the orthonormal basis spanned by the local states
$|i_k\rangle$ and the left and right Schmidt vectors given by products of
the other MPS tensors. With these definitions, we can derive the overlap from
Eq.~\eqref{eq:overlap}. We assume that we truncate singular values at the
bond of the sites $k$ and $k + 1$ and the sites up to and including site $k$
are of the form $\mpst{L}{}{}{}{j \le k}$ and the tensors beginning on site
$k + 1$ are of type $\mpst{R}{}{}{}{j \ge k + 1}$. The truncation does not
affect any of the tensors $\mpst{L}{}{}{}{j \le k}$ or
$\mpst{R}{}{}{}{j \ge k + 1}$. Contracting these tensors for the overlap with
their complex conjugated counterparts, we obtain identities on all sites and
$\braket{\psi''}{\psi}$ simplifies to
\begin{eqnarray}
  \braket{\psi'}{\psi}
 &=& \sum_{\alpha_{k-1},\alpha_{k}} \lambda_{\alpha_{k-1}, \alpha_{k}}
     \lambda_{\alpha_{k-1}, \alpha_{k}}' \, ,
\end{eqnarray}
where $\ket{\psi''}$ is the unnormalized truncated state and $\ket{\psi'}$
the truncated normalized state. The diagonal
structure of the matrices $\lambda$ containing the singular values leads
to $\braket{\psi'}{\psi} = \sum_{\alpha} \lambda_{\alpha} \lambda_{\alpha}'$.
Since the smallest singular values in $\lambda'$ are set to zero, the result
is the sum of the squared singular values in $\lambda'$. The additional term
in the denominator in Eq.~\eqref{eq:overlap} originates in the normalization
of $\ket{\psi''}$. We emphasize that this procedure only works if the sites
are completely in the form of $\mpst{L}{}{}{}{j \le k}$ and
$\mpst{R}{}{}{}{j \ge k + 1}$ since otherwise, the contraction with the
complex conjugated tensor does not lead to an identity.

Moreover, we introduce the following actions on tensors in our MPS library:

\begin{itemize}
\item{\textbf{Contractions} over two tensors are defined as the summation
  over one (or more) common indices, and hence generalize matrix-matrix
  multiplication to higher-rank tensors. A commonly used example would be to
  contract two neighboring tensors of an MPS,
  $\mpst{A}{\alpha_{k-1}}{i_k}{\alpha_{k}}{k}$ and
  $\mpst{A}{\alpha_{k}}{i_{k+1}}{\alpha_{k+1}}{k+1}$, to one tensor
  representing the sites $k$ and $k+1$. The summation is in this case over the
  index $\alpha_{k}$ and we obtain a tensor
  $\mpst{\Theta}{\alpha_{k-1}}{i_{k}, i_{k+1}}{\alpha_{k+1}}{k, k+1}$.}
\item{\textbf{Splitting} of a tensor is the reverse action of a contraction.
  The indices of the tensor form two subgroups where the splitting is
  enacted between those two groups. Taking the two~site tensor
  $\mpst{\Theta}{\alpha_{k-1}}{i_{k}, i_{k+1}}{\alpha_{k+1}}{k, k+1}$ as an
  example, we group $\alpha_{k-1}, i_{k}$ together and $i_{k+1}, \alpha_{k+1}$
  in order to obtain two single site tensors, up to a possible
  truncation. The splitting can be achieved via three possibilities:
  \begin{itemize}
  \item[$\circ$]{An SVD splits the tensor directly into two unitary tensors
    and the singular values, described by
    \begin{eqnarray}
      \mpst{\Theta}{\alpha_{k-1}}{i_{k}, i_{k+1}}{\alpha_{k+1}}{k, k+1}
      &=& \mpst{L}{\alpha_{k-1}}{i_{k}}{\alpha_{k}}{k} \lambda_{\alpha_{k}}
          \mpst{R}{\alpha_{k}}{i_{k + 1}}{\alpha_{k + 1}}{k + 1} \, ,
    \end{eqnarray}
    where the singular values $\lambda_{\alpha_{k}}$ allow us to truncate the
    state to a certain $\chi$. The maximal bond dimension is defined as
    $\min\left(\chi_{k-1}d_k, d_{k+1}\chi_{k+1}\right)$.
  }
  \item[$\circ$]{
    The eigenvalue decomposition is related to the singular value
    decomposition, which is the reason the eigenvalue decomposition
    can replace the SVD. If the SVD decomposes $A$ into $U \lambda V$,
    the eigendecomposition $\mathcal{E}(\; \cdot \;)$ is set up as follows:
    \begin{eqnarray}                                                            \label{eq:AAD_ULLUD}
      \mathcal{E}(A A^{\dagger})
      = \mathcal{E} \left(U \lambda V V^{\dagger} \lambda U^{\dagger} \right)
      = \mathcal{E} \left(U \lambda^2 U^{\dagger} \right)
      = U \lambda^2 U^{\dagger} \, .
    \end{eqnarray}
    The eigendecomposition of $A A^{\dagger}$, which is built from a
    matrix-matrix multiplication, returns
    a unitary matrix and the singular values squared. To obtain the
    right matrix, we multiply $U^{\dagger}$ with the original matrix
    $A$ leading to
    \begin{eqnarray}
      U^{\dagger} A = U^{\dagger} U \lambda V = \lambda V \, ,
    \end{eqnarray}
    which already contains the singular values. After completing the
    series of steps, we obtain

    \begin{eqnarray}                                                            \label{eq:ThetaAL}
      \mpst{\Theta}{\alpha_{k-1}}{i_{k}, i_{k+1}}{\alpha_{k+1}}{k, k+1}
      &=& \mpst{L}{\alpha_{k-1}}{i_{k}}{\alpha_{k}}{k}
          \mpst{A}{\alpha_{k}}{i_{k + 1}}{\alpha_{k + 1}}{k + 1} \, ,
    \end{eqnarray}
    where the truncation is possible due to the knowledge of $\lambda^2$ in
    the intermediate step of Eq.~\eqref{eq:AAD_ULLUD}, although the
    singular values do not appear in the previous equation~\eqref{eq:ThetaAL}.
    As in the case of the SVD, the maximal bond dimension is
    $\min(\chi_{k - 1} d_{k}, d_{k+1} \chi_{k + 1})$. The unitary matrix
    can be obtained for the right side starting with $A^{\dagger} A$. This
    procedure is generally faster than the SVD.
  }
  \item[$\circ$]{The QR decomposition decomposes a matrix into a unitary
    matrix and an upper triangular matrix $T$. If the unitary matrix is on the
    right side, it may be referred to as RQ decomposition. It does not allow
    for truncation as the singular values are not calculated. The example
    for the QR is
    \begin{eqnarray}
      \mpst{\Theta}{\alpha_{k-1}}{i_{k}, i_{k+1}}{\alpha_{k+1}}{k, k+1}
      &=& \mpst{L}{\alpha_{k-1}}{i_{k}}{\alpha_{k}}{k}
          \mpst{T}{\alpha_{k}}{i_{k + 1}}{\alpha_{k + 1}}{k + 1} \, .
    \end{eqnarray}
    Therefore, the new bond dimension is the maximal one,
    $\chi = \chi_{\mathrm{max}} = \min(\chi_{k - 1} d_{k},$ $d_{k+1} \chi_{k + 1})$.
    The fact that the QR scenario is not rank revealing is the reason
    for not using it in the splitting of two sites in the library, but it is used
    for shifting as explained in the following.
  }
\end{itemize}
The different options for splitting a tensor are summarized in
Fig.~\ref{fig:splitting}. We choose
the SVD to obtain the singular values and two unitary matrices. In contrast,
the eigenvalue decomposition yields a unitary matrix and the singular values.
The QR decomposition differs from the first two methods as it does not reveal
the singular values and returns only one unitary matrix. Therefore, the QR is
computationally less expensive than the approach with the eigenvalue
decomposition. The SVD is computationally more costly than both other
algorithms.

  \begin{minipage}{0.98\linewidth}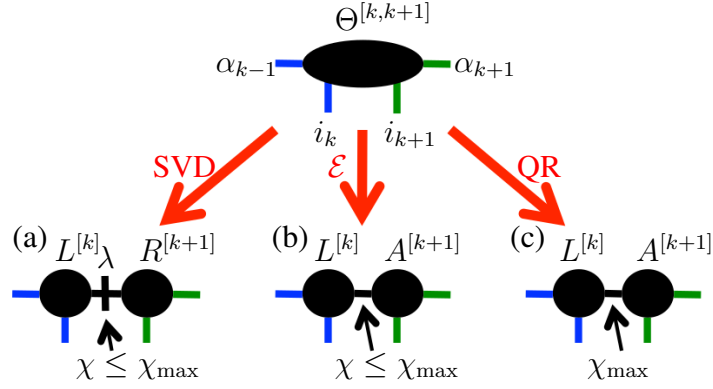
\begin{figure}[H]
    \begin{center}
      \begin{overpic}[width=0.75 \columnwidth,unit=1mm]{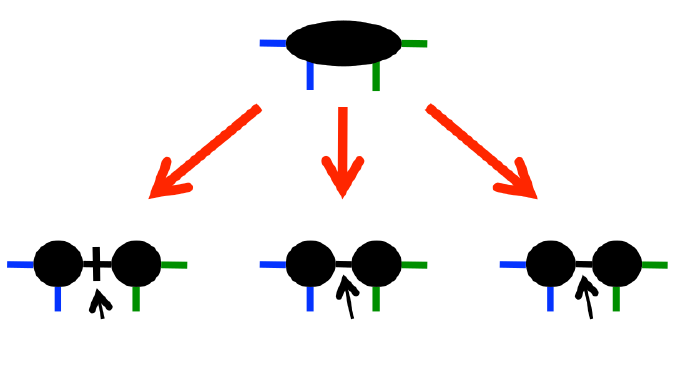}
        \put( 7,20){$L^{[k]}$}
        \put(13,19){$\lambda$}
        \put(10,4){$\chi \le \chi_{\max}$}
        \put(19,20){$R^{[k+1]}$}
        \put(21,32){{\color{red}SVD}}
        \put(44,20){$L^{[k]}$}
        \put(47, 4){$\chi \le \chi_{\max}$}
        \put(54,20){$A^{[k+1]}$}
        \put(46,32){{\color{red}$\mathcal{E}$}}
        \put(79,20){$L^{[k]}$}
        \put(83, 4){$\chi_{\max}$}
        \put(90,20){$A^{[k+1]}$}
        \put(73,32){{\color{red}QR}}
        \put(30,47){$\alpha_{k-1}$}
        \put(44,37){$i_{k}$}
        \put(47,53){$\Theta^{[k,k+1]}$}
        \put(54,37){$i_{k+1}$}
        \put(64,47){$\alpha_{k+1}$}
        \put( 1,22){(a)}
        \put(38,22){(b)}
        \put(72,22){(c)}
      \end{overpic}\vspace{-0.5cm}
      \caption{\emph{Methods for splitting a two site tensor} into two one site
        tensors include (a) an SVD decomposition, (b) an
        eigenvalue decomposition $\mathcal{E}$ in combination with matrix multiplications,
        and (c) a QR decomposition.
                                                                                \label{fig:splitting}}
    \end{center}
  \end{figure}\end{minipage}

}
\item{\textbf{Shifting} the orthogonality center can be done with local
  operations, meaning that the operations act only at one site at a time and do not use
  any two site tensors. Running an SVD or QR (RQ) decomposition for site $k$ on a
  single rank-3 tensor with dimensions $\chi_{k-1}$, $d$, and $\chi_{k}$, we reshape
  the tensor as a $\chi_{l} d \times \chi_r$ ($\chi_l \times d \chi_r$) matrix and
  obtain a left-canonical (right-canonical) unitary tensor for site $k$ and an
  additional matrix. The additional matrix consists of the singular values
  contracted into a unitary matrix when choosing an SVD. For the QR (RQ)
  decomposition we obtain a left (right) canonical unitary and an additional
  upper triangular matrix. This additional matrix can be contracted
  to the corresponding neighboring site $k \pm 1$, resulting in that site
  becoming the new orthogonality center. We note in this case that the QR and RQ
  decomposition does not change the ranks of the matrices, and is roughly a
  factor of two faster than the SVD.}
\end{itemize}

With the knowledge of the basic features of an MPS, we introduce in the
next chapter how a model is defined in the \OSMPS{} library and how we
obtain results as in Fig.~\ref{fig:01_Ising_BE3d}.

\section{Defining systems and variational ground state search                            \label{sec:design}}

\begin{figure}[t]
 \begin{center}
   \vspace{0.7cm}
   \begin{minipage}{0.25\linewidth}
      \vspace{0.99cm}
      \begin{overpic}[width=1.0 \columnwidth,unit=1mm]{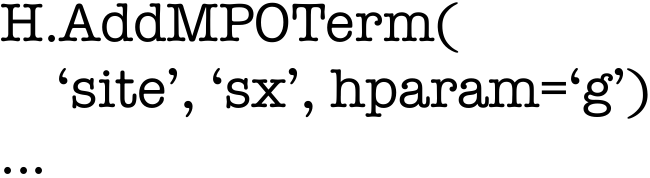}
        \put(2, 78){(a)}
        \put(15,45){{\Large \color{red} Python}}
      \end{overpic}
    \end{minipage}\hfill
    \begin{minipage}{0.33\linewidth}
      \vspace{0.99cm}
      \begin{overpic}[width=1.0 \columnwidth,unit=1mm]{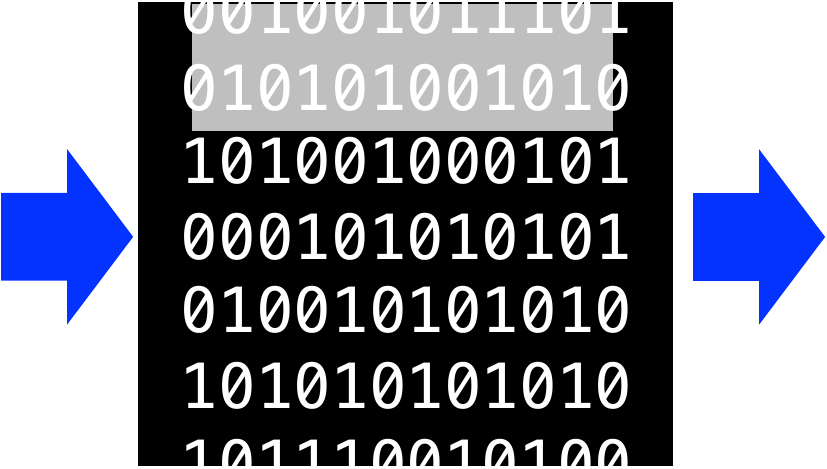}
        \put(8,77){(b)}
        \put(28,45){{\Large \color{red} Fortran}}
      \end{overpic}
    \end{minipage}\hfill
    \begin{minipage}{0.41\linewidth}
      \vspace{0.99cm}
      \begin{overpic}[width=1.0 \columnwidth,unit=1mm]{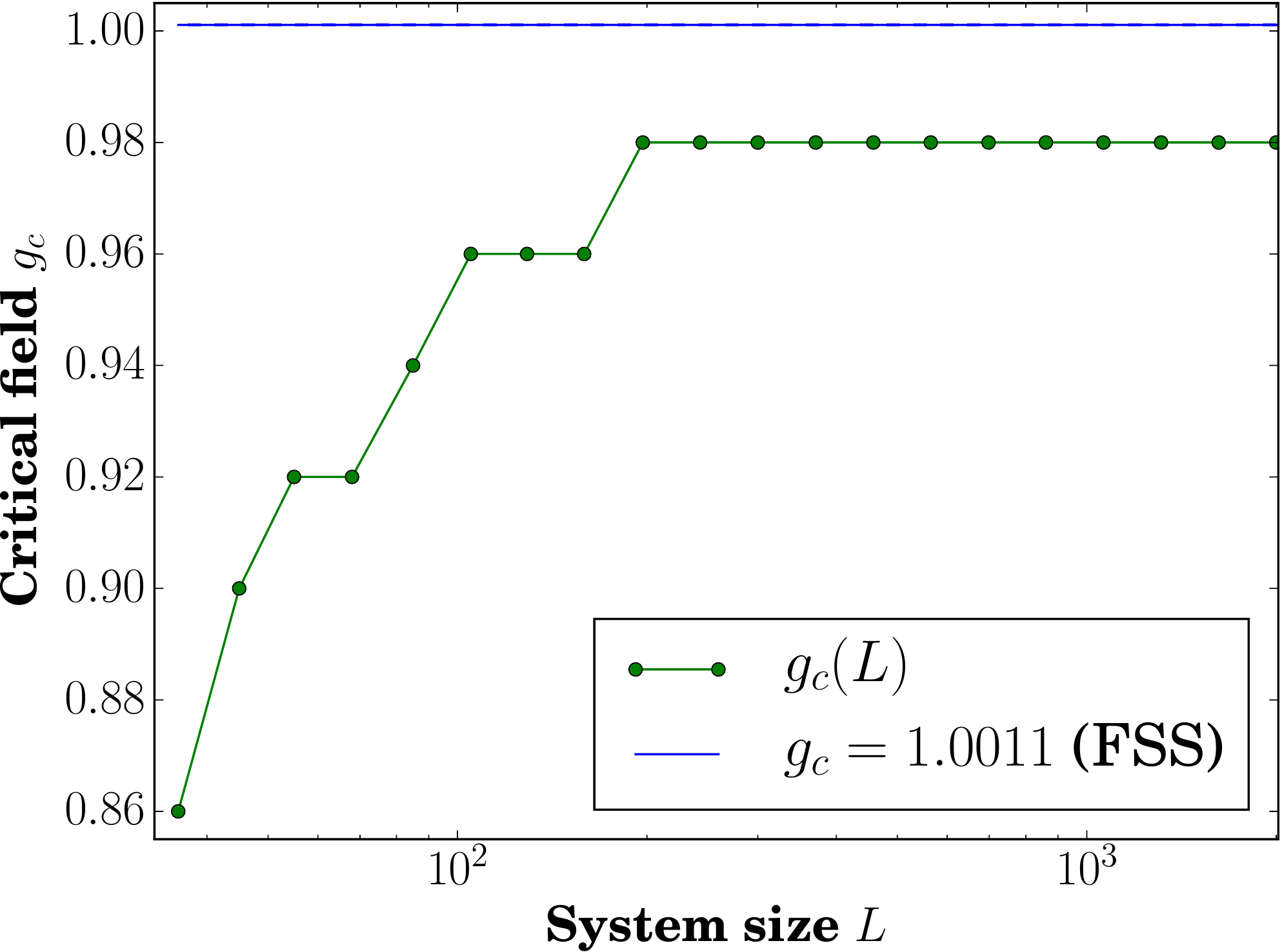}
        \put(2,78){(c)}
        \put(15,45){{\Large \color{red} Python}}
      \end{overpic}\vspace{-0.2cm}
    \end{minipage}
   \caption{\emph{\OSMPS{} flow chart for a simulation.} The \OSMPS{} library combines a
     user-friendly interface in \py{} with a computationally powerful core
     written in \fort{}. (a) The simulation setup is done in \py{}. (b) A write
     function provides the files for \fort{} and a corresponding read function
     imports the results from \fort{} to \py{} (blue arrows). (c) The \py{} front end then
     takes care of the evaluation of the data. The plot in the flow charts
     shows the critical value of the external field in the Ising model as a
     function of the system size $L$ evaluated via the maximum of the bond
     entropy. Finite size scaling (FSS) delivers the critical field in the
     thermodynamic limit for $L \rightarrow \infty$.
                                                                                \label{fig:Interfaces}}
 \end{center}
\end{figure}
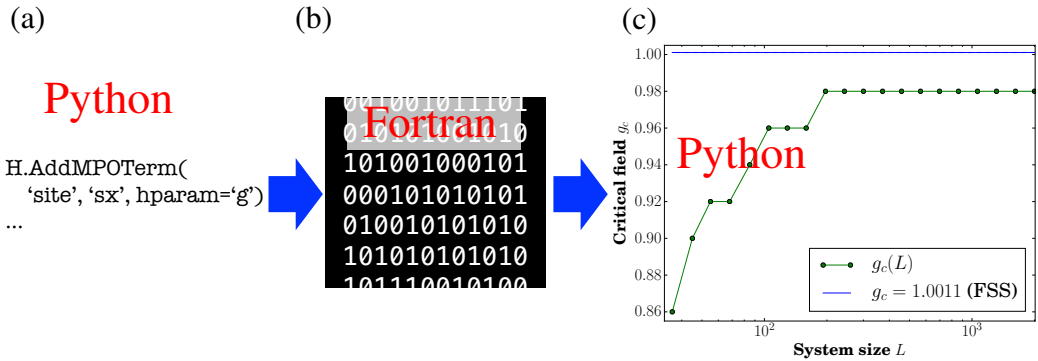

We now outline the definition of systems in \OSMPS{}. As an example we
consider finding the ground state of the finite size quantum Ising model.
The 1D long-range transverse field Ising Hamiltonian is \cite{Ising1925,
Dutta2001}
\begin{eqnarray}                                                                \label{eq:LRIsing}
  H &=& -J \sum_{i<j \le L} \frac{\sigma_{i}^{z} \sigma_{j}^{z}}{(j - i)^{\alpha}}
    - J g \sum_{i=1}^{L} \sigma_{i}^{x} \, ,
\end{eqnarray}
where the operators are defined over the Pauli matrices
$\left\{ \sigma_i^{x}, \sigma_i^{y}, \sigma_i^{z} \right\}$
acting on a site $i$ in the system. The interactions between the spins at
different sites decay
following a power-law introduced in the first term of the Hamiltonian
governed by $\alpha$, the distance $| j - i |$, and the overall energy scale
$J$. The external field is governed by the dimensionless $g$ appearing in the second term of
the Hamiltonian. The number of sites is $L$. We focus in this section on the
nearest neighbor case $H_{\mathrm{NN}}$ obtained for the limit
$\alpha{} \rightarrow{} \infty{}$, commonly called the transverse quantum Ising
model,                                                                          
\begin{eqnarray}                                                                \label{eq:Ising}
  H_{\mathrm{NN}} = -J \sum_{i=1}^{L-1} \sigma_{i}^{z} \sigma_{i+1}^{z}
    - J g \sum_{i=1}^{L} \sigma_{i}^{x} \, .
\end{eqnarray}
The overall approach to \OSMPS{} is a user-friendly \py{} environment
calling a \fort{} core for the actual calculations. This scheme
is depicted in Fig.~\ref{fig:Interfaces}. Thus, we guide the reader through the
simulation with a corresponding summary of the \py{} files; the
complete files are contained in supplemental material, see
Appendix~\ref{app:suppl}.

From a quantum mechanical point of view the following steps are necessary
to describe a system. First, we have to generate the operators which are acting
on the local Hilbert space, described in Sec.~\ref{sec:subops}. Once we have
the operators, we
build the Hamiltonian out of rule sets in Sec.~\ref{sec:subham}. Then,
we set up the measurements to be carried out in Sec.~\ref{sec:subobs}. This
procedure completes the definition of the quantum system, but we have two more
tasks with regards to the numerics. In the fourth step we define the convergence
parameters of the algorithm, where Sec.~\ref{sec:variational} describes this
step for the variational ground state search. Finally, the simulation is set
up and executed in Sec.~\ref{sec:setupandrun}.

One general comment remains before starting with a detailed description
of the simulation setup. Every simulation in \OSMPS{} is represented by a
\py{} dictionary, which contains observables and convergence
parameters as well as general parameters such as the system size.

\subsection{Operators                                                          \label{sec:subops}}

\OSMPS{} comes with predefined sets of operators for three different physical
systems to facilitate the setup of simulations. These predefined sets of
operators are returned by the corresponding functions for bosonic systems
such as the Bose-Hubbard model, fermionic systems including their phase
operators originating from the Jordan-Wigner transformation, or spin
operators. We use the last set in the example for the quantum Ising model.
The corresponding function \pyfunc{BuildSpinOperators} returns the set of
operators $\{ \sigma^{+}, \sigma^{-}, S^{z}, \1 \}$. In order to obtain
the Pauli operators $\sigma^{x}$ and $\sigma^{z}$ from the spin lowering and
raising operators $\sigma^{\pm}$ and the spin operator $S_{z}$, we suggest
following the prescription in Listing~\ref{py:01_Ising_op}.

\pyfrag{./code/01_Ising.py}
       {Defining the operators of the quantum Ising model overwriting $S^Z$
        with $\sigma^Z$.}
       {py:01_Ising_op}{38}{44}

\subsection{Hamiltonians                                                       \label{sec:subham}}

Through these operators, stored in a dictionary-like \py{} class, we define
the Hamiltonian and later on the observables. The Hamiltonian is described
as a matrix product operator (MPO), which is effectively an MPS
with rank-four tensors instead of rank-three tensors:
\begin{eqnarray}                                                                 \label{eq:mpo}
  H = \sum_{i_1, \ldots, i_L} \sum_{i_1', \ldots, i_L'}
      \sum_{\alpha_{1}, \ldots, \alpha{L-1}} \! \! \!
      \mpot{M}{}{i_1'}{i_1}{\alpha_1}{1}
      \mpot{M}{\alpha_1}{i_2'}{i_2}{\alpha_2}{2} \cdots
      \mpot{M}{\alpha_{L-1}}{i_L'}{i_L}{}{L}
      \ket{i_1'} \bra{i_{1}} \cdots \ket{i_L'} \bra{i_{L}}
      \, .
\end{eqnarray}
A key property is that an MPO acting on an MPS can be written as another
MPS, generally with a larger bond dimension. The physical indices $i_{k}$
act on the physical indices of an MPS and
take them to new physical indices $i_{k}'$. The auxiliary
indices $\alpha$ of the MPO are fused with the corresponding auxiliary
indices in the MPS. For most physical operators, this structure of rank-4
tensors is sparse and therefore we rather seek for an efficient
implementation in terms of MPO-matrices $M^{[k]}$ than in rank-4 tensors.
The MPO matrix $M^{[k]}$ including the iteration over all indices for one
site representing the rule sets~\cite{Crosswhite2008,Wall2012} for local terms,
bond terms for the interaction of nearest neighbors, and exponential rules
for long-range interactions between two sites takes the form
\begin{eqnarray}
  \label{eq:MPOdef} M^{[k]} \! \! \! &=& \! \! \! \begin{pmatrix}
    A_{\alpha_{k-1}=1,\alpha_{k}=1}^{[k]} & 0                                     & 0                                     & 0 \\
    A_{\alpha_{k-1}=2,\alpha_{k}=1}^{[k]} & A_{\alpha_{k-1}=2,\alpha_{k}=2}^{[k]} & 0                                     & 0 \\
    A_{\alpha_{k-1}=3,\alpha_{k}=1}^{[k]} & 0                                     & A_{\alpha_{k-1}=3,\alpha_{k}=3}^{[k]} & 0 \\
    A_{\alpha_{k-1}=4,\alpha_{k}=1}^{[k]} & A_{\alpha_{k-1}=4,\alpha_{k}=2}^{[k]} & A_{\alpha_{k-1}=4,\alpha_{k}=3}^{[k]} & A_{\alpha_{k-1}=4,\alpha_{k}=4}^{[k]}
  \end{pmatrix} \, ,
\end{eqnarray}
where $k = 2, 3, \ldots, L-1 $.
The matrix structure in Eq.~\eqref{eq:MPOdef} corresponds to the auxiliary
indices. Each element within this structure $A_{\alpha_{k-1}, \alpha_{k}}$
is a matrix acting on the local Hilbert space of site $k$, e.g. the Pauli
matrix $\sigma_{k}^{z}$. Thus the auxiliary indices $\alpha$ of
the rank-4 tensor encode the row and column of $M^{[k]}$, while the
indices $i_{k}$ and $i_{k}'$ are related to the local Hilbert space are
located in the rows and columns of the matrices $A_{\alpha_{k-1}, \alpha_{k}}$.
We now illustrate the meaning of the matrices depending on their position in
$M^{[k]}$. In order to build the MPO for the Hamiltonian for the long-range
Ising model, we only need the first column, last row, and the diagonal of
$M^{[k]}$ and store it as a sparse structure. Matrices in the first column
(last row) of $M^{[k]}$ are multiplied with identity operators on the right
(left) side of site $k$, i.e., they do not interact with any sites right
(left) of themselves.
Diagonal elements propagate operators through the system and are completed
by other operators to the left and right of site $k$, and hence represent
long-range interactions. For the first and last site we define the MPO-matrix
as vectors
\begin{eqnarray}
  M^{[1]} &=& \begin{pmatrix}
    A_{1,1}^{[1]} & A_{1,2}^{[1]} & A_{1,3}^{[1]} & A_{1,4}^{[1]}
  \end{pmatrix} \, , \quad
  M^{[L]} = \begin{pmatrix}
    A_{1,1}^{[L]} \\ A_{2,1}^{[L]} \\ A_{3,1}^{[L]} \\ A_{4,1}^{[L]}
  \end{pmatrix} \, ,
\end{eqnarray}
corresponding to the auxiliary rank-one structure for the MPO matrices
at the boundary in Eq.~\eqref{eq:mpo}. Note
that the first MPO matrix is a row vector and the last is a column vector,
resulting in the contracted MPO object Eq.~\eqref{eq:mpo} being a $1\times 1$
matrix in the auxiliary indices.

%
%
\begin{figure}[t]
  \begin{center}
    \begin{overpic}[width=0.65 \columnwidth,unit=1mm]{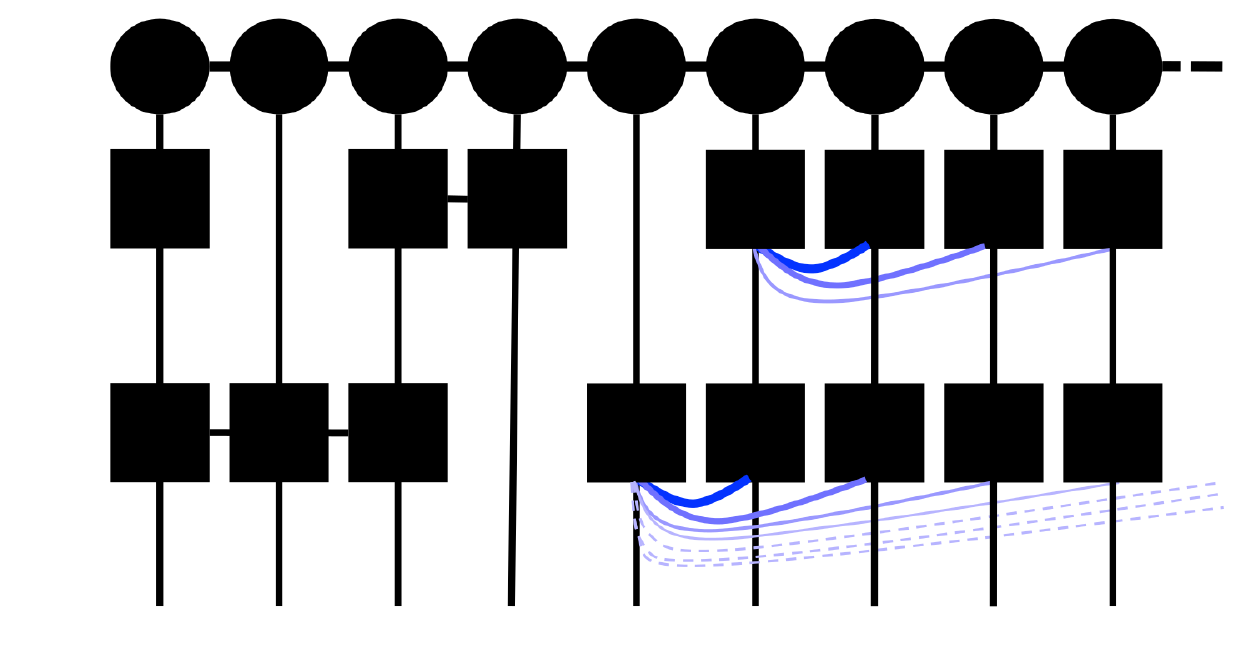}
      \put( 5,42){(a)}
      \put(25,42){(b)}
      \put(53,42){(c)}
      \put( 5,24){(d)}
      \put(43,24){(e)}
      \put(73.5,27){{\color{blue}$f(|i-j|, \; O_1, O_2)$}}
      \put(50, 0){{\color{blue}$f(|i-j|, O_1, O_2)$}}
      \put(11,17.5){{\color{white}$O_1$}}
      \put(21,17.5){{\color{white}$O_2$}}
      \put(31,17.5){{\color{white}$O_3$}}
    \end{overpic}
    \caption{\emph{Rules for building a Hamiltonian.} (a) Local terms;
      (b) bond terms acting on two neighboring sites; (c) finite range terms
      of two operators $O_{1}, O_{2}$ and a coupling depending
      on the distance. The coupling function $f(x)$ is a finite sum of a
      limited number of neighboring sites. (d) String of arbitrary operators,
      e.g., a three-body term built from $O_{1}$, $O_{2}$, and $O_{3}$ and
      (e) infinite terms of two operators $O_1$ and $O_2$ with a distance
      depending on decaying coupling either as a general function
      $f(|j - i|)$ (InfiniteFunction) or as an exponential (Exponential).
      Any infinite function is expressed as a sum of exponentials within
      \OSMPS{}. The coupling function $f(x)$ is extended to all sites.
                                                                                \label{fig:Modeling}}
  \end{center}
\end{figure}
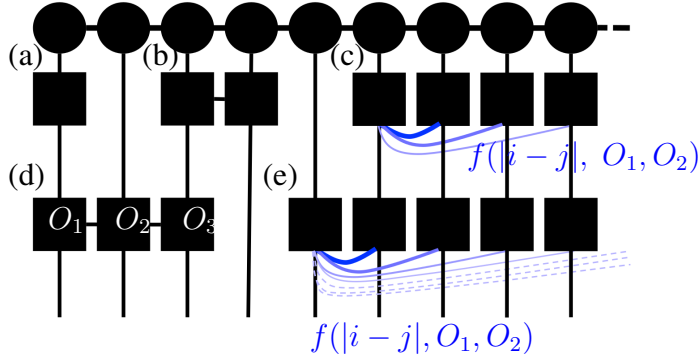

We now build the nearest neighbor Hamiltonian $H_{\mathrm{NN}}$ from
Eq.~\eqref{eq:Ising} for the quantum Ising model to continue with our
example. Figure~\ref{fig:Modeling} shows the possible rule sets provided through
\OSMPS{}. We need the local site term depicted in Fig.~\ref{fig:Modeling}(a) and
the bond term from Fig.~\ref{fig:Modeling}(b). In general these operators are
filled with identities on all other sites and act on each possible site. The
corresponding MPO-matrices depending on the site index $k$ are then
\begin{eqnarray}
  M^{[1]} &=& \begin{pmatrix} -J g \sigma_{1}^{x} & - J \sigma_{1}^{z} & \1
  \end{pmatrix} \, , \; M^{[k]} = \begin{pmatrix}
    \1                & 0                 & 0 \\
    \sigma_{k}^{z}    & 0                 & 0 \\
    -J g \sigma_{k}^{x} & -J \sigma_{k}^{z} & \1
  \end{pmatrix} \, ,                                                            \nonumber \\
  M^{[L]} &=& \begin{pmatrix}
    \1 \\ \sigma_{L}^{z} \\ -J g \sigma_{L}^{x} \end{pmatrix} \, ,
\end{eqnarray}
where $k = 2, 3, \ldots, L-1$.
To see how this MPO structure results in the proper many-body operator, we
will explicitly build the Hamiltonian for three sites, i.e.,
$H = M^{[1]} \times M^{[2]} \times M^{[3]}$, where $\times$ is understood
to be ordinary matrix multiplication in the auxiliary indices together with
tensor products on the physical indices. We start by multiplying the row
vector for the first site with the matrix for the second site leading to
the first line of the following equation.
%
%
The multiplication of this result with the column vector for the last and
third site results then in the Hamiltonian
\begin{eqnarray}
  M^{[1]} \times M^{[2]} \times M^{[3]}
 &=& \begin{pmatrix}
    \left(- J g \sigma_{1}^{x} - J \sigma_{1}^{z} \sigma_{2}^{z} - J g \sigma_{2}^{x} \right)&
    - J \sigma_{2}^{z} & \1 \end{pmatrix} \times \begin{pmatrix}
    \1 \\ \sigma_{3}^{z} \\ -J g \sigma_{3}^{x} \end{pmatrix}                     \nonumber \\
 &=& - J g \sigma_{1}^{x} - J \sigma_{1}^{z} \sigma_{2}^{z} - J g \sigma_{2}^{x}
    - J \sigma_{2}^{z} \sigma_{3}^{z} - J g \sigma_{3}^{x} \, .                   
\end{eqnarray}                                                                  
The MPO matrices have more entries for models beyond nearest neighbor
interactions. The local terms remain in the last row of the first
column as the identities stay in their places. The diagonal is set in the
case of long-range interactions with an identity times a decay factor.
The larger the distance between two sites becomes, the higher the
contribution of the decay multiplied at each site in between. Elements
in the lower triangular part of the matrix besides the first column and
last row are used e.g. in the FiniteTerm rule set.

Independent of the system, we first initialize an instance of the
\pyfunc{MPO} class. The different types of terms are specified via a string
argument in the class function \pyfunc{AddObservable}. Keyword arguments to
any MPO terms are the weight and Hamiltonian parameters \texttt{hparam}.
Further arguments specific for the rule can be found in the documentation,
e.g. the infinite function can take the function as an additional keyword
argument.

\pyfrag{./code/01_Ising.py}
       {Defining the Hamiltonian of the quantum Ising model.}
       {py:01_Ising_ham}{49}{53}

In the code we have given a string variable name for the coupling of the bond
and site term. The energy scale (\pyline{J=1}) and the different values for
$h$ are specified later in the \py{} setup script inside the dictionary
representing the simulation allowing for flexibility.

\subsection{Observables                                                        \label{sec:subobs}}

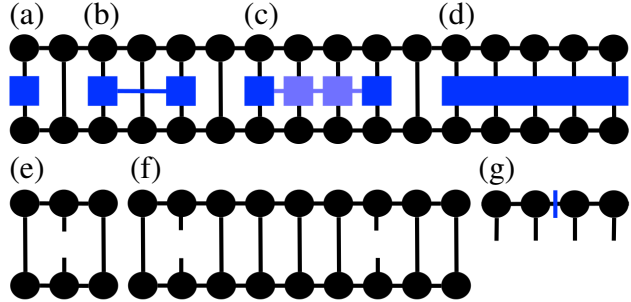
\begin{figure}[t]
  \begin{center}
    \begin{overpic}[width=0.65 \columnwidth,unit=1mm]{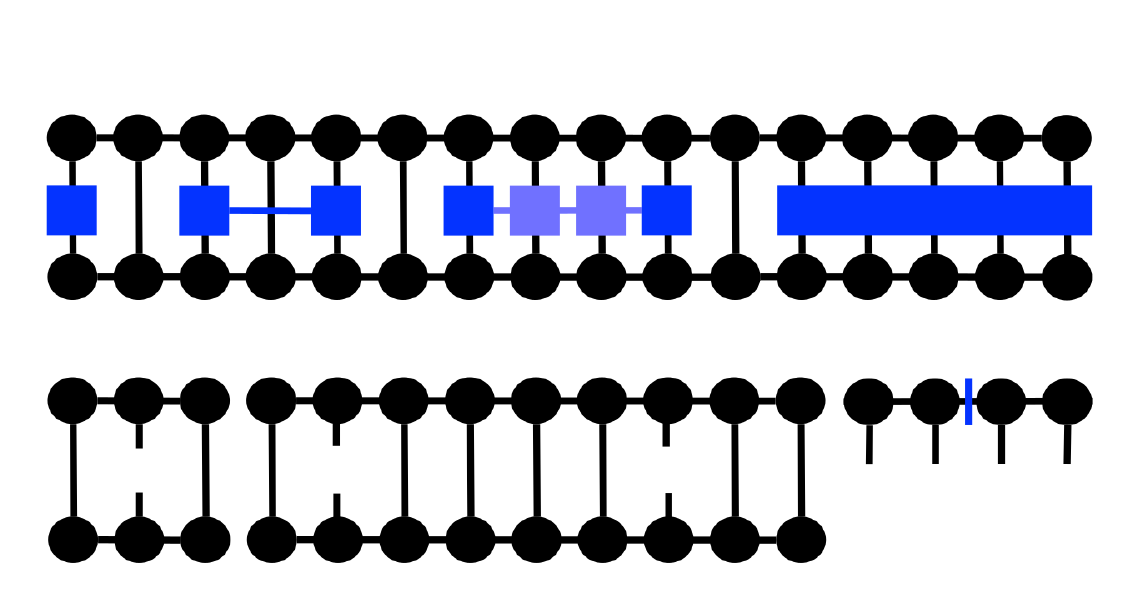}
      \put( 4,44){(a)}
      \put(15,44){(b)}
      \put(39,44){(c)}
      \put(68,44){(d)}
      \put( 4,21){(e)}
      \put(22,21){(f)}
      \put(74,21){(g)}
    \end{overpic}
    \caption{\emph{\OSMPS{} measurements} can be selected from the following
      options: (a) Local terms. (b) Two-site correlators including
      correlations for fermionic systems. (c) String operators of
      type $\langle O_{A} O_{B} O_{B} \ldots O_{B} O_{B} O_{C} \rangle$.
      (d) MPO as used in the default measurement of the energy
      (Hamiltonian). (e) Single site density matrices tracing over
      the remaining system. (f) Two-site density matrices tracing
      out everything but two sites as defined in Eq.~\eqref{eq:reducedrho}.
      (g) Singular values between left and right part of the MPS.
                                                                                \label{fig:Measuring}}
  \end{center}
\end{figure}

In order to evaluate the behavior of the system, we have to define the
observables. Figure~\ref{fig:Measuring} shows the possible measurements:
local site terms including site and bond entropy, correlations, MPOs,
string operators and one or two-site reduced density matrices where the
reduced density matrices $\rho_{i}$ and $\rho_{i,j}$ are defined as
\begin{eqnarray}                                                                \label{eq:reducedrho}
  \rho_{i} = \tr_{k \neq i}\left(\rho \right) \, , \qquad
  \rho_{i, j} = \tr_{k \neq i,j}\left(\rho \right) \, ,
\end{eqnarray}
where the density matrix on the complete system is defined as $\rho = \ket{\psi}
\bra{\psi}$.
The energy as an MPO measurement of the Hamiltonian, the bond dimension, the
variance within variational algorithms, or the overlap between the initial state
and the time evolved state (Loschmidt echo) are measured by default. For the Ising
example, we measure $\langle \sigma_i^{z} \rangle$ and
$\langle \sigma_i^{z} \sigma_j^{z} \rangle$. Due to the local observable
we gain as well the bond entropy shown in Fig.~\ref{fig:01_Ising_BE3d}.
The following Listing~\ref{py:01_Ising_obs} shows the necessary code for
measuring these observables.

\pyfrag{./code/01_Ising.py}
       {Defining the observables of the quantum Ising model.}
       {py:01_Ising_obs}{71}{78}

\subsection{Fundamentals of the library: Variational ground state search       \label{sec:variational}}

The previous steps completed the setup of the physical system, and we continue
with  the specification of the convergence parameters. Therefore, we explain
the variational algorithm used to find the ground state which serves as
input for the algorithms for excited state search and real time evolution.

\begin{figure}[t]
  \begin{center}
    \begin{overpic}[width=0.80 \columnwidth,unit=1mm]{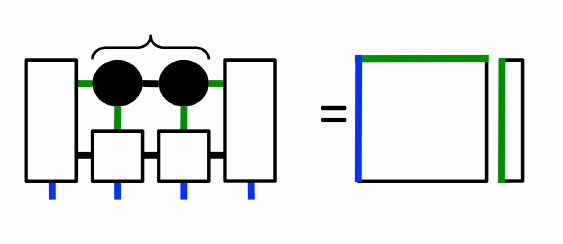}
      \put( 8,21){L}
      \put(19.5,14.5){H}
      \put(31,14.5){H}
      \put(43,21){R}
      \put(71,22){LHHR}
      \put(26,39){$\psi$}
      \put(90,22){$\psi$}
      \put(65,18){$\in \mathbb{C}^{\chi^2 d^2 \times \chi^2 d^2}$}
      \put(94,22){$\in \mathbb{C}^{\chi^2 d^2}$}
    \end{overpic}\vspace{-1cm}
    \caption{\emph{Effective Hamiltonian for the Lanczos algorithm} is built
      via the contractions of all MPO matrices with tensors for the sites
      $k' \neq k, k+1$.
                                                                                \label{fig:LanczosMPS}}
  \end{center}
\end{figure}
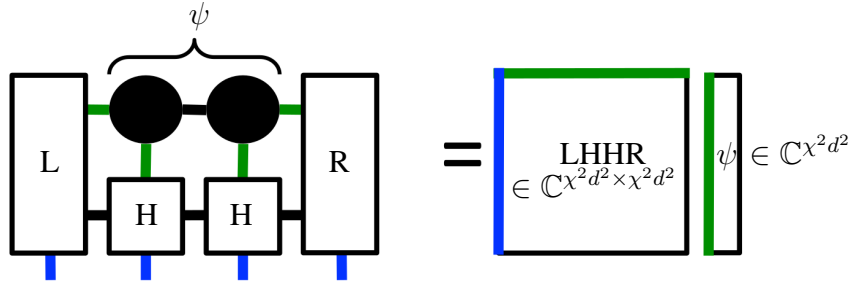

From exact diagonalization, we know how to find the ground state via solving
the eigenequation, which is optimally done with sparse methods such as the
Lanczos algorithm~\cite{Golub_VanLoan_96}. The same procedure cannot be used
in the same way beyond a few tens of particles due to the exponentially growing
Hilbert space in the many-body system, but the variational ground state search
adapts the eigenvalue problem to an effective eigenvalue problem for a few
neighboring sites. In principle, the eigenequation
$H | \psi \rangle = E | \psi \rangle$ can be solved for the ground state on
the complete Hilbert space using imaginary time evolution, e.g. with the
Krylov method presented for the dynamics later, using the equation
$\e^{- \beta H}$ with $\beta \rightarrow \infty$. The thermodynamic beta
approaches infinity as the system approaches the ground state at zero
temperature. Instead of searching
for the global minimum, we seek for local minima transferring the problem
to an effective eigenequation for $n$ neighboring sites in the \OSMPS{}
algorithm. The other $L - n$ sites are kept fixed while finding the effective
ground state of the $n$ sites. This effective eigenproblem does not grow
exponentially with the system size, but depends on the local dimension and the
bond dimension of the constant parts of the system, i.e. $d^n \chi^2$. The
number of these effective eigenvalue problems grows linearly with the system
size. In the following for simplicity we set $n = 2$, which corresponds to
the value used in \OSMPS{}:
\begin{eqnarray}
  \varepsilon[| \psi \rangle ] =
  \langle \psi | H | \psi \rangle - E \langle \psi | \psi \rangle \quad
  \overset{\mathrm{local}}{\underset{\mathrm{global}}\rightleftharpoons} \quad
  H_{\mathrm{eff}} A^{[k, k+1]} = E A^{[k, k+1]} \, .
\end{eqnarray}
The local minimization over $n$ neighboring sites is done iteratively moving
through the neighboring pairs of sites until convergence
is reached (see details in Appendix~\ref{app:convergence}). We point out the
role of the effective Hamiltonian in more detail with regards to the Lanczos
algorithm. The Lanczos algorithm finds the eigenvalues and vector of a problem
using only matrix vector multiplications, in our case $H \ket{v}$ for some
vector $\ket{v}$. Because we restrict ourselves to the sites $k, k+ 1$,
the tensors of the other sites remain constant and we can contract them
with their MPO matrices.
These fixed sites form an environment which acts as part of the
total matrix vector multiplication. The contraction can be continued until
we only have one tensor $L$ to the left and one tensor $R$ to the right
representing those contractions.\footnote{In this context the symbol $L$
represents the left tensor and not the system size.}
Together with the MPO matrices $M^{[k]}$ and $M^{[k + 1]}$, the tensors
$L$ and $R$ build the effective Hamiltonian. We now find the
minimum in energy for this effective Hamiltonian with regards to sites $k$ and
$k + 1$. This effective Hamiltonian is resumed in Fig.~\ref{fig:LanczosMPS}.
%
In the case of matrices, the Lanczos algorithm is ideal for sparse problems
and calculating only a few eigenvectors. In the tensor network scenario it
provides a considerable speedup in contrast to dense methods due to the
tensor network structure: contracting the tensors $L$, $R$, and the
MPO matrices $\mpot{M}{}{}{}{}{k}$ and $\mpot{M}{}{}{}{}{k+1}$
step-by-step
to $\ket{v}$ is more efficient than building $H_{\mathrm{eff}}$ of dimension
$\chi^2 d^2 \times \chi^2 d^2$ and multiplying it with $\ket{v}$ or solving
the eigenvalue problem, i.e. $\mathcal{O}(\chi^3)$ versus
$\mathcal{O}(\chi^6)$ \cite{Wall2012}. The two-site eigenvalue problem
corresponds to finding the stationary point of the energy functional through
the equation
\begin{eqnarray}
  \frac{\partial}{\partial \left( A^{[k, k+1]} \right)}
  \left( \langle \psi | H | \psi \rangle -
  E \langle \psi | \psi \rangle \right) = 0 \, .                                \label{eq:mps_lagrange}
\end{eqnarray}
where $E$, the energy eigenvalue, is a Lagrange multiplier enforcing
normalization, and the derivative with respect to a tensor is defined to be a
tensor of the same shape whose elements are the derivatives with respect to
the individual tensor elements.

\pyfrag{./code/01_Ising.py}
       {Defining the two sets of convergence parameters for the quantum Ising
        simulation.}
       {py:01_Ising_conv}{84}{90}


The key to obtaining meaningful results are the convergence parameters of
the variational algorithm. The convergence parameters are stored in a
corresponding \py{} object which is shown in Listing~\ref{py:01_Ising_conv}
and the different parameters are defined in the following. For example,
the variance $V_{\psi} = \langle H^2 - \langle H \rangle^2 \rangle$ indicates
the distance from an eigenstate. Table~\ref{tab:conv_mps} presents out of the
analysis in Appendix~\ref{app:convergence} the parameters to obtain ground
states with a variance tolerance $\pevar = 10^{-12}$,
effectively $L \times 10^{-12}$ for the whole system,
for different models. Here, we concentrate on the values of the Lanczos
tolerance $\pelanc$ and bond dimension where other parameters are kept constant.
Those are especially interesting because the bond dimension defines the
fraction of the Hilbert space which can be captured. On the other hand, the
Lanczos tolerance determines the accuracy of the eigenvector solved in
Eq.~\eqref{eq:mps_lagrange}. The parameters
kept constant are the number of Lanczos iterations. If the number of Lanczos
iterations is sufficiently high the accuracy of the Lanczos tolerance is met,
otherwise not. The local tolerance $\peloc$, defining the cutoff of the singular
values in the Schmidt decomposition of Eq.~\eqref{eq:schmidt}, guarantees
that we do not use the full bond dimension if the sum of the singular values
squared are below the local tolerance. It relates to the variance
tolerance and defaults to
\begin{eqnarray}
  \peloc = \frac{\pevar}{4L} \, .
\end{eqnarray}
The motivation to choose this value is that the local error made during one sweep
consists of the approximately $2 L$ splittings, where the additional factor of
two is a safety factor to ensure good convergence. The number of sweeps through
the system, optimizing each pair of two sites twice, is specified with the
number of inner sweeps $N_{\mathrm{i}}$, which is bounded between
\texttt{min\_}\texttt{num\_}\texttt{sweeps} and \texttt{max\_}\texttt{num\_}
\texttt{sweeps}, and $N_{\mathrm{o}}$, the parameter for the outer sweeps
\texttt{max\_}\texttt{outer\_}\texttt{sweeps}. The maximal number of overall
sweeps $N_{\mathrm{sweep}}$ is then
\begin{eqnarray}
  N_{\mathrm{sweep}} = N_{\mathrm{i}} \cdot N_{\mathrm{o}} \, .
\end{eqnarray}
Convergence is checked after every inner sweep. One outer sweep is completed
after the set of $N_{\mathrm{i}}$ inner sweeps followed by an adjustment of the local
tolerances. The new local tolerance $\peloc'$ is decreased according to
\begin{eqnarray}                                                                \label{eq:linearExtrap}
  \peloc'
  = \peloc \frac{\pevar}{V_{\psi}} \, ,
\end{eqnarray}
where $V_{\psi}$ is the actual variance on the current MPS.
Equation~\eqref{eq:linearExtrap} assumes a linear
connection between the local tolerance and the variance fulfilled for small
local tolerances \cite{McCulloch2007}. Moreover, we have two more
parameters to grow the system up to $L$ sites with the same algorithm later
explained in Sec.~\ref{sec:imps} for the infinite system. The local tolerance
(\pyline{warmup\_tol}) and the maximal bond dimension
(\pyline{warmup\_bond\_dimension}) during that warmup phase can be tuned
individually. These values provided in Table~\ref{tab:conv_mps}
provide a first overview of how models behave within \OSMPS{}. We choose
points with high entanglement within each model. The parameters are either
close to a critical point or in a phase which has high entanglement such as
the superfluid phase of the Bose-Hubbard model.

\begin{table}[t]
  \centering
  \begin{tabular}{@{} ccccc @{}}
    \toprule
    Parameter
    & QI
    & LRQI
    & Bose
    & Fermi\\
    \cmidrule(r){1-1} \cmidrule(rl){2-2} \cmidrule(rl){3-3} \cmidrule(rl){4-4}
    \cmidrule(l){5-5}
    $\chi$                    & $46$    & $100$   & $261$   & $177$              \\
    $\pelanc$                 & $2.6 \cdot 10^{-9}$         & $2.15 \cdot 10^{-8}$
                              & $1.1 \cdot 10^{-10}$    & $5.1 \cdot 10^{-7}$  \\
    Number of inner sweeps    & 2       & 2             & 2       & 4           \\
    Number of outer sweeps    & 1       & 1             & 1       & 1           \\
    System size $L$           & 128     & 128           & 32      & 65          \\
    Lanczos iterations        & 500     & 500           & 500     & 500         \\
    \bottomrule
  \end{tabular}
  \caption{\emph{Empirically Determined Convergence Parameters.} The
    convergence parameters for the bond dimension $\chi$ and
    Lanczos tolerance $\pelanc$ for four different models achieve a variance
    tolerance $10^{-12}$ using a state with high entropy. The quantum Ising model (QI)
    and quantum long-range Ising model (LRQI) are evolved close to the critical point. The
    Bose-Hubbard model (Bose) is considered in the superfluid phase and a spinless
    Fermi model (Fermi) with nearest neighbor repulsive interaction $W$ and
    nearest neighbor tunneling $J$ is again in a region with high entanglement with
    $J = 1.04$. Details on the study are in Appendix~\ref{app:convergence}.
                                                                                \label{tab:conv_mps}}
\end{table}

\begin{figure}[t!]
  \begin{center}
    \begin{minipage}{0.3\linewidth}
      \begin{overpic}[width=1.0\columnwidth,unit=1mm]{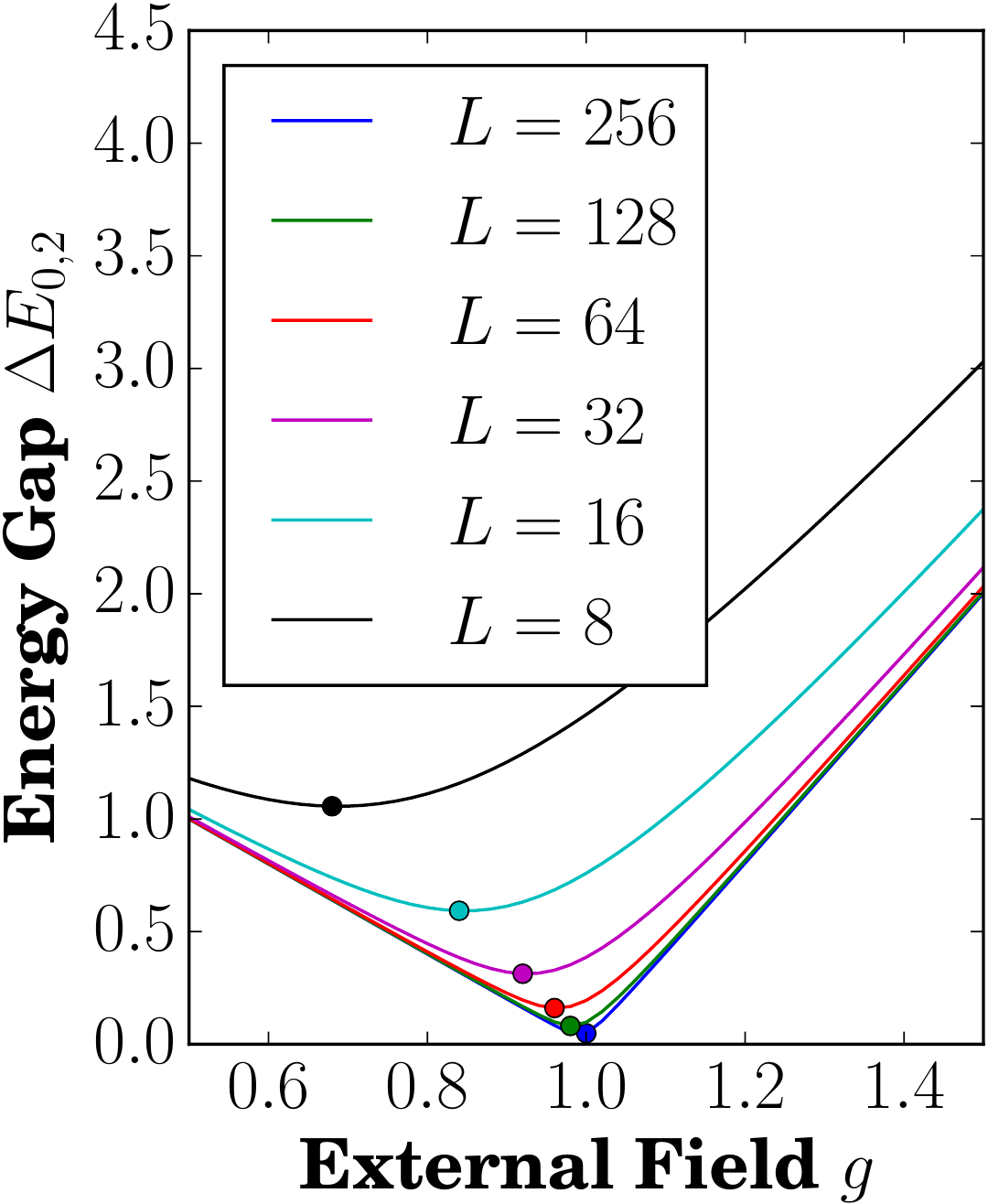}
        \put( 1,102){(a)}
      \end{overpic}
    \end{minipage}\hfill
    \begin{minipage}{0.3\linewidth}
      \begin{overpic}[width=1.0 \columnwidth,unit=1mm]{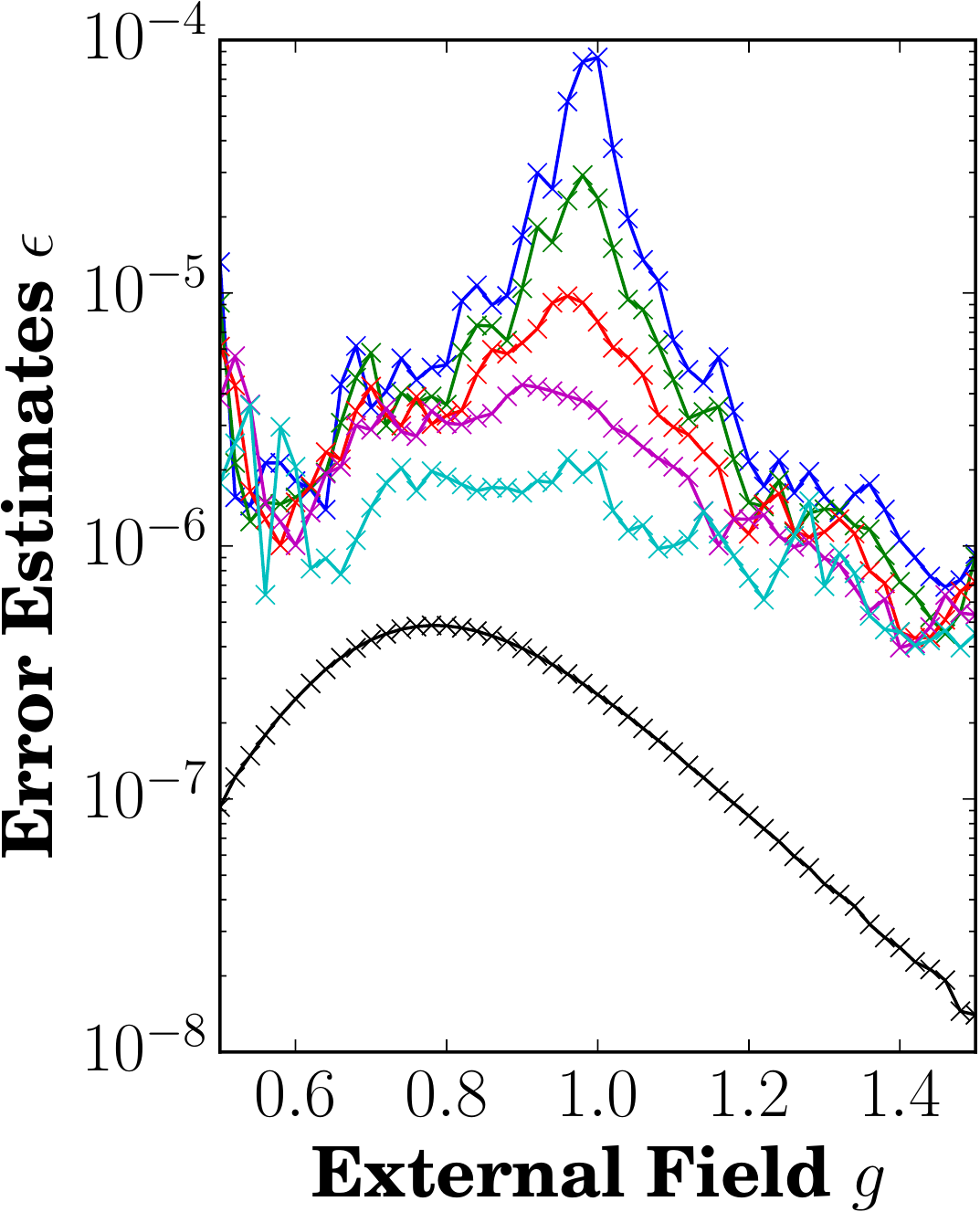}
        \put( 1,102){(b)}
      \end{overpic}
    \end{minipage}\hfill
    \begin{minipage}{0.3\linewidth}
      \begin{overpic}[width=1.0\columnwidth,unit=1mm]{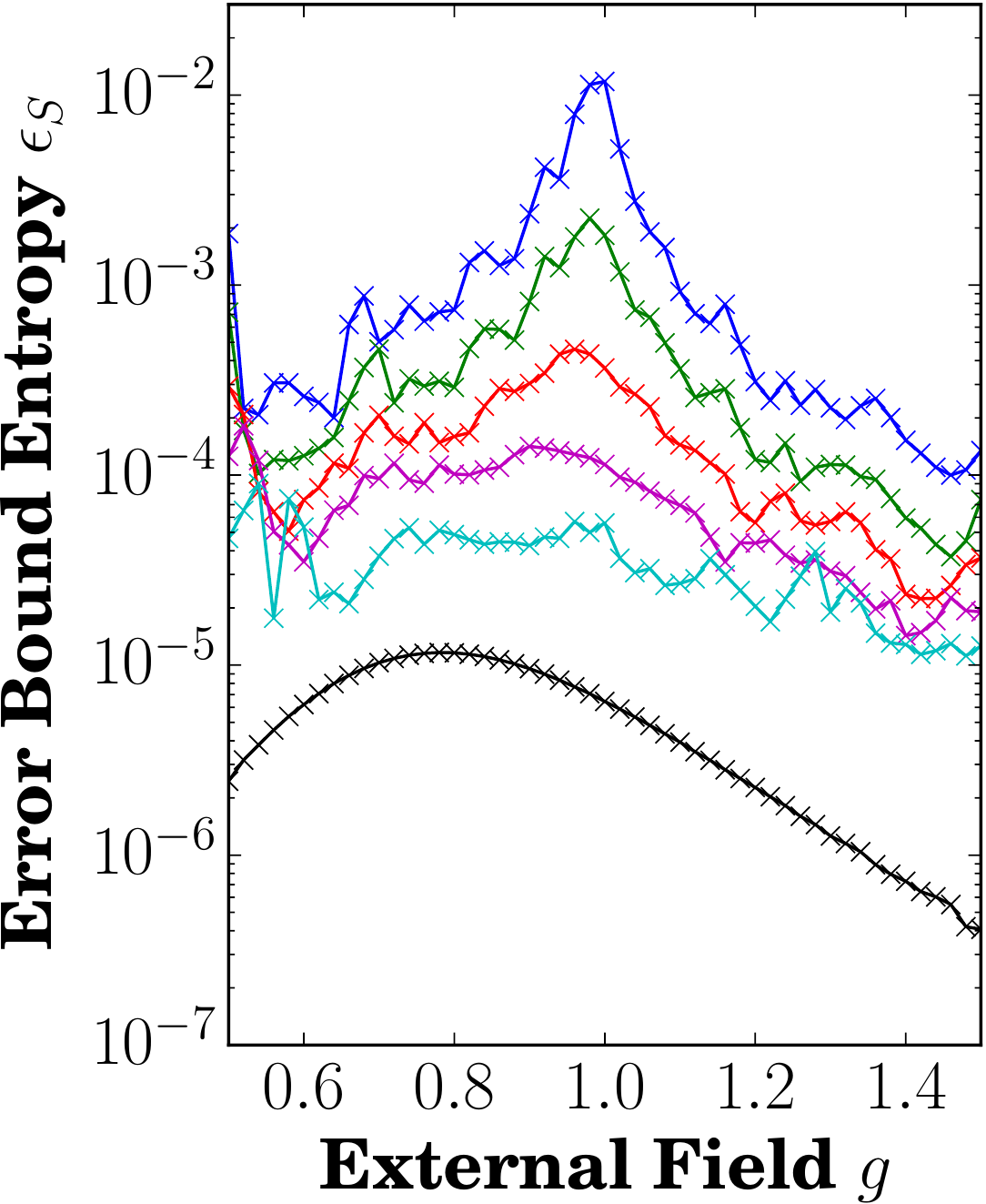}
        \put( 1,102){(c)}
      \end{overpic}
    \end{minipage}
    \caption{\emph{Error bounds applied to the quantum Ising model.} We
      consider the Ising model with $\mathbb{Z}_2$ symmetry for the
      demonstration of error bounds.
      (a) The energy gap $\Delta E_{0,2}$ between the ground state and the
      second excited is shown, as the error depends on the inverse of the
      gap. The gap decreases towards the critical points and in the
      thermodynamic limit $L \to \infty$. The first excited state is located in
      the odd symmetry sector and therefore not relevant for this calculation.
      (b) The error bound for $\epsilon$, where $\epsilon$ is the contribution
      orthogonal to the ground state. Observables such as the spin measurements,
      e.g. $\sigma^x$ with a maximum of $1$, are bounded by $3 \epsilon$.
      (c) The bound for the error in the bond entropy $\eentr$ depends furthermore
      on the dimension of bipartition, which increases the bound especially
      for larger systems.
                                                                                \label{fig:ErrScaling}}
  \end{center}
\end{figure}

Finally, we present arguments as to why the variance tolerance is a convenient
convergence criterion. In Appendix~\ref{app:boundstatic} we derive the
bounds of multiple variables, we provide a short summary of those bounds here.
The variance of the ground state $V_{\psi}$ determines a bound on $\epsi$,
where $\epsi$ is the contribution for $\ket{\psi_{\perp}}$ of all states
orthogonal to the true ground state $\ket{\psi_{0}}$ in the result
$\ket{\psi}$ returned from \OSMPS{}:
\begin{eqnarray}
  \ket{\psi}
  = f \ket{\psi_0} + \epsi \ket{\psi_{\perp}} \, , \quad
  |\epsi| \le \frac{\sqrt{V_{\psi}}}{\Delta_{0,1}} \, .
\end{eqnarray}
The value of $\Delta_{0,1}$ is the energy gap between the ground state
and the first excited state. Starting from there, we derive in
Appendix~\ref{app:boundstatic} bound on an observable $O$ acting on
$\ket{\psi}$ as well as the bond entropy $S$:
\begin{eqnarray}
  | \bra{\psi_0} O \ket{\psi_0} - \bra{\psi} O \ket{\psi} |
  &\le&  3 \epsi \mathcal{M} \, , \quad
  \mathcal{M} = \max_{\ket{\phi}, \ket{\phi'}} | \bra{\phi} O \ket{\phi'} \, ,            \\
  \eentr = | S(\rho_0) - S(\rho) |
  &\le& \frac{\sqrt{2 V_{\psi}}}{\Delta_{0, 1}} \log(D)
      - \frac{\sqrt{2 V_{\psi}}}{\Delta_{0, 1}}
        \log \left(\frac{\sqrt{2 V_{\psi}}}{\Delta_{0, 1}}\right) \, .          \label{eq:mpserrboundentr}
\end{eqnarray}
In Eq.~\eqref{eq:mpserrboundentr} the dimension of the density matrix,
$D$, appears in addition to the variance and the gap. We pick as an example
for the error bounds the nearest neighbor quantum Ising model with
$\mathbb{Z}_2$ symmetry. Due to the symmetry, the first excited states lies,
in contrast to the ground state, in the odd sector. Therefore, the relevant
gap is the energy difference between the ground state and second excited state.
The value of the gap for different systems sizes and the upper error bound for
$\epsilon$ and $\eentr$ are shown in Fig.~\ref{fig:ErrScaling}.

\subsection{Running the simulations                                           \label{sec:setupandrun}}

Finally, we discuss how to set up simulations and execute them on the
computer. Each simulation is contained in a dictionary and we can create a
list of dictionaries to run multiple simulations at the same time. While certain
parameters such as the measurement setup stay the same for a set of simulations,
other parameters may be varied. In this example we create an empty list
\texttt{params} and add the dictionaries to the list looping over the system
size $L$ and the external field $h$. The dictionary is shown in
Listing~\ref{py:01_Ising_append}.

\pyfrag{./code/01_Ising.py}
       {Appending different simulations looping over $L$ and $g$.}
       {py:01_Ising_append}{102}{126}

In the following we generate a submit script for our simulations by
writing the files for \fort{} with a call to\newline
%
\pyline{MainFiles = mps.WriteMPSParallelFiles(params, Operators, H, hpcsetting,}\newline
\pyline{PostProcess=False)}\newline
and the simulations are executed when submitted to the computing cluster.
The fourth argument \texttt{hpcsetting} is a dictionary with various
settings such as the number of nodes requested on the cluster.
As an alternative, the user may call the parallel executable on a local
machine and can find the corresponding call inside the submit script. We
do not cover the post-processing itself here, but the sample scripts
presented in the supplemental materials in
Appendix~\ref{app:suppl} provide guidance on how to read the results with the
corresponding \OSMPS{} functions and access the measurements inside the
dictionaries.

\section{Highlights of static algorithms                                       \label{sec:algorithms}}

In the previous section we presented static simulations for the
ground state, which builds a basis for other algorithms within \OSMPS{}.
The next algorithm searches for the excited state obtained through variational
means. It can find sequentially ascending excited states above the ground state. In
addition, we highlight our infinite size statics as a method to
calculate properties for the ground state in the thermodynamic limit with an example.

\subsection{Excited state search                                                \label{sec:emps}}

The search for excited states, eMPS, is implemented in a successive fashion
after the ground state has been obtained. The algorithm is based on the
variational procedure now introducing additional Lagrange multipliers
to Eq.~\eqref{eq:mps_lagrange} to enforce orthogonality with previously
obtained eigenstates. If the ground state is now labeled as
$\ket{\psi_{0}}$ and the $i^{\mathrm{th}}$ excited state as
$| \psi_{i} \rangle$, we then need $i$ additional Lagrange multipliers $\mu_{i}$
corresponding to the number of states with lower energy. These Lagrange
multipliers enforce orthogonality between the eigenstates,
$\langle \psi_{i} | \psi_{j} \rangle = \delta_{i,j} \;$:
\begin{eqnarray}                                                                \label{eq:emps_lagrange}
  \varepsilon[| \psi_{i} \rangle ] =
  \langle \psi_i | H | \psi_{i} \rangle - E \langle \psi_i | \psi_i \rangle
  - \sum_{j=0}^{i-1} \mu_{j} \langle \psi_{i} | \psi_{j} \rangle \, .
\end{eqnarray}
We base our example for the excited state search on the previous study of the
quantum Ising model. We introduce long-range interactions following a power
law decay in this example. The corresponding Hamiltonian of the long-range
Ising model was presented in Eq.~\eqref{eq:LRIsing}, and we recall that it was
defined as
\begin{eqnarray}
  H &=& - J \sum_{i < j \le L} \frac{\sigma_{i}^{z} \sigma_{j}^{z}}{(j - i)^{\alpha}}
        - J g \sum_{i=1}^{L} \sigma_{i}^{x} \, .                                \nonumber
\end{eqnarray}
The update to the Hamiltonian due to the long-range interaction is reflected
in Listing~\ref{py:02_Ising_ham}. Since we loop over different
$\alpha$, we generate the Hamiltonian as well inside the loop over the
different parameters and generate a list of them; in the previous example
we could use a single MPO because only the couplings changed, but not the
function describing the coupling. For the complete file see the supplemental
material, Appendix~\ref{app:suppl}.

\pyfrag{./code/02_LRIsing.py}
       {Adapting the MPO for the long-range Ising model.}
       {py:02_Ising_ham}{94}{100}

\begin{figure}[t]
  \begin{center}
    \begin{minipage}{0.47\linewidth}
      \vspace{0.99cm}
      \begin{overpic}[width=1.0 \columnwidth,unit=1mm]{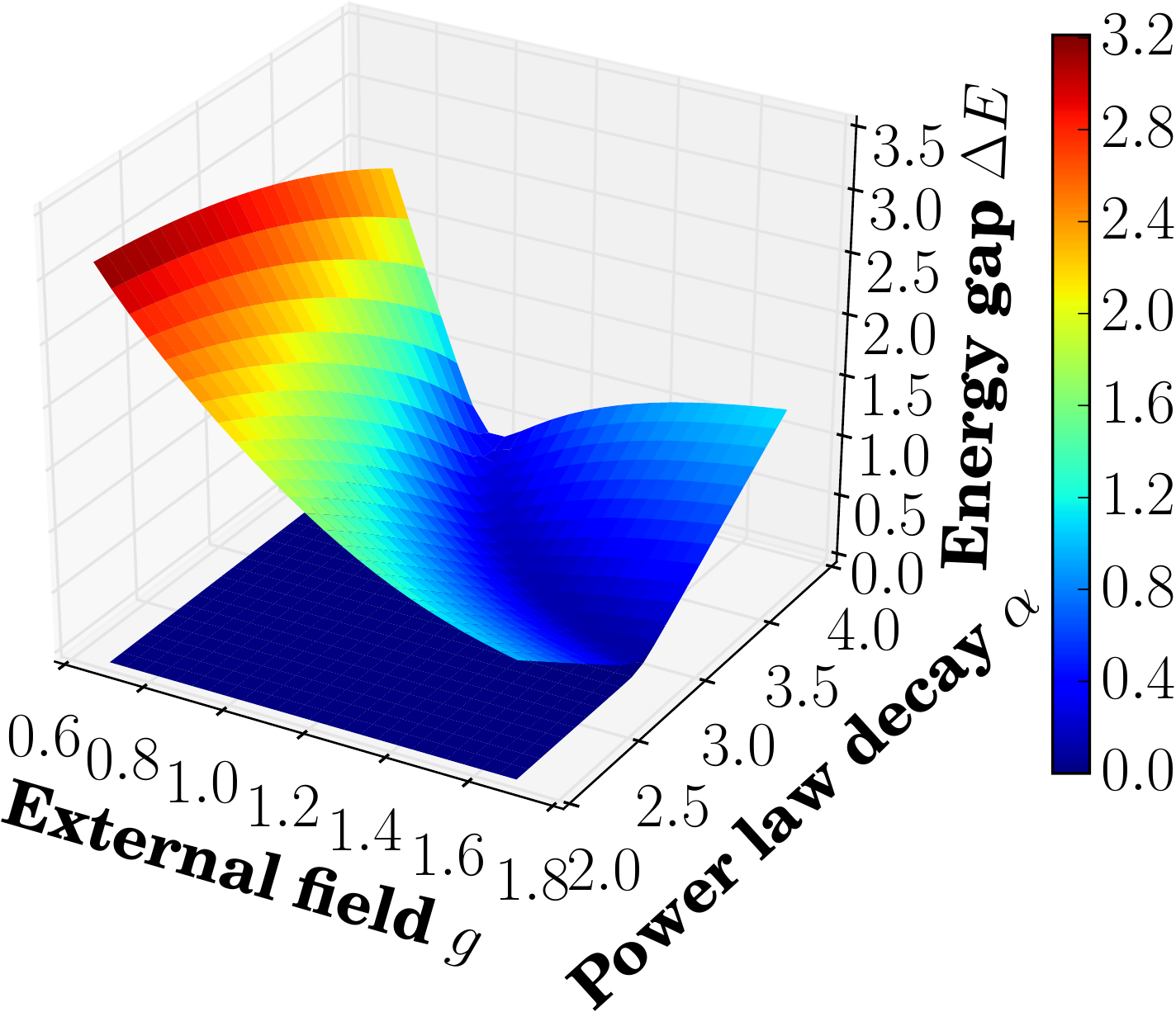}
        \put(47,91.5){(a)}
      \end{overpic}
    \end{minipage}\hfill
    \begin{minipage}{0.47\linewidth}
      \vspace{0.99cm}
      \begin{overpic}[width=1.0 \columnwidth,unit=1mm]{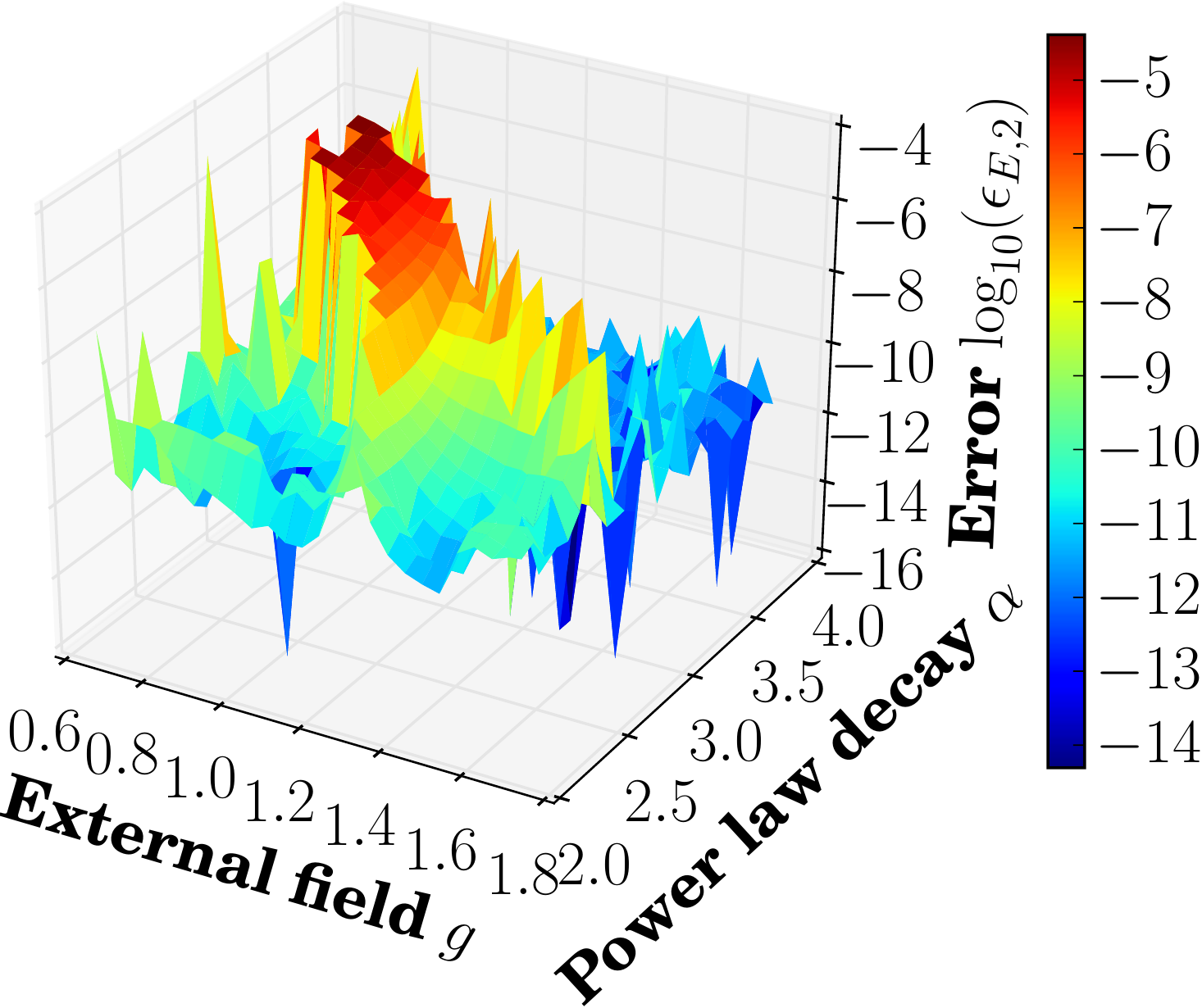}
        \put(45,90){(b)}
      \end{overpic}
    \end{minipage}
    \caption{\emph{Energy gap for the long-range quantum Ising model} as a
      function of the interaction strength $\alpha$ and the external field
      $g$. (a) The energy gap between the ground state
      and the first excited state is close to degeneracy in the ferromagnetic
      phase of the long-range quantum Ising model. The gap between ground
      and second excited states closes toward the quantum critical point,
      e.g. for $\alpha = 4$ around $g \approx 1$. The eigenenergies
      for the second excited state are compared to the eigenenergies
      of the ground state and the second excited state of the combined
      odd and even symmetry sector in (b) to estimate the error.
      (Same labels apply to color bar and $z$-axis.)
                                                                                \label{fig:02_LRIsing}}
    \end{center}
\end{figure}

In order to calculate the excited states, we add the information showed in
Listing~\ref{py:02_Ising_dict} to the simulation dictionary. In general it is
possible to define different observables or convergence parameters for the
ground state and the excited state, although it is not possible to have
different settings for each excited state. We present the results of the
excited states of the long-range Ising model in Fig.~\ref{fig:02_LRIsing}.
The excited states can reveal physical phenomena or support theory, e.g.
we deduct from Fig.~\ref{fig:02_LRIsing} the close to degenerate ground and
first excited state in the ferromagnetic phase. Both the ground and
second excited state in the ferromagnetic phase belong to the even
sector of the $\mathbb{Z}_{2}$ symmetry, and their closing gap indicates
the position of the quantum critical point. This closing gap can be seen as
valley starting around $\alpha = 4$ and $g = 1.0$. We use the symmetry conserving
simulation to calculate the ground state and first excited state and show
the errors in energy in Fig.~\ref{fig:02_LRIsing}(b) and (c). In the latter
one we see as well the growing error around the critical point. Because the
variational state can only guarantee to find an eigenstate, but might end up
in a minimum corresponding to a higher excited state, it is useful to resort
the energies and converge results with more excited states than actually
desired.

\pyfrag{./code/02_LRIsing.py}
       {Specify the number of excited states to be calculated and the
        settings for convergence and measurements.}
       {py:02_Ising_dict}{124}{127}

\subsection{Infinite systems in the thermodynamic limit}                       \label{sec:imps}

The \OSMPS{} library possesses another tool to obtain information about the
ground state of a quantum system, which is applicable in the thermodynamic
limit. The iMPS algorithm searches for the ground state of
a translationally invariant Hamiltonian \cite{McCulloch2008}. The core idea of
the algorithm is based on a unit cell of $L$ sites. The Hamiltonian is
translationally invariant in the sense that we consider an infinite sequence of
these unit cells with the same Hamiltonian. Within the $L$ sites, the
Hamiltonian can depend on the site, creating a sublattice or similar features.

\begin{figure}[t]
 \begin{center}
   \begin{minipage}{0.315\linewidth}
     \vspace{0.32cm}
     \begin{overpic}[width=1.0 \columnwidth,unit=1mm]{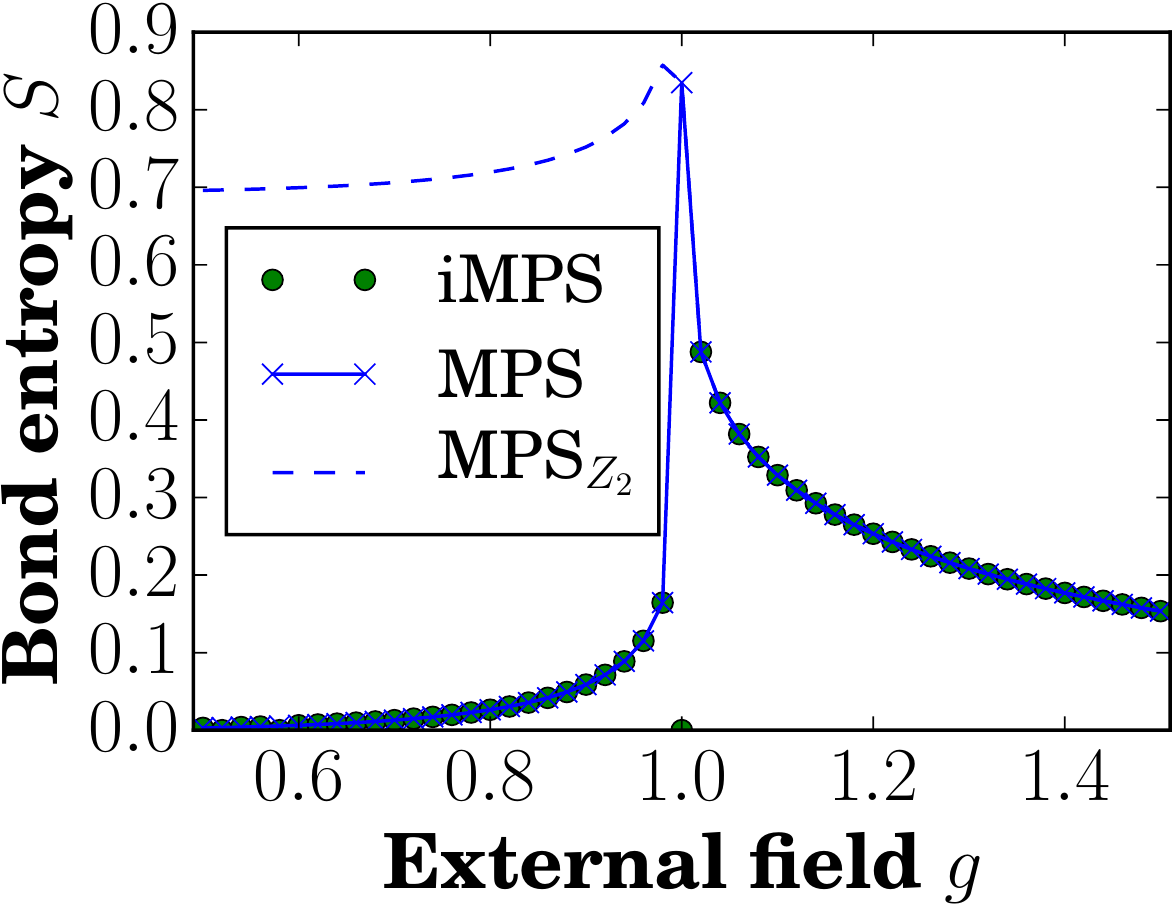}
       \put( 0,88){(a)}
     \end{overpic}
   \end{minipage}\hfill
   \begin{minipage}{0.33\linewidth}
     \begin{overpic}[width=1.0 \columnwidth,unit=1mm]{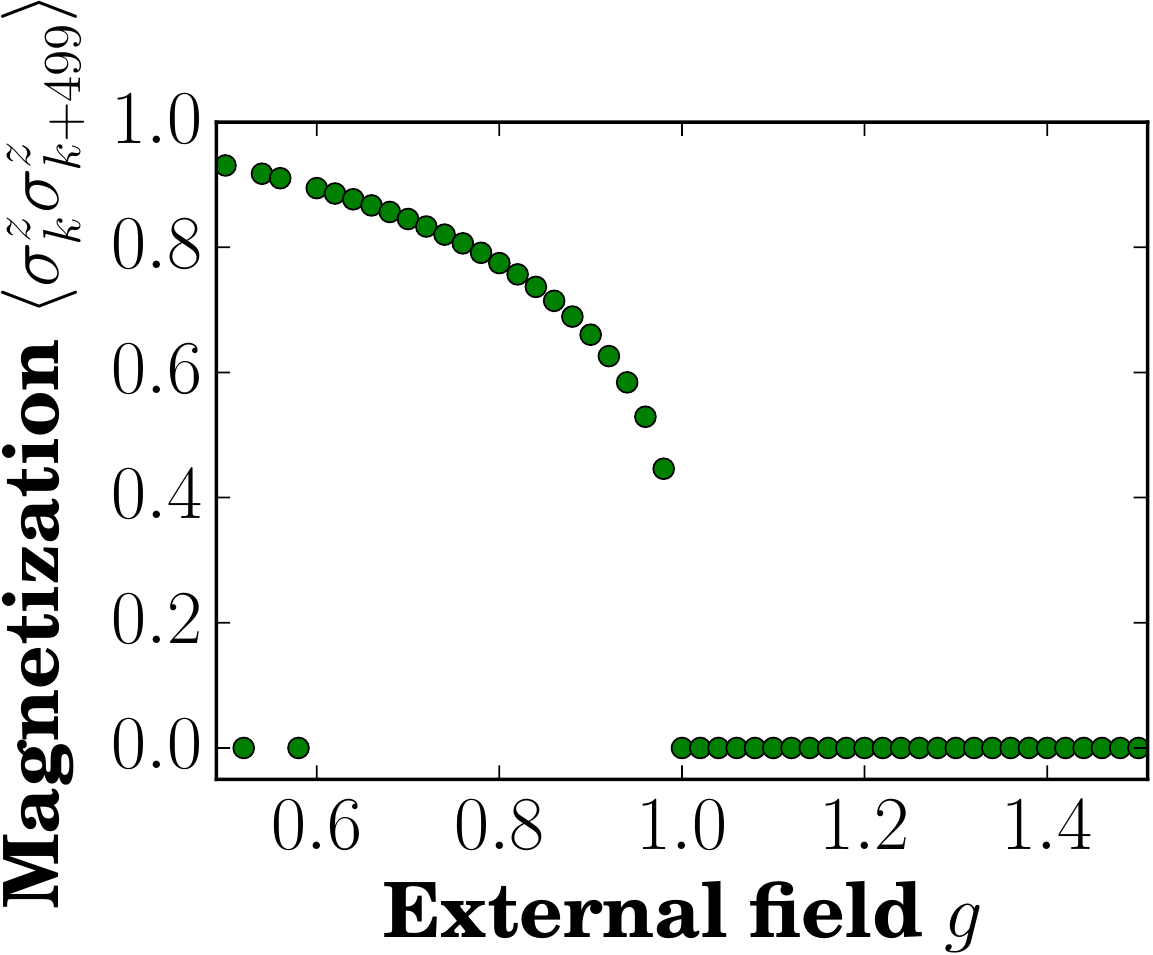}
       \put( 0,85){(b)}
     \end{overpic}
   \end{minipage}\hfill
   \begin{minipage}{0.315\linewidth}
     \vspace{0.33cm}
     \begin{overpic}[width=1.0 \columnwidth,unit=1mm]{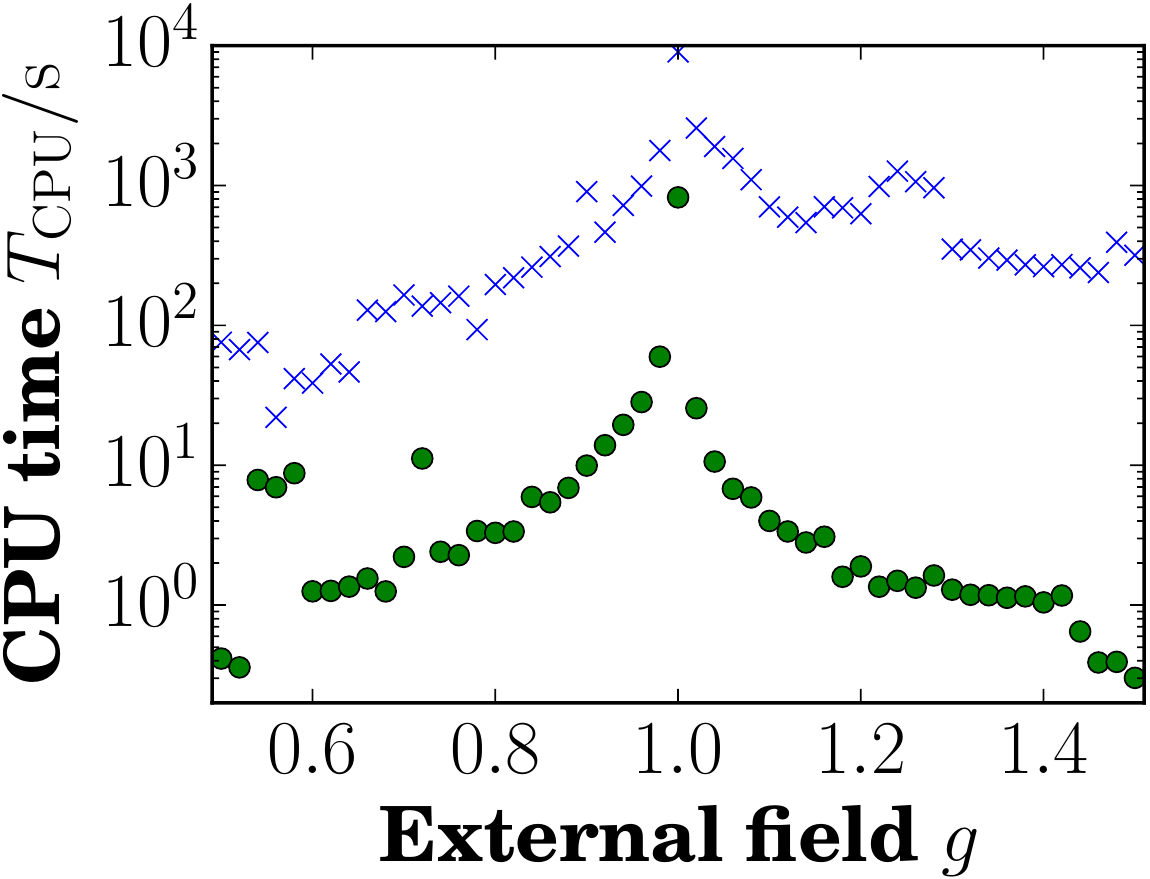}
       \put( 0,88){(c)}
     \end{overpic}
   \end{minipage}
   \caption{\emph{The quantum Ising model in the infinite MPS simulations.}
     (a) The bond entropy of the iMPS peaks at the critical point
     $g_c = 1.0$ which reproduces the results of the finite size MPS simulation
     for $L = 2000$ without $\mathbb{Z}_{2}$ symmetry. In general, the bond
     entropy of the iMPS in the ferromagnetic phase can lie in between the
     results with and without the $\mathbb{Z}_{2}$ symmetry. (b) The
     magnetization based on the correlation $\langle \sigma_{i}^{z}
     \sigma_{i+499}^{z}\rangle$ dies away at $g = 1.0$. (c) The compute times of
     the iMPS simulations in comparison with the finite size algorithm at
     $L=2000$. iMPS can give a quick estimate of the behavior in the
     thermodynamic limit.
                                                                                \label{fig:04_Ising_iMPS}}
 \end{center}
\end{figure}

The final state is obtained when the state of the unit cell is converged by
parameters discussed in the following. Starting with the first unit cell the
ground state of the system is obtained. The system size is increased by
inserting another unit cell in the middle of the system and summarizing the
previous result in an environment. The new ground state of the unit cell is
computed under the action of the environment. Subsequent steps of inserting
cells while growing the environment lead to the result.
The class \texttt{iMPSConvParam} comes with the convergence parameters for
the bond dimension $\chi$, and the local tolerance and the settings for the
Lanczos algorithm keep their meaning in regard to previous algorithms. We
introduce the maximal number of iMPS iterations determining how often a
new unit cell is introduced into the iMPS state before stopping the
algorithm. To break out of the algorithm before reaching the maximal
number of iMPS iterations we consider the orthogonality fidelity $\mathcal{F}$
(\texttt{variance\_tol}). In order to define this orthogonality fidelity
$\mathcal{F}$, we introduce the density matrices $\rho_{n-1}$, i.e., the
density matrix of the previous step, and $\rho_{n}^{R}$ as the density matrix
of the present step without the new unit cell introduced in step $n$. The
overlap or fidelity serves then as a convergence criterion:
\begin{eqnarray}
  \mathcal{F}(\rho_{n-1}, \rho_{n}^{R})
 &=& \sqrt{\sqrt{\rho_{n}^{R}} \rho_{n-1} \sqrt{\rho_{n}^{R}}} \, .
\end{eqnarray}
We can use the algorithm to compare the results of the first study of the
nearest neighbor limit with those of iMPS. In Fig.~\ref{fig:04_Ising_iMPS}
we show the bond entropy of the iMPS which peaks at the critical point.
We point out that of all simulations, three fail being the second,
fifth data point and the bond entropy at the critical point. In
comparison we show the largest finite size system with $L = 2000$ with
and without using the $\mathbb{Z}_{2}$ symmetry. Both lines show good
agreement. Possible disagreement in the bond entropy may arise from
the actual ground state, i.e. $\left( | \uparrow \cdots \uparrow
\rangle + | \downarrow \cdots \downarrow \rangle \right) / \sqrt{2}$. But
any other superposition of all spins up plus all spins down fulfills the
minimization of the energy as well. Furthermore, the output of the
infinite system is partly different from the finite size algorithm, e.g.
the maximal distance for the correlation is specified. More of those
differences are described in detail in the manual.

\section{Time evolution methods                                                \label{sec:timeevo}}

The only missing piece to complete the library at this point is the time
evolution of quantum states. In total we provide four different algorithms:
Krylov time evolution \cite{Saad1992} by default, the Time-Dependent
Variational Principle algorithm (TDVP) \cite{Haegeman2016}, a local
Runge-Kutta method (LRK) \cite{Zaletel2015}, and a Sornborger-Stewart
decomposition \cite{NielsenChuang,Sornborger1999}. The Sornborger-Stewart
decomposition is an alternate decomposition to the Trotter decomposition and
is used to implement the Time-Evolving Block Decimation (TEBD)
\cite{Vidal2003}. The first three methods support long-range interactions,
and align with our motivation to support such systems.

\subsection{Computational error and convergence                                \label{sec:tconv}}

\begin{figure}[t!]
 \begin{center}
    \includegraphics[width=0.65 \columnwidth]{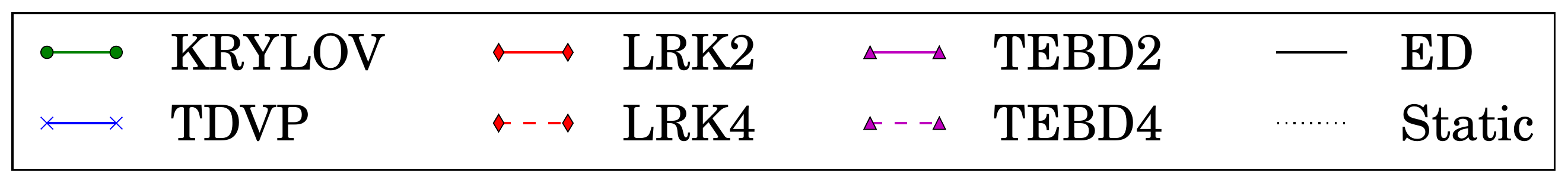}
    \begin{minipage}{0.48\linewidth}
      \begin{overpic}[width=0.9\columnwidth,unit=1mm]{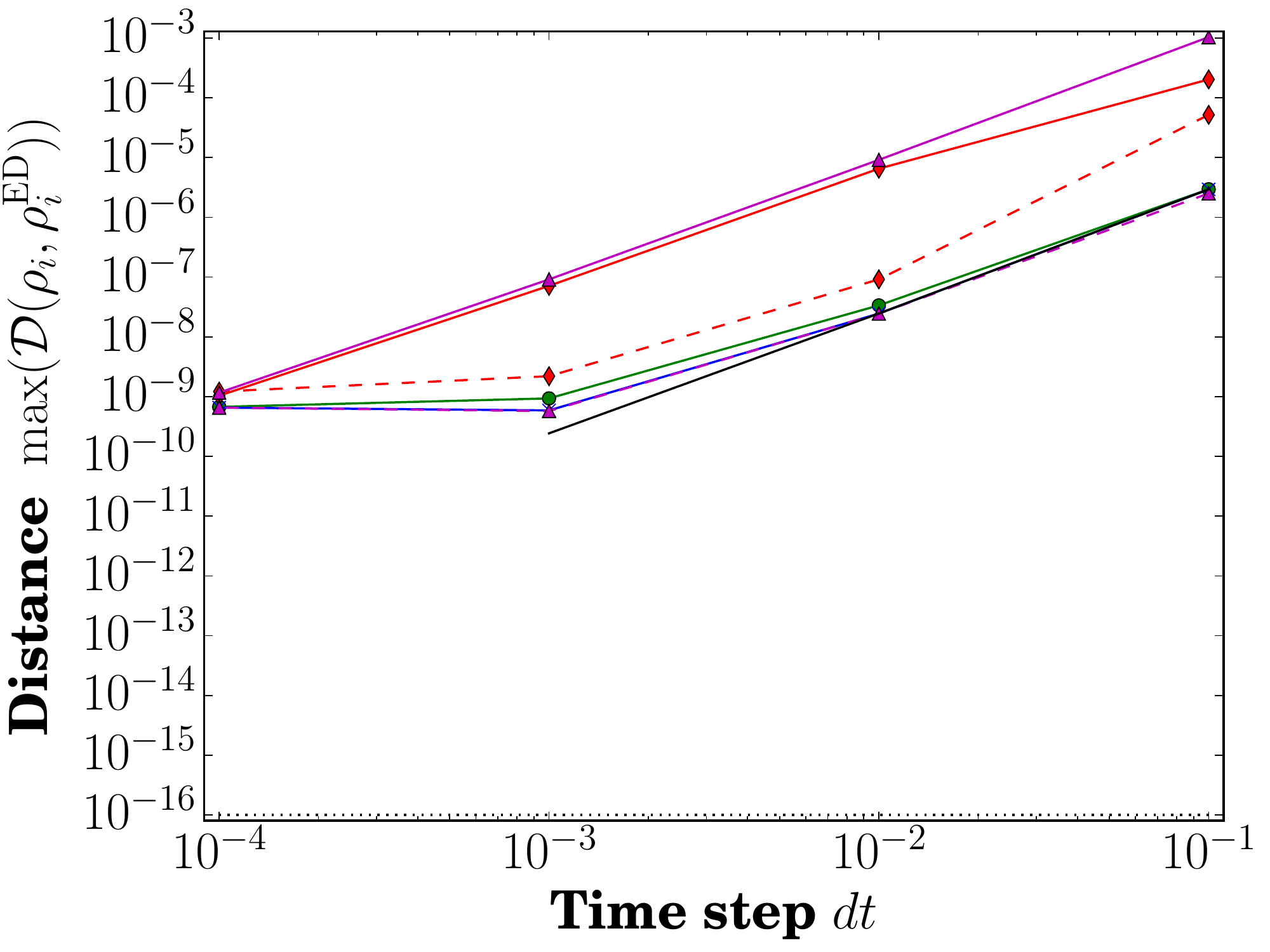}
        \put(18,67){(a)}
      \end{overpic}
    \end{minipage}\hfill
    \begin{minipage}{0.48\linewidth}
      \begin{overpic}[width=0.9 \columnwidth,unit=1mm]{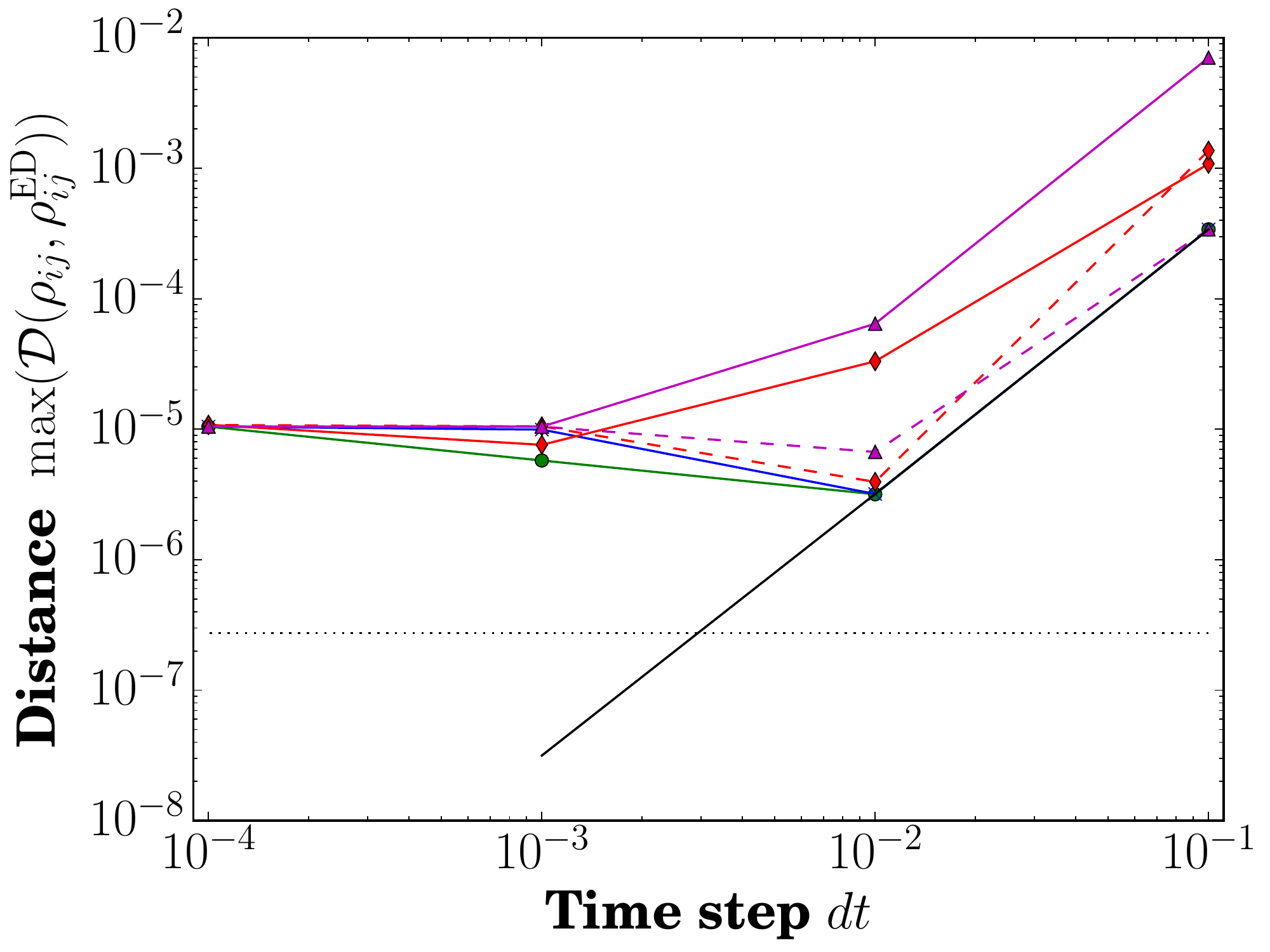}
        \put(18,67){(b)}
      \end{overpic}
    \end{minipage}

    \vspace{0.1cm}

    \begin{minipage}{0.48\linewidth}
      \begin{overpic}[width=0.9\columnwidth,unit=1mm]{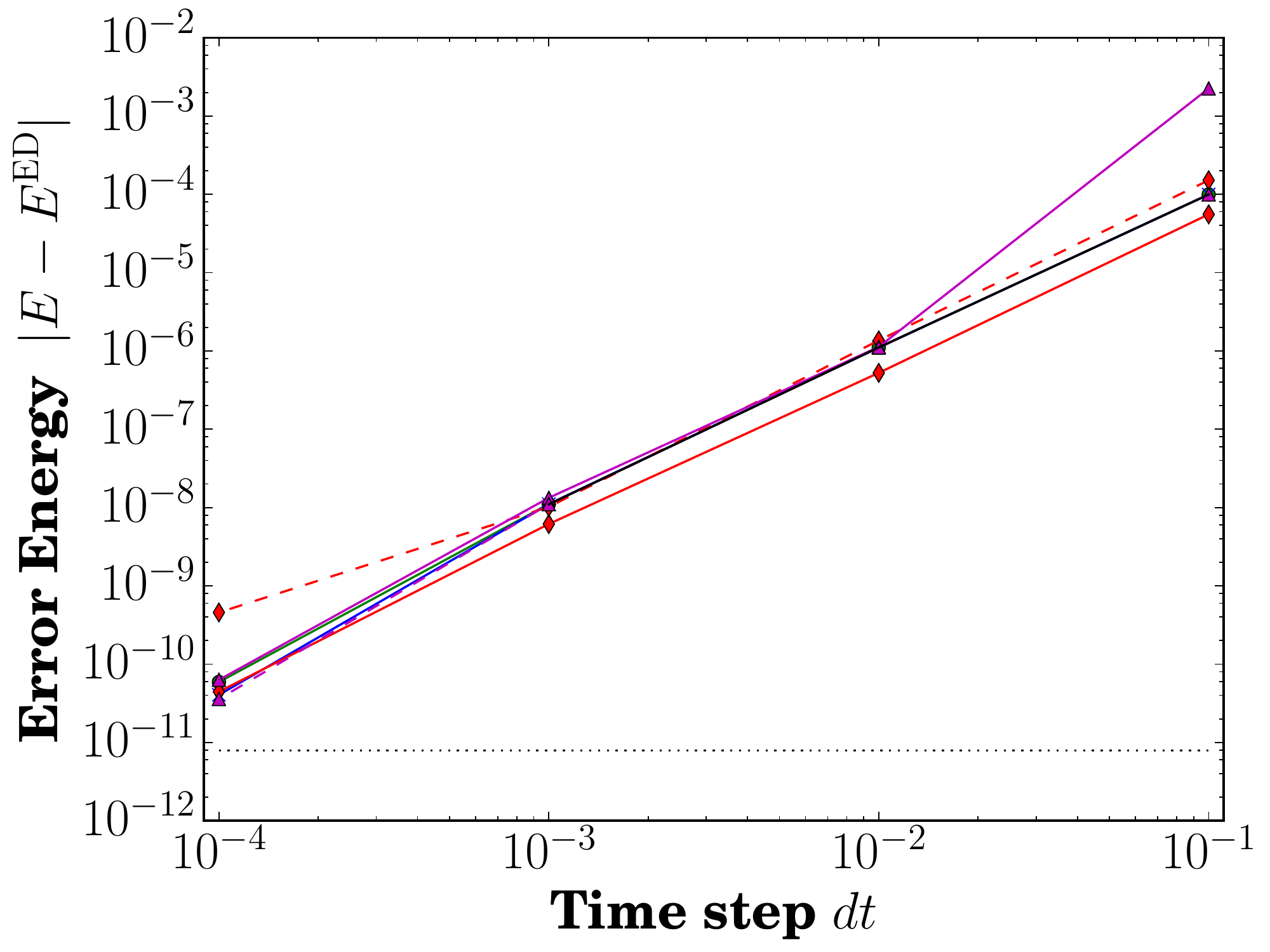}
        \put(18,67){(c)}
      \end{overpic}
    \end{minipage}\hfill
    \begin{minipage}{0.48\linewidth}
      \begin{overpic}[width=0.9 \columnwidth,unit=1mm]{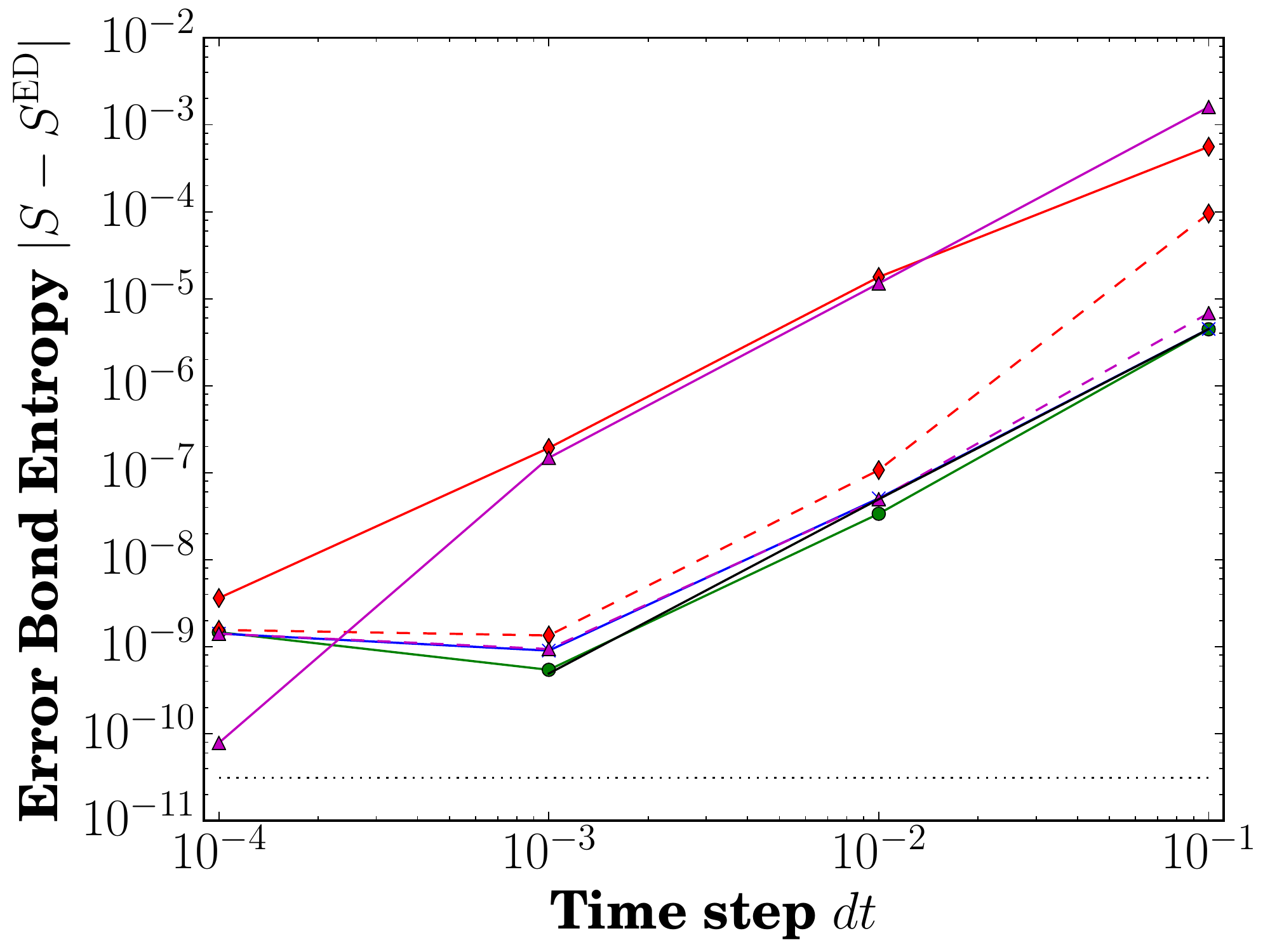}
        \put(18,67){(d)}
      \end{overpic}
    \end{minipage}

    \vspace{0.1cm}

    \begin{minipage}{0.48\linewidth}
      \begin{overpic}[width=0.9\columnwidth,unit=1mm]{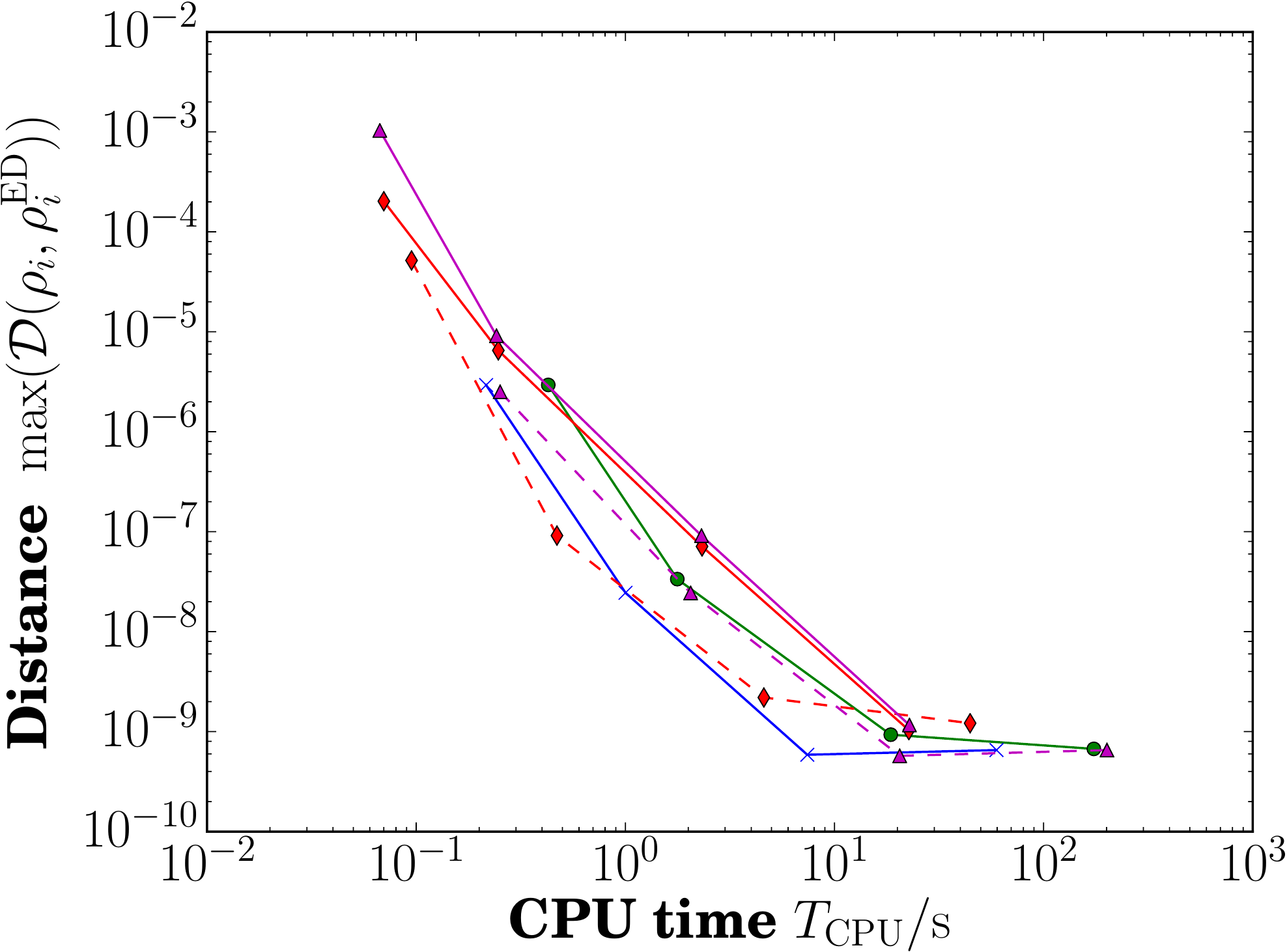}
        \put(18,66){(e)}
      \end{overpic}
    \end{minipage}\hfill
    \begin{minipage}{0.48\linewidth}
      \begin{overpic}[width=0.9 \columnwidth,unit=1mm]{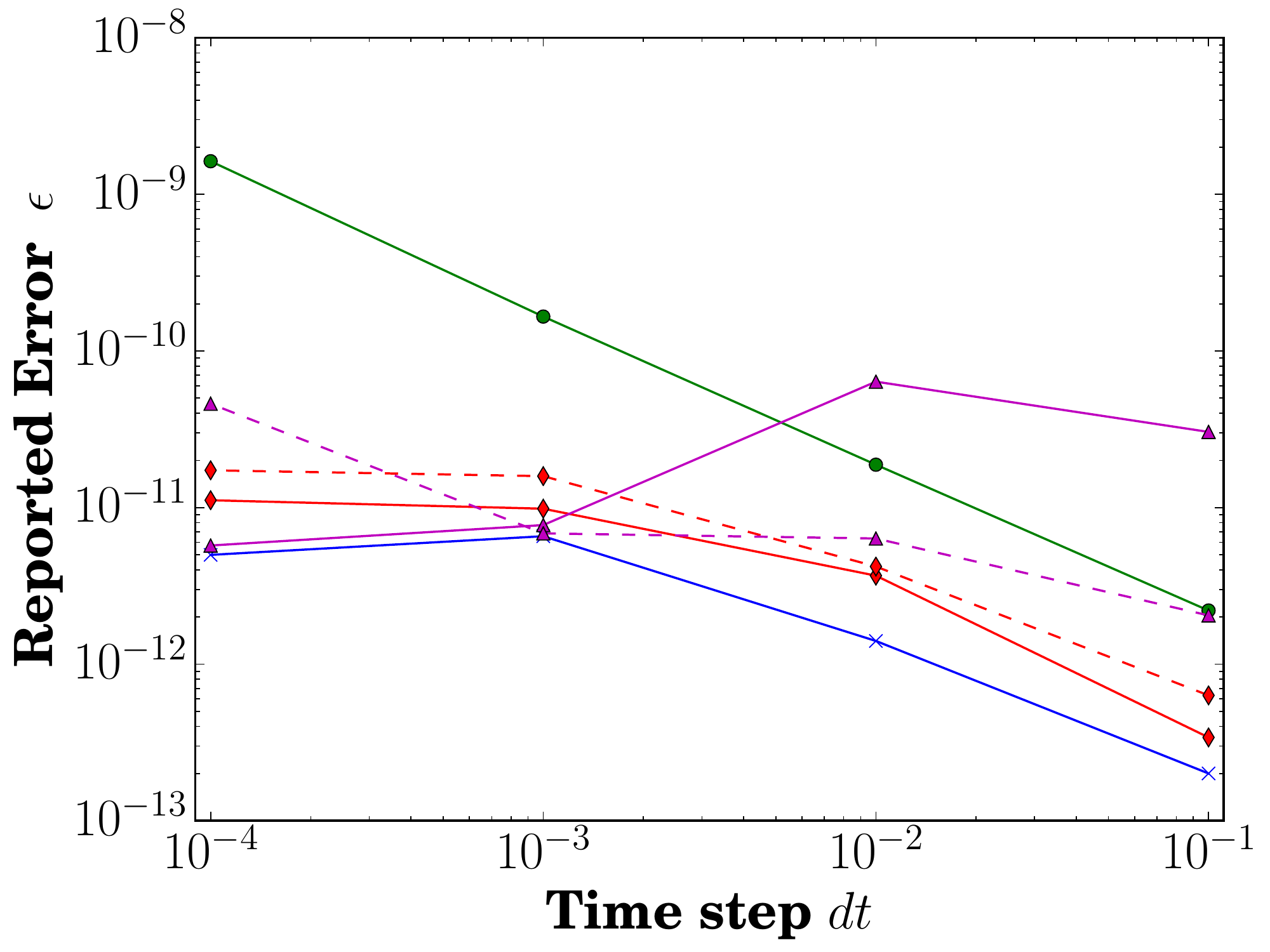}
        \put(18,67){(f)}
      \end{overpic}
   \end{minipage}
   \caption{\emph{Scaling of the error in time evolution methods} decreases as
     expected with the size of the time step. This example shows a quench
     of the Ising model in the paramagnetic phase. The error decreases as
     $\mathcal{O}(dt^2)$, the leading order of the error due to
     time-slicing the time-dependent Hamiltonian $H(t)$ with a CFME. The error
     for the exact diagonalization method is not plotted for $dt = 10^{-4}$,
     because this is the result used as a reference and naturally leads to zero
     error. The time step $dt$ is in units of $\hbar / J$. Curves
     are a guide to the eye; points represent actual data.
                                                                                \label{fig:03_Ising}}
 \end{center}
\end{figure}

We first provide an overview of each method's behavior in terms of convergence
in Fig.~\ref{fig:03_Ising}. The figure compares the \OSMPS{} algorithms to
analogous exact diagonalization time propagation schemes, focusing on four key
measures:
\begin{enumerate}
\item{\emph{The maximum trace distance of all local reduced density matrices},
  \begin{eqnarray}                                                              \label{eq:erhoi}
    \erhoi &=& \max_{i \in \{1, \ldots L \}}
               \mathcal{D}(\rho_i, \rho_i^{\mathrm{ED}}), \quad
    \mathcal{D}(\rho_{A}, \rho_{B}) \equiv
        \frac{1}{2} | \rho_{A} - \rho_{B} | \, , \quad                          \nonumber \\
    |A| &\equiv& \sqrt{A^{\dagger} A} \, .
  \end{eqnarray}
  The superscript \emph{ED} refers to the results of exact diagonalization
  methods. Equation~\eqref{eq:erhoi} can be used to bound any local
  observable, as explained in Appendix~\ref{app:boundtracedist}.}
\item{The error of \emph{correlation measurements.} We consider the
  maximal trace distance $\erhoij$, here on all two-site density matrices, to
  bound the error for the correlation measurements:
  \begin{eqnarray}
      \erhoij &=& \max_{(i, j)} \mathcal{D}(\rho_{i,j}, \rho_{i,j}^{\mathrm{ED}}), \quad
      i,j \in \left\{ 1, \ldots L \mid| i < j \right\} \, .
  \end{eqnarray}
}
\item{The \emph{energy} of the system:
  \begin{eqnarray}
    \eener &=& | E - E^{\mathrm{ED}} | \, .
  \end{eqnarray}}
\item{As the maximal bond dimension is one of the key parameters in MPS
  algorithms, we finally compare the \emph{bond entropy} defined over the
  von Neumann entropy $S$ of singular values squared obtained by
  cutting the system in half:
  \begin{eqnarray}                                                              \label{eq:eentr}
    \eentr &=& | S(L / 2) - S^{\mathrm{ED}}(L / 2) | \, .
  \end{eqnarray}
  The maximal bond dimension necessary in small systems which can still be
  studied in exact diagonalization is unfortunately limited. Therefore, we
  cannot study the effects of truncation in comparison to exact
  diagonalization.}
\end{enumerate}

We compare the \OSMPS{} results with the data from exact diagonalization.
Therefore, we take the simulation with the smallest time step corresponding to
the most accurate result. In addition, we display the error from the static
simulation as a lower bound for the error. The static simulation serves as an
input state for the dynamics and sets the lower bound for the error.
Figure~\ref{fig:03_Ising} shows the
error at the end of the time evolution for a quench in the Ising model. The
quench starts at $g(t=0) = 5.0$ and ends at $g(t=0.5) = 4.5$ for a system
size of $L = 10$. The time is in units of $\hbar / J$. The exact
diagonalization method, which is always at time-ordering
$\mathcal{O}(dt^2)$, takes the whole Hamiltonian to the exponent evaluating
the coupling at $t + 0.5 \dt$ resulting in $\ket{\psi(T)} =
\exp(- \mathrm{i} H(0.5 dt) dt) \exp(- \mathrm{i} H(1.5 dt) dt) \cdots
\exp(- \mathrm{i} H(T - 0.5 dt) dt) \ket{\psi(0)}$. In contrast, the
MPS time evolution methods support higher order time ordering, which is not
used for the studies within this work. We briefly point out the
trends within this Fig.~\ref{fig:03_Ising}.
We defined the first error as the maximal trace distance over all single
site density matrices $\erhoi$, which is shown in
Fig.~\ref{fig:03_Ising}(a). We see two major trends for $\erhoi$. First, there
is a clear difference between the second and fourth order methods in the case
of TEBD and LRK algorithms, labeled as TEBD2 and TEBD4 as well as LRK2
and LRK4, respectively. The fourth order algorithms and the TDVP and Krylov
algorithms nearly match the result of exact diagonalization for
$dt \ge 10^{-3}$ in comparison to the ED result with $dt = 10^{-4}$.
Figure~\ref{fig:03_Ising}(b) analyzes the second error,
i.e., the error in the reduced two site density matrices $\erhoij$. This error
is much larger, which is already present in the ground and therefore initial
state with an order of magnitude of $10^{-7}$.
The third kind of error, the error in energy $\eener$, is shown in
Fig.~\ref{fig:03_Ising}(c). $\eener$ decreases for all the methods
with the same overall trend. We recall that the Hamiltonian in this case
contains single site terms and nearest neighbor terms, so large errors in
two site reduced density matrices with sites far apart would not contribute
to the error in the energy.
The error in the bond entropy $\eentr$, the fourth value considered for
the estimate of the error, follows the behavior of the previous measures,
see Fig.~\ref{fig:03_Ising}(d). All methods except
TEBD2 and LRK2 do not improve from $dt = 10^{-3}$ to $dt = 10^{-4}$ as the
entropy is already close to the static result.
In general in order to tackle a given problem, we seek for the method which
consumes the least resources to achieve a certain error
Fig.~\ref{fig:03_Ising}(e) answers this question. We take the error
$\erhoi$ as example and plot it as a function of the CPU time. This
allows us for example to compare the second order methods versus their fourth
order algorithm. Both the TEBD and LRK fourth order methods use less resources
at a bigger $dt$ to achieve the same error in comparison to their second order
equivalent.
The error reported back from \OSMPS{} is analyzed in
Fig.~\ref{fig:03_Ising}(f). We consider the Krylov method first: the reported
error is bigger for a smaller time step despite the results clearly getting
better in the previous plots. The reported Krylov error contains errors from
state fitting and cutting the bond dimension. Although it is not necessary to
cut the bond dimension, there still remains the possibility for truncation due
to the local tolerance criteria. Since the reported error is accumulated during
the evolution, the small contributions add up due to the multiple time steps.
Considering the error for $dt = 0.0001$ of the order $10^{-9}$ with $5000$
time steps, each time step adds about $10^{-13}$ to the total error.
The other time evolutions report purely
the error from truncation of singular values restricted to the local tolerance
in this evolution. Increasing errors for decreasing time steps follow the same
arguments as for the Krylov time evolution.
In Appendix~\ref{app:convergence} we discuss in detail the sudden quench and the
following evolution under a time-independent Hamiltonian.

The conclusion drawn from this first study are that the total error is bounded by
of
\begin{eqnarray}                                                                \label{eq:toterr}
  \etot \le \edt + \emeth + \echi \, ,
\end{eqnarray}                                                                  
where $\edt$ is due to evaluating a time-dependent Hamiltonian at
discrete points in time. The second error originates from the specific method
used to evolve the quantum system in time; an example for $\emeth$
would be the approximation in the Sornborger-Stewart decomposition. Finally,
all methods have in common that they truncate singular values to remain with
a certain $\chi$ leading to $\echi$. This source of error will dominate
the error even for well-chosen settings, since the entanglement grows over
time and therefore saturates the bond dimension for long enough evolution time.
Figure~\ref{fig:R3_ForthBack} in the appendix is one example for this behavior.
Equation~\eqref{eq:toterr} is an upper bound
for the error. A detailed analysis of the error goes beyond the scope of this
work. We emphasize that all three kind of error sources
manifest in the four different error measures defined in Eqs.~\eqref{eq:erhoi}
to \eqref{eq:eentr}.
The scaling of the first error source $\edt$ can be obtained through exact
diagonalization since there is neither an error depending on the method nor
a truncation of the Hilbert space. $\edt$ appears in the same way for all
MPS methods using the same time-ordering as done in this error study.
We remain with the estimate of the errors due to the method $\emeth$ and
the truncation in the bond dimension $\echi$. The Hilbert space for
the MPS simulations with $L = 10$ is small enough to capture all singular
values and $\echi$ is not present. Therefore, $\emeth$ can be seen in
Fig.~\ref{fig:03_Ising} if at the same order of magnitude as $\edt$. For
example, the TEBD2 method has an additional error introduced through the
Sornborger-Stewart decomposition making it less accurate than the exact diagonalization
result with the same time step. We discuss a time-independent Hamiltonian in
Appendix~\ref{app:convergence} with $\edt = 0$. Therefore, $\emeth$ can be
estimated independent of the other errors in this case.

Before we discuss particular time-propagation methods, we first discuss how
\OSMPS{} accounts for time-ordering of propagators for time-dependent
Hamiltonians. When considering time-ordering in MPS algorithms, we want to
apply as few operators as possible to avoid increasing the bond dimension,
and would like all operators to be easily and efficiently constructed from
the MPO form of the Hamiltonian. Therefore, \OSMPS{} uses Commutator-free
Magnus expansions (CFMEs) \cite{Alvermann2011,Wall2012}. CFMEs are
advantageous over other expressions such as the Dyson series or the
original formulation of the Magnus series due to explicit unitarity and
the avoidance of nested integrals and/or commutators. \OSMPS{} implements
a few different CFMEs with orders of error $N$, defined such that the
propagator is accurate to $\mathcal{O}\left(\delta t^{N+1}\right)$,
and numbers of exponentials $s$. The default settings are $N=2$ and $s=1$,
where the CFME amounts to evaluating the Hamiltonian in $[t,t+\delta t]$
using the midpoint rule for integration. Having introduced the general
convergence of the methods, we now look at each method individually and
discuss their principles.

\subsection{Krylov time evolution                                              \label{sec:krylov}}

The Krylov method \cite{Manmana2005time,Garcia2006time,Wall2012} is the
default option for real time evolution in the \OSMPS{} code. The main point of
using this technique is the support of long-range interactions. The Krylov
method applies the exponential of an operator expressed as an MPO to an MPS
\begin{eqnarray}                                                                \label{eq:propSE}
  \ket{\psi(t+dt)} \approx \exp(- \i H dt) \ket{\psi(t)} \, ,
\end{eqnarray}
using the method of Krylov subspace approximations~\cite{Saad1992,%
Gallopoulos1992}. In the time-in\-de\-pend\-ent case $H$ is the Hamiltonian,
while in the time-dependent case it is an operator constructed by the
particular CFME used.


The Krylov algorithm is not limited to the MPS algorithm, but it is commonly
used to obtain the new vector of a matrix exponential acting on a vector.
The idea \cite{Moler2003} is to change into a truncated basis (the Krylov
subspace) $V' = [ v_1, v_2', \ldots, v_n' ]$ in order to calculate the
propagated state $\ket{\psi(t + dt)} = \exp(-\i A dt) \ket{\psi(t)}$. The
vectors $v_j'$ are chosen as $A^{j-1} \ket{\psi}$ and are orthonormalized
to the basis $V = [v_1, v_2, \ldots, v_n ]$. In particular, the first Krylov
vector is
$v_1 = \ket{\psi} = V \ket{e_1}$ with $\ket{e_1} = (1, 0, 0, 0, \ldots, 0)^T$.
Approximating the dot product between the exponential and the state vector
leads to
\begin{eqnarray}
  \exp(-\i A dt) \ket{v_1}
 &\approx& V V^{\dagger} \exp(- \i A dt) \ket{v_1}
  =  V V^{\dagger} \exp(- \i A dt) V \ket{e_1}                                   \nonumber \\
 &=& V \exp(- \i V^{\dagger} A V dt) \ket{e_1} \, .
\end{eqnarray}
Calculating the exponential $M$ of the matrix
$V^{\dagger} A V \in \mathbb{R}^{n \times n}$ is a numerically feasible task as
long as the number of basis vectors $n$ is much smaller than the dimension of
the Hilbert space $D$. Furthermore, the relation simplifies to a real
tridiagonal matrix for Hermitian matrices, which is satisfied by the
Hamiltonian. This leads to
\begin{eqnarray}                                                                \label{eq:krylov:approx}
  \exp(-\i A dt) \ket{v_1}
  &\approx& V M \ket{e_1} = \sum_{i=1}^{n} M_{1, i} \ket{v_i} \, .
\end{eqnarray}

While in many applications where the state is represented as a vector this is
enough to obtain the approximation of $\ket{\psi(t + dt)}$, the problem in
the case of the MPS is that the summation over $\ket{v_i}$ can not be
exactly carried out, as the set of MPSs with fixed bond dimension do not form
a vector space. Based on the previous approaches using variational
algorithms, we instead find $\ket{\psi(t + dt)}$ by variationally optimizing
the overlap
\begin{eqnarray}                                                                \label{eq:krylov:fit}
  \ket{\psi(t + dt)} = \sum_{i=1}^{n} m_{1, i} \ket{v_i}
  \quad \Leftrightarrow \quad
  1 =  \bra{\psi(t + dt)} \left( \sum_{i=1}^{n} m_{1, i} \ket{v_i} \right)
  \, .
\end{eqnarray}
This procedure is done optimizing local tensors as in the ground state search.
However, instead of solving an eigenvalue problem at each iteration, this
optimization takes the form of a linear system of equations as shown on
the left part of Eq.~\eqref{eq:krylov:fit}. By exploiting the isometrization
of MPSs (see Eq.~\eqref{eq:oc}), this linear system of equations
is transformed into an inequality keeping the distance between the new
state $\ket{\psi(t + dt)}$ and its Krylov representation
$\sum m_{1,i} \ket{v_{i}}$ below a specified tolerance. The interested reader
can find further details on this algorithm in
Refs.~\cite{Wall2012,Schollwoeck2011}.

\subsection{Sornborger-Stewart decomposition                                   \label{sec:sornborger}}

This implementation inside the \OSMPS{} library is suitable for
nearest-neighbor Hamiltonians. The Sornborger-Stewart decomposition
\cite{Sornborger1999} used in the \OSMPS{} algorithms sweeps through the
system acting on every site, instead of every second pair of sites as in a
more common alternative
Suzuki-Trotter decomposition \cite{NielsenChuang}. The second order expansion
takes the form
\begin{eqnarray}                                                                \label{eq:sornborger2}
  \exp \left(\!- \i \dt \sum_{i=1}^{L-1} H_{i,i+1}\! \right)
  \!\!\!\! &=& \!\!\!\! \prod_{i=1}^{L-1} \exp \left(\!- \i \frac{\dt}{2} H_{i, i+1} \!\right)
      \prod_{i=1}^{L-1} \exp \left(\!- \i \frac{\dt}{2} H_{L - i, L - i + 1}\!
      \right) \, .                                                              \nonumber \\
\end{eqnarray}
We again follow the Krylov approach to propagate the quantum state under the
given MPO taking the exponential in the Krylov subspace. The essential
difference is the local characteristics of the Hamiltonian in the
Sornborger-Stewart decomposition. With the orthogonality center at one of the
sites $i$ and $j = i + 1$ being acted on, the overlap from the left and right
is the identity
operator. If we denote the two sites acted on with $\ket{C}$ and the parts
to the left of $i$ and right of $j$  with $\ket{L_i}$ and $\ket{R_j}$, the
actual state vector $\ket{\psi}$ can be derived from
Eq.~\eqref{eq:krylov:approx},
\begin{eqnarray}
  \ket{\psi} &=&
  \sum_{i,j} \ket{L_i} \ket{C_{i,j}} \ket{R_j} \, .
\end{eqnarray}
Within the construction of the Krylov basis $[v_1, \ldots, v_n]$ the states
$\ket{L_i}$ and $\ket{R_j}$ remain unchanged as shown in
Appendix~\ref{app:krylov}. The same applies to the sum of the propagated
state
\begin{eqnarray}
  \ket{\psi(t+dt)} = \sum_{k} c_k v_k \, ,
\end{eqnarray}
leading to the following construction of the state in the MPS picture:
\begin{eqnarray}
  \ket{\psi(t+dt)}
  = \sum_{i, j} \ket{A_i} \left( \sum_{k} c_{k} \ket{C_{i,j}(v_k)} \right)
               \ket{B_j} \, .
\end{eqnarray}
That means that we can sum locally over the two site tensors, as they form a
vector space, in contrast to the long-range case where we had to variationally
find the MPS closest to the summation.

In order to specify the error due to the method, we have to consider the
decomposition of the exponential. By separating non-commuting terms in matrix
exponential, we get an error of order $dt^2$ in the first order
approximation for a single time step:
\begin{eqnarray}
  \exp\left[(A + B) dt \right] =
  \exp(A dt) \exp(B dt) + \mathcal{O}(dt^2) \, .
\end{eqnarray}
Having implemented the second and fourth order approximations we obtain
methodical errors $\emeth$ for the whole time evolution as follows.
$\emeth$ is defined as part of the total error in Eq.~\eqref{eq:toterr}.
\begin{eqnarray}
  \etebdt = \mathcal{O}(dt^2) \, , \qquad
  \etebdf = \mathcal{O}(dt^4) \, .
\end{eqnarray}
In addition to the error of the Sornborger-Stewart decomposition, we have as
well an error from the Krylov subspace approximation to the exponential for
the local two-site propagators. The error bound for a single step is derived in
\cite{Gallopoulos1992}.

The convergence parameters necessary to set up time evolution with TEBD methods
reflect the simplicity of the approach. The parameter \texttt{psi\_local\_tol}
determines the local truncation on the singular values, while the pair
(\texttt{lanczos\_} \texttt{tol}, \texttt{max\_num\_lanczos\_iter}) provides the tolerance
for the Krylov approximation and the maximal number of Krylov vectors. In
addition, the maximum bond dimension $\chi$ can be defined.

With regards to the convergence study for the quench in the Ising model in
Fig.~\ref{fig:03_Ising} we make two observations. The fourth order
Sornborger-Stewart method is better than second order implementation of
Sornborger-Stewart, as expected due to the smaller error at each time
step. The rate of convergence as a function of the time step $dt$, that
is, the slope of the
line, is equal for both implementations. The error $\edt$ from the time-ordering
of the time-dependent Hamiltonian governs both implementations with
$\mathcal{O}(dt^2)$ and the better convergence of the fourth order
Sornborger-Stewart cannot be observed. In contrast, if the Hamiltonian is
time-independent, there is no error from the time-ordering.
Figure.~\ref{fig:B8_10ParaStepZ2} in Appendix~\ref{app:convergence} indicates that
TEBD4 has a higher rate of convergence in this case.
The total error has to be considered in comparison to the Krylov time
evolution in Sec.~\ref{sec:krylov} or the TDVP discussed in Sec.~\ref{sec:tdvp}.
We remark that the fourth order TEBD method has a comparable error to the Krylov
and TDVP method within one order of magnitude for the smallest time step
$dt = 10^{-4}$. Larger time steps and the second order method TEBD2 introduce
errors which are sometimes two orders of magnitude larger than the errors
introduced through Krylov or TDVP, especially for large time steps.

\subsection{Time-dependent variational principle                                \label{sec:tdvp}}

As a third option for the time evolution of a quantum system, we provide the
time-dependent variational principle (TDVP) \cite{Haegeman2016}. This is another
method supporting long-range interactions. In brief, all other previous methods
apply the propagator, which is an entangling many-body operator, to a state
represented as an MPS, and produces a new state obtained as an MPS. The updated
MPS has a larger bond dimension in general, so then we must variationally
project this new MPS onto the set of states with reduced computational
resources. The approach of the TDVP method is instead to project the
time-dependent Schr\"odinger equation onto the manifold of MPSs with fixed
bond dimension, and then integrate this equation directly within this manifold.
\OSMPS{} has implemented the two-site version of this
algorithm~\cite{Haegeman2016}, in which the bond dimension is still allowed
to grow over the course of time evolution, but the propagator is determined
from a projection of the full many-body operator onto a more local subspace.
We remark that TDVP performs well on the convergence study against exact
diagonalization methods in the example of Fig.~\ref{fig:03_Ising}. All four
estimates for the error defined in the Eqs.~\eqref{eq:erhoi} to \eqref{eq:eentr}
and shown in the four upper panels of Fig.~\ref{fig:03_Ising}
are close to exact diagonalization result with the same time step used as
reference. The maximal distance of all reduced two-site density matrices
does not reach the reference due to initial errors in the static results
for the ground state.

\subsection{Local Runge-Kutta propagation                                       \label{sec:lrk}}

Another option for time evolution with long-range interactions available in
\OSMPS{} is the local Runge-Kutta method proposed in \cite{Zaletel2015}. The
basic idea can be summarized as using the Runge-Kutta method on the local MPO
matrices instead of the whole propagator. The new MPO representing the
propagator $U_{\mathrm{LRK}}$ has a compact representation, i.e., it does not
increase in bond dimension beyond that of the Hamiltonian MPO. Since the
method is defined on the MPO, it is the third method supporting long-range
interactions.

A detailed overview on how to build the MPO for the propagator,
$U_{\mathrm{LRK}}$, as a second order approximation is beyond the scope of
this description; we suggest \cite{Zaletel2015} to the interested reader. But
we do provide here a short description of how the application of the MPO to
the MPS is implemented in \OSMPS{}. We fit the product of the
MPO and the MPS to a new MPS, where the new MPS represents the propagated
state,
\begin{eqnarray}
  \min_{\ket{\psi(t + dt)}}
  \| \ket{\psi(t+dt)} - U_{\mathrm{LRK}} \ket{\psi(t)} \|.
  \, .
\end{eqnarray}
This fit employs the same method used to fit the next Krylov vector
$\ket{v_{i+1}} = H \ket{v_{i}}$ in the Krylov method. The four upper panels
of Fig.~\ref{fig:03_Ising} show a similar behavior with regards to the
convergence of the LRK method as for the TEBD method. The
higher order variant of the LRK method reduces the error, but the rate of
convergence is not better due to the time-dependent Hamiltonian and the
particular choice of CFME.
Within the implementation of this time evolution method we use the EXPOKIT
package \cite{Expokit,Sidje1998} to calculate matrix exponentials.


\subsection{Time evolution case study: Bose-Hubbard model in a rotating
            saddle point potential                                             \label{sec:rotatingsaddle}}

\begin{figure}[t]
 \begin{center}
   \vspace{0.8cm}
   \begin{minipage}{0.48\linewidth}
     \begin{center}
     \begin{overpic}[width=0.9 \columnwidth,unit=1mm]{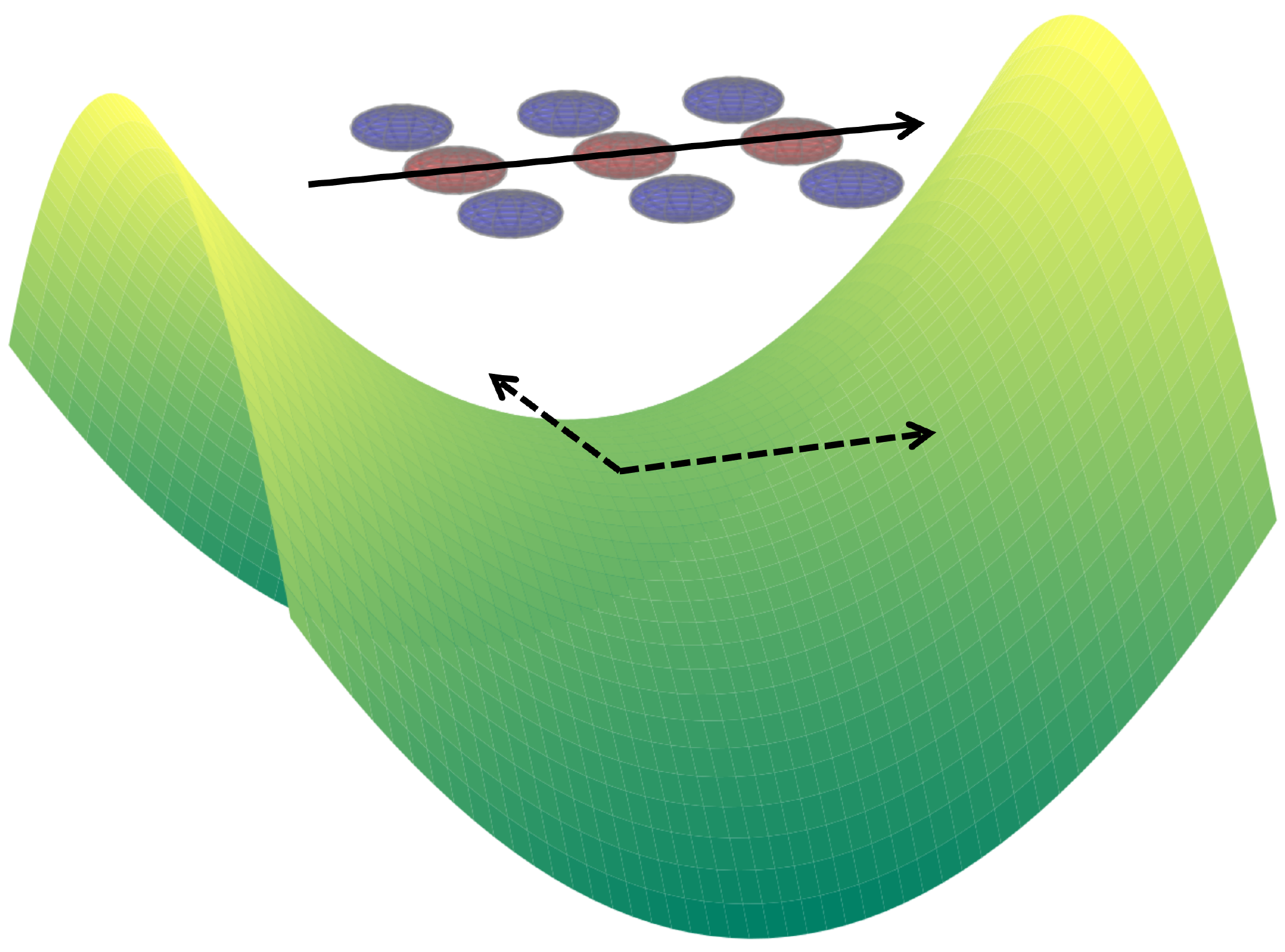}
       \put( 1,82){(a)}
       \put(67,44){$x(t)$}
       \put(33,47){$y(t)$}
       \put(68,67){$r$}
     \end{overpic}
     \end{center}
   \end{minipage}\hfill
   \begin{minipage}{0.48\linewidth}
     \begin{overpic}[width=1.0 \columnwidth,unit=1mm]{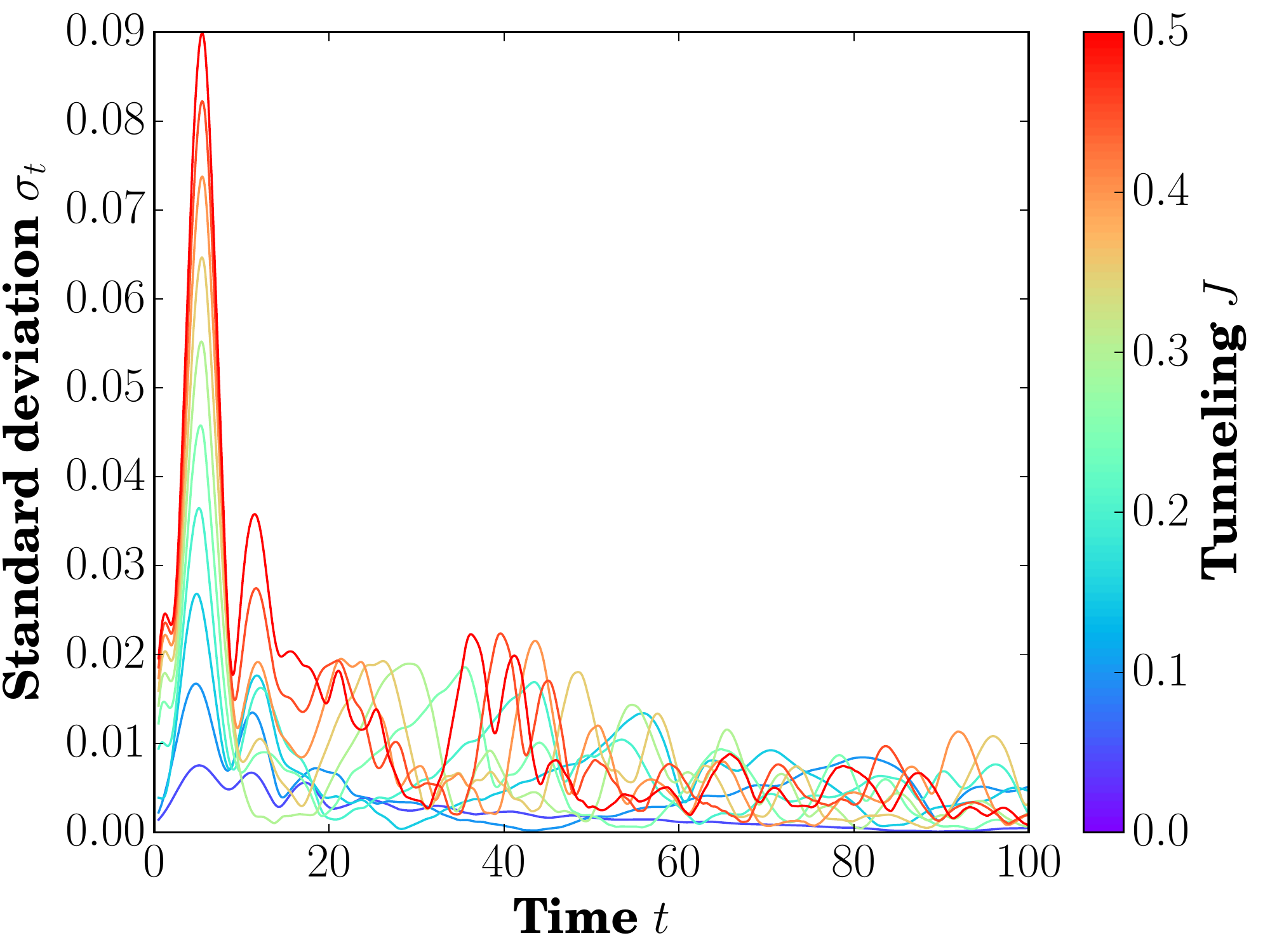}
       \put( 1,77.5){(b)}
     \end{overpic}
   \end{minipage}\vspace{0.5cm}
   \caption{\emph{Bose-Hubbard model in a rotating saddle-point potential.} The
     system shows an increasing stability for faster frequencies. (a)~In order
     to simulate a one dimensional system from the two dimensional optical
     lattice, we consider one slice of sites marked in red with its coordinate
     system $r$. The saddle point potential in green and yellow rotates under the
     system characterized by the coordinates $x(t), y(t)$. (b)~The
     standard deviation over time intervals of $\tau = 5$ is measured for the
     spreading in $x$, i.e. the standard deviation of the number operator. This
     observable can be expressed as $\sigma_{t}(\sigma_{x}(n), \tau=5)$.
     The time $t$ is in units of $\hbar / U$ and the tunneling $J$ is in units
     of the on-site interaction $U$.
                                                                                \label{fig:03_RotatingSaddlePointab}}
 \end{center}
\end{figure}
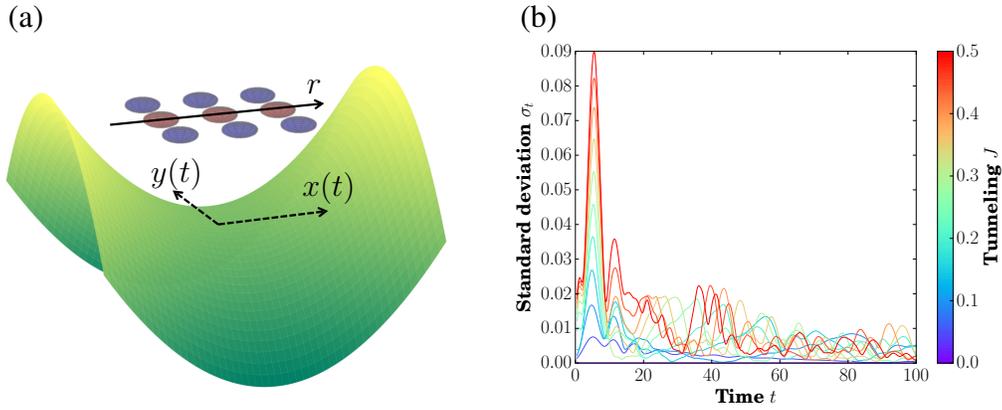

We consider bosons in an optical lattice confined in a rotating potential
as an example of the setup of a time evolution in \OSMPS{}. We consider the
potential $V_{xy}$ with a saddle point at $x = y = 0$ and a weight $c$,
\begin{eqnarray}
  V_{xy} &=& c \left(x^2 - y^2 \right) \, .
\end{eqnarray}
Figure~\ref{fig:03_RotatingSaddlePointab}(a) shows this potential for
a two dimensional lattice. We consider a one-dimensional optical lattice in this
potential marked by the red sites and start to rotate $V_{xy}$; the
one-dimensional system sees the following potential $V_{S}$ depending
on the angle of rotation $\phi \;$ and the distance $r$ from the center of
the potential and integrate it into the Bose-Hubbard Hamiltonian:
\begin{eqnarray}                                                                \label{eq:Vt}
  V_{S}
 &=& c \left(r \cos(\phi(t)) \right)^2
     - c \left(r \sin(\phi(t)) \right)^2 \, ,                                   \\
  H &=& - J \sum_{i=1}^{L-1} b_{i} b_{i+1} + h.c.
        + \frac{U}{2} \sum_{i=1}^{L} n_{i} (n_{i} - \1)
        + \sum_{i=1}^{L} V_{S}(t, i) n_{i} \, .
\end{eqnarray}
We slowly ramp up the frequency of the rotations with an acceleration
$\alpha$ starting with a flat potential
\begin{eqnarray}
  \phi(t) &=& \frac{\pi}{4} + \frac{1}{2} \alpha t^2 \, .
\end{eqnarray}
In order to estimate the stability of the system, we first calculate the
standard deviation of the number operator with regards to spatial dimension $x$
for every measurement in time, denoted by $\sigma_x(n)$ \footnote{We emphasize
that this is not the Pauli operator, but the standard derivation using in
contrast to the Pauli operators the subscript}. We calculate
$\sigma_t(\sigma_x(n, \tau = 5))$, that is the standard deviation of
$\sigma_x(n)$ for time intervals $t - \tau$ to $t$ for $\tau = 5$, in a second
step. For fast enough frequencies the rotating potential
stabilizes the system, which can be seen in
Fig.~\ref{fig:03_RotatingSaddlePointab}(b). The other trend is that
towards the superfluid regime, with higher tunneling, the stability decreases.
The first peak shows this trend.

The setup of the dynamics has many common steps with the statics. The
definition of the operators, the MPO representing the Hamiltonian, and
the observable class do not change. In regard to the time-dependent potential
$V_{S}$ in Eq.~\eqref{eq:Vt}, we define the weight $c$, the acceleration
$\alpha$ (\texttt{alpha}), the system size $L$, and the function returning
the corresponding potential at a point in time $t$. The latter step is
shown in Listing~\ref{py:03_potential}.

\pyfrag{./code/05_RotSaddle.py}
       {Define a time-dependent function for the rotating saddle potential.}
       {py:03_potential}{76}{85}

Next, we build the time evolution. As before with MPOs, observables, and
convergence parameters, the dynamics are contained in an instance of a
\py{} class: \texttt{QuenchList}. Once created, we can add multiple quenches
which are executed sequentially. We only add one quench in our example in
Listing~\ref{py:03_quench}. Different convergence parameters are supported
for each quench you add.

\pyfrag{./code/05_RotSaddle.py}
       {Creating the class object containing one or more quenches.}
       {py:03_quench}{87}{91}

As a last step we have to specify the dynamics in the dictionary for the
simulation, including the quenches and the observables to be measured
during the dynamics. The additional
lines are shown in Listing~\ref{py:03_dict}. The initial state of the
time evolution is the ground state if not specified otherwise.

\pyfrag{./code/05_RotSaddle.py}
       {Additional dictionary entries for the time evolution.}
       {py:03_dict}{114}{115}

The post processing works similar to the statics. A listed list contains the
dictionaries with the results of the simulation. The outer list contains
the different simulations, and the inner list contains the measurements for
each time. The complete \py{} code may be found in the supplemental
material described in Appendix~\ref{app:suppl}.

\section{Future developments                                                   \label{sec:future}}

The present \OSMPS{} library has a wide field of possible applications
including two-component mixtures \cite{Bellotti2017}, topological phases
\cite{Gong2016B}, and complexity in the quantum mutual information
\cite{Vargas2015}. Still,
we are planing to enhance the code to solve other sets of problems. The need
for enhancements is driven by our research in atomic, molecular, and
optical physics, macroscopic quantum phenomena, and complexity, but
we are open to suggestions from the community for extensions to the
\OSMPS{} library. For instance, the feature to provide the singular values
of a bipartition in addition to its bond entropy began as suggestions made
on our developer's site \cite{ForumRequestSVDSpectrum}.
We already have a broad suite of tools for the current MPS
algorithms for finite systems; therefore, we are focusing on implementing
additional measurements or new MPO rule sets to target Hamiltonians which
are impossible or difficult to build with present rule sets. For the
measurements, observables such as unequal time correlations are a possible
add-on. A user-friendly support for ladder systems or rectangular systems
with $L_x \gg L_y$ could be one focus for new types of Hamiltonians, either
as new rule sets or with convenient interfaces in \py{}. In contrast, other
tensor network structures such as PEPS are not being considered as an
extension to the library at this point. Periodic boundary conditions are
on the agenda for selected MPO rule sets.

On the other hand, we intend to extend \OSMPS{} for a new set of problems in
the future, namely open quantum systems. We anticipate them to be key
component to establish a more realistic picture of simulations with regards
to experiments, for example, future quantum computers will suffer effects
such as decoherence from the coupling to the environment; including those
effects in simulations fosters the understanding and addresses problems
induced by open systems. The standard approach is the Lindblad master
equation describing a system coupled weakly to a Markovian environment.
Several approaches have been proposed to implement the Lindblad master
equation, i.e. quantum trajectories, MPDOs, and Locally Purified Tensor
Networks (LPTNs) \cite{Werner2016}. We are developing as well techniques to
evolve open quantum systems with a non-Markovian environment within the
tensor network algorithms.

Finally, small features are on the list of future developments, e.g.
providing optional support for results in HDF5 file format for more convenient
data post processing, improving the speedup when using shared-memory
parallelization with openMP, and a TEBD algorithm which is not based on the
Krylov approximation.

\section{Conclusions                                                           \label{sec:conclusion}}

In this paper we have presented
a description of the MPS library \OSMPS{}
containing a set of powerful tools to study static and dynamic properties of
entangled one-dimensional many body quantum systems, where the initial focus is
on a broad set of methods for quantum simulators based on atomic, molecular and optical physics architectures.
These algorithms and interfaces are widely generalizable to
other fields of quantum physics, e.g. condensed matter and materials modeling.
The usefulness of our methods is underscored by the rapid community adaption and subsequent publications in
various areas of quantum physics
making use of the \OSMPS{} library or a derivative based on \OSMPS{}
\cite{Anisimovas2016,Bellotti2017,Dhar2016,Dolfi2014,Gardas2016,Gong2016,
Gong2016B,JaschkeLRQIC,Koller2016,Maghrebi2015,Russomanno2016,Vargas2015,Wall2012,
Wall2013,Wall2013NJP,Weimer2014}. \OSMPS{} has been downloaded more than 2300
times from over 55 countries since its initial release on SourceForge in January
2014.

We combine a \fort{} core with an easy to use \py{} front end. By introducing
the \py{} interface we hide the complex structures of the core algorithms
from the end user. Moreover, \py{} is a simpler programming language lowering
the barrier to start simulations for non-specialists and students.
\OSMPS{} provides fast and easy-to-access numerical predictions for 1D quantum
many-body systems. We facilitate this learning process by the recent integration
of a small exact diagonalization package inside \OSMPS{} written purely in \py{}
\cite{OSMPSED}. On the other hand, we also support sophisticated
enough features to provide helpful tools for quantum many-body theorists and
specialists in quantum computational physics. We presented
these tools throughout the paper describing the features of the library, which
include ground states of infinite systems and ground states plus low lying
excited states in finite size systems. For the dynamics of finite size
systems, we provided four different tools starting from a nearest-neighbor
Sornborger-Stewart decomposition, a modified Trotter approach, to methods supporting long-range
interactions such as Krylov subspaces, local Runge-Kutta, and the time
dependent variational principle. We gave a detailed description of the nuanced
convergence properties of these methods.

We anticipate that many unexplored problems in quantum simulators, materials
modeling, and other areas are suitable for \OSMPS{}, starting
with long-range quantum physics and its still unexplored corners
reaching to less studied models, e.g. facets of the XYZ model, and the
vast variety of untouched far from equilibrium dynamics. The development of
new features in \OSMPS{}, as described in Sec.~\ref{sec:future}, follows new
emerging fields in quantum physics which come naturally with explorations
and requests by the community of both end users and developers.

Providing the library as an open source package including a dedicated forum
fosters continued community development and research in many-body entangled
quantum physics with a transparent tool, much as density function theory
(DFT) was instrumental to the materials genome initiative. Especially the
aspect of modifying the core
code to integrate tailored tools on top or integrate modules into other open
source packages strengthens the idea of cooperative research. The version
2.0 of our library described in this paper is available under the
\emph{GNU General Public License} (GPL3) \cite{GPL3}
on our homepage \url{http://sourceforge.net/projects/openmps/}.

\section*{Acknowledgments}

We gratefully appreciate contributions from and discussions with A.~Dhar,
B.~Gardas, A.~Glick, W.~Han, D.~M.~Larue, and D.~Vargas during the
development of \OSMPS{}. We are equally thankful to the ALPS collaboration
\cite{Bauer2011,ALPS} and to C.~W.~Clark, I.~Danshita, R.~Mishmash,
B.~I.~Schneider, and J.~E.~Williams who contributed heavily to the predecessor
of \OSMPS{}, OpenTEBD \cite{OpenTEBDPackage}. The calculations were
carried out using the high performance computing resources provided by the
Golden Energy Computing Organization at the Colorado School of Mines. This
work has been supported by the NSF under the grants PHY-120881, PHY-1520915,
and OAC-1740130, and the AFOSR under grant FA9550-14-1-0287.


\begin{thebibliography}{10}
\bibitem{TDDFTNote}
This algorithm is not appropriate for capturing entanglement dynamics and
  strong correlations.
\end{thebibliography}

\section*{Appendices                                                           }\appendix
\renewcommand*{\thesection}{\Alph{section}}

\section{Convenient features                                                   \label{app:features}}

\OSMPS{} contains several features which exploit physical considerations to
increase the power and reach of the algorithms or are not intrinsically based
on physics but simplify the handling of simulations for the user. These
convenient features are covered in this part of the appendix.

\begin{itemize}
\item{\textbf{Symmetry conservation:} \OSMPS{} supports an arbitrary number of
  $\mathcal{U}(1)$ symmetries. In order to employ symmetries, the user
  has to provide the symmetry generator for each $\mathcal{U}(1)$ symmetry in
  a diagonal form. For the convergence study on the Bose-Hubbard model in
  Appendix~\ref{app:convergence}, the symmetry generator is the number
  operator $n$ (which is diagonal). Therefore, simulations can be easily
  adapted to number conservation if possible. The advantage of the
  $\mathcal{U}(1)$ symmetries is the ensuing numerical speedup.
  It is comparable to the use of block-diagonal matrices versus the full matrix
  with the block diagonal structure. Decompositions are carried out on smaller
  subspaces. Taking advantage of the block diagonal structure leads to a
  speedup of more than an order of magnitude for
  the example simulation treated in Table~\ref{tab:runtime_qmps}.

  Consider the one-dimensional Bose Hubbard model
  \begin{eqnarray}                                                              \label{eq:bosehubbard}
    H =
    - J \sum_{i=1}^{L-1} b_{i} b_{i+1}^{\dagger} + h.c.
    + \frac{1}{2} U \sum_{i=1}^{L} n_i (n_i - \1)
    - \mu \sum_{i=1}^{L} n_{i} \, ,
  \end{eqnarray}
  where the chemical potential $\mu$ regulates the average number of particles
  $\bar{n} = \frac{1}{L} \sum_{i=1}^{L} \langle n_i \rangle$ in the non-number
  conserving case. We scan for unit filling with $\bar{n}(\mu=-0.22) \approx 1$
  and use this chemical potential for the comparison together with $U=1.0$ and
  $J=0.5$. The compute times are determined from a \mioseventeenx{} node.
  The \fort{} library is compiled with \emph{ifort} and optimization flag
  \texttt{O3} and using \emph{mkl}. We see that the unit filling case can
  improve the simulation by an order of magnitude or more.

  \begin{minipage}{0.98\linewidth} 
    \begin{table}[H]
      \centering
      \begin{tabular}{@{} ccc @{}}
        \toprule
        Settings
        & $\Tcpu / \mathrm{s}$ (no $\mathcal{U}(1)$)
        & $\Tcpu / \mathrm{s}$ with $\mathcal{U}(1)$                            \\
        \cmidrule(r){1-1}  \cmidrule(rl){2-2} \cmidrule(l){3-3}
        $\chi=10,    \pevar = 10^{-4}, \pelanc =10^{-4}$
        & $42$     & $22$                                               \\
        $\chi=60,    \pevar = 10^{-7}, \pelanc = 10^{-6}$
        & $654$    & $21$                                              \\
        $\chi=320,   \pevar =10^{-12}, \pelanc = 10^{-10}$
        & $19838$  & $150$                                             \\
        \bottomrule
      \end{tabular}
      \caption{Comparison of the CPU time $\Tcpu$ in seconds for number
        conserving and
        non-number conserving MPS algorithm in the case of the Bose Hubbard
        model with local dimension $d=6$ and $L=32$ for different
        convergence parameters.
                                                                                \label{tab:runtime_qmps}}
    \end{table}
    \vspace{0.1cm}
  \end{minipage}

  In addition, we study the Ising model defined in Eq.~\eqref{eq:Ising} where
  we can make use of the
  $\mathbb{Z}_{2}$ symmetry, which reduces the dimension of the Hilbert space
  by half. In the block diagonal structure, we never have more than two blocks
  and we expect that using a symmetry-conserving MPS does not lead to the
  same speedup as in the Bose-Hubbard model. In Table~\ref{tab:runtime_zmps}
  we find that for small bond dimension the overhead due to symmetry
  conservation is larger than the actual gain. This is in agreement with
  previous results in \cite{Singh2011}. Starting with $\chi = 320$ in
  this table, the simulation with $\mathbb{Z}_{2}$ is faster than the one
  without symmetry. The simulations were run with the same setting as
  the Bose-Hubbard model simulations on a \mioseventeenx{} node. The times
  include the ground state calculation, one local measurement, and one
  correlation measurement where the value of the external field
  $g=0.98$ is close to the critical point.

  \begin{minipage}{0.98\linewidth} 
    \begin{table}[H]
      \centering
      \begin{tabular}{@{} ccc @{}}
        \toprule
        Settings
        & $\Tcpu / \mathrm{s}$ (no $\mathbb{Z}_{2}$)
        & $\Tcpu / \mathrm{s}$ with $\mathbb{Z}_{2}$                                        \\
        \cmidrule(r){1-1}  \cmidrule(rl){2-2} \cmidrule(l){3-3}
        $\chi=10,    \pevar = 10^{-4}, \pelanc = 10^{-4}$
        & $0.15$     & $0.79$                                              \\
        $\chi=60,    \pevar = 10^{-7}, \pelanc = 10^{-6}$
        & $0.57$     & $1.94$                                              \\
        $\chi=320,   \pevar = 10^{-12}, \pelanc = 10^{-10}$
        & $6.31$     & $5.81$                                              \\
        \bottomrule
      \end{tabular}
      \caption{Comparison of the CPU time $\Tcpu$ in seconds for MPS
        ground states
        with and without conservation of $\mathbb{Z}_{2}$ symmetry in the
        quantum Ising model. The system size is $L=128$, the interaction
        $J = 1$, the external field $g = 0.98$, and default
        convergence parameters are used if not specified in the table.
                                                                                \label{tab:runtime_zmps}}
    \end{table}
    \vspace{0.1cm}
    \end{minipage}
}
\item{\textbf{Support for fermionic systems:} Fermionic systems obey
  nonlocal anticommutation relations different from the local commutation
  relations of bosons. In order to represent fermionic operators in \OSMPS{},
  we use a Jordan-Wigner transformation \cite{Jordan1928,SachdevQPT} which
  uses locally anticommuting
  operators together with strings of phase operators $(-1)^{n}$, with
  $n$ the number operator. The routines within MPS, for example, the
  measurement of correlation functions and construction of Hamiltonian terms,
  have a flag \texttt{Phase} for whether a string of phase operators is
  needed to enforce proper fermionic anticommutation relations.

  Such fermions can be described via the Fermi-Hubbard
  model~\cite{Hubbard1963}. We consider in the following an example of
  spinless fermions using the Hamiltonian from \cite{Wang2014} in a
  one-dimensional system
  \begin{eqnarray}                                                              \label{eq:Hspinlessfermihubbard}
    H_{F}
    &=& W \sum_{i=1}^{L-1} \left( n_{i} - \frac{1}{2} \right)
                           \left( n_{i + 1} - \frac{1}{2} \right)
        - J \sum_{i=1}^{L-1} \left( c_{i}^{\dagger} c_{i+1} + h.c. \right) \, .
  \end{eqnarray}
  We briefly introduce the phase terms for fermionic systems and show
  that it affects the system. Instead of comparing a single
  correlation measurement or the single particle density matrix built
  from the correlations $\langle c_{i}^{\dagger} c_{j} \rangle$, we use
  the quantum depletion $\xi$ from the eigenvalues $\Xi_i$ of the
  single particle density matrix, yielding a single value. Furthermore,
  the quantum depletion $\xi$ is constant for spinless fermions and
  errors can be detected more easily. If $\Xi_{i}$ are the eigenvalues
  of the single particle density matrix, the depletion is defined as
  \begin{eqnarray}                                                              \label{eq:defdepletion}
    \xi = 1 - \frac{\max_{i} \Xi_{i}}{\sum_{i} \Xi_{i}} \, .
  \end{eqnarray}
  Figure~\ref{fig:A2_FermiHubbard} shows multiple aspects of the model
  with a system size of $L=65$ and approximately half-filling with $33$
  fermions in the system. The deviation from the average filling at each
  site
  \begin{eqnarray}
    \delta
   &=& \frac{1}{65} \sum_{i=1}^{65} \left| n_{i} - \frac{33}{65} \right| \, ,
  \end{eqnarray}
  and the bond entropy indicate the phase transition from a charge
  density wave to a superfluid phase dominated by the tunneling
  term in the Hamiltonian. Those values are plotted in
  Figure~\ref{fig:A2_FermiHubbard}(a). In Figure~\ref{fig:A2_FermiHubbard}(b) we
  show the quantum depletion in three different scenarios including
  calculating it correctly from the correlation with the phase terms,
  and incorrectly from the correlation without phase terms and the
  reduced density matrices. We emphasize that the reduced density
  matrices never contain phase terms and cannot be used to calculate
  correlation in fermionic systems which need a phase term.
}
\end{itemize}

\begin{figure}[t]
  \begin{center}
    \begin{minipage}{0.48\linewidth}
      \begin{overpic}[width=1.0\columnwidth,unit=1mm]{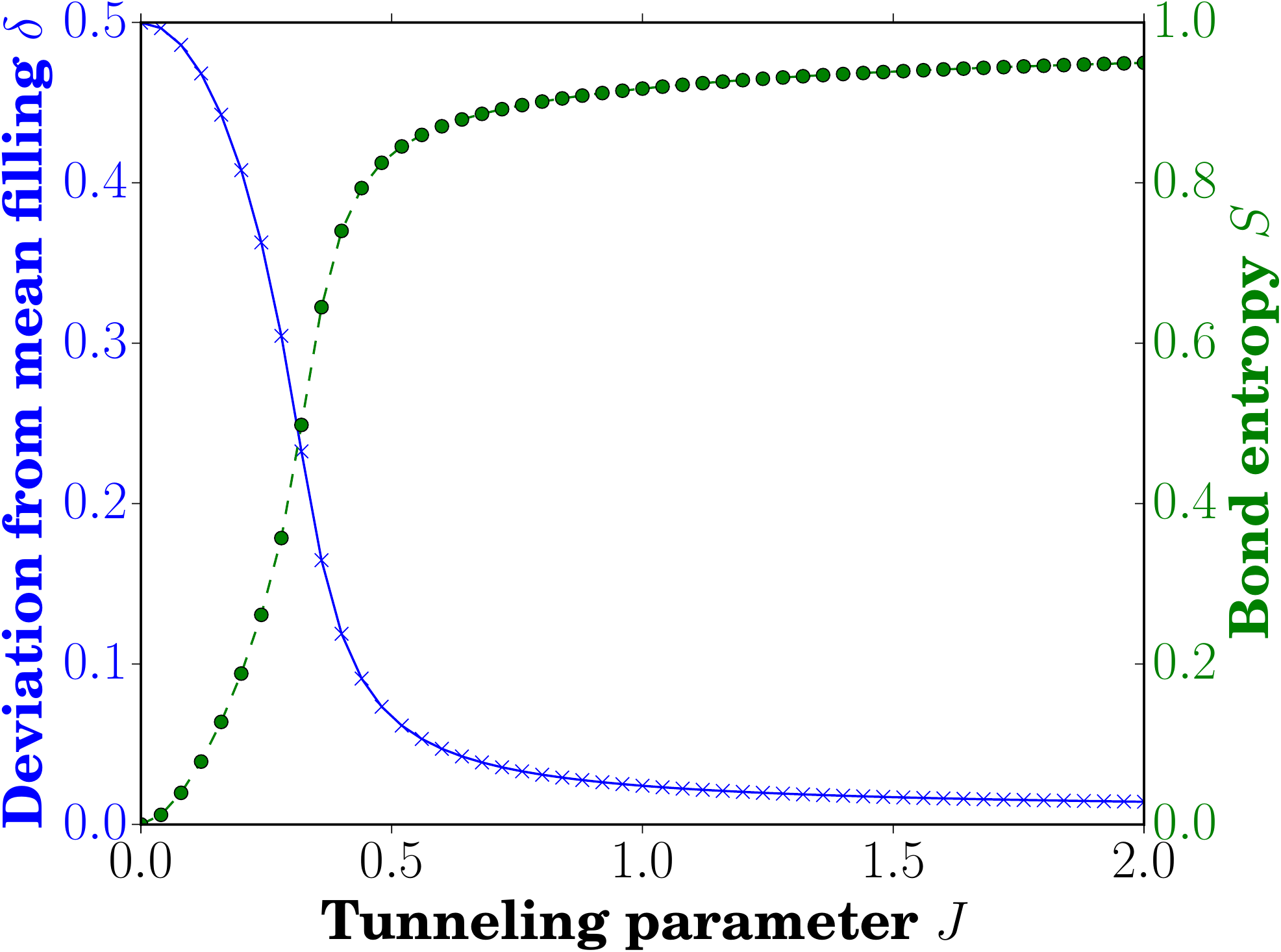}
        \put( 1,77){(a)}
      \end{overpic}
    \end{minipage}\hfill
    \begin{minipage}{0.48\linewidth}
      \begin{overpic}[width=1.0 \columnwidth,unit=1mm]{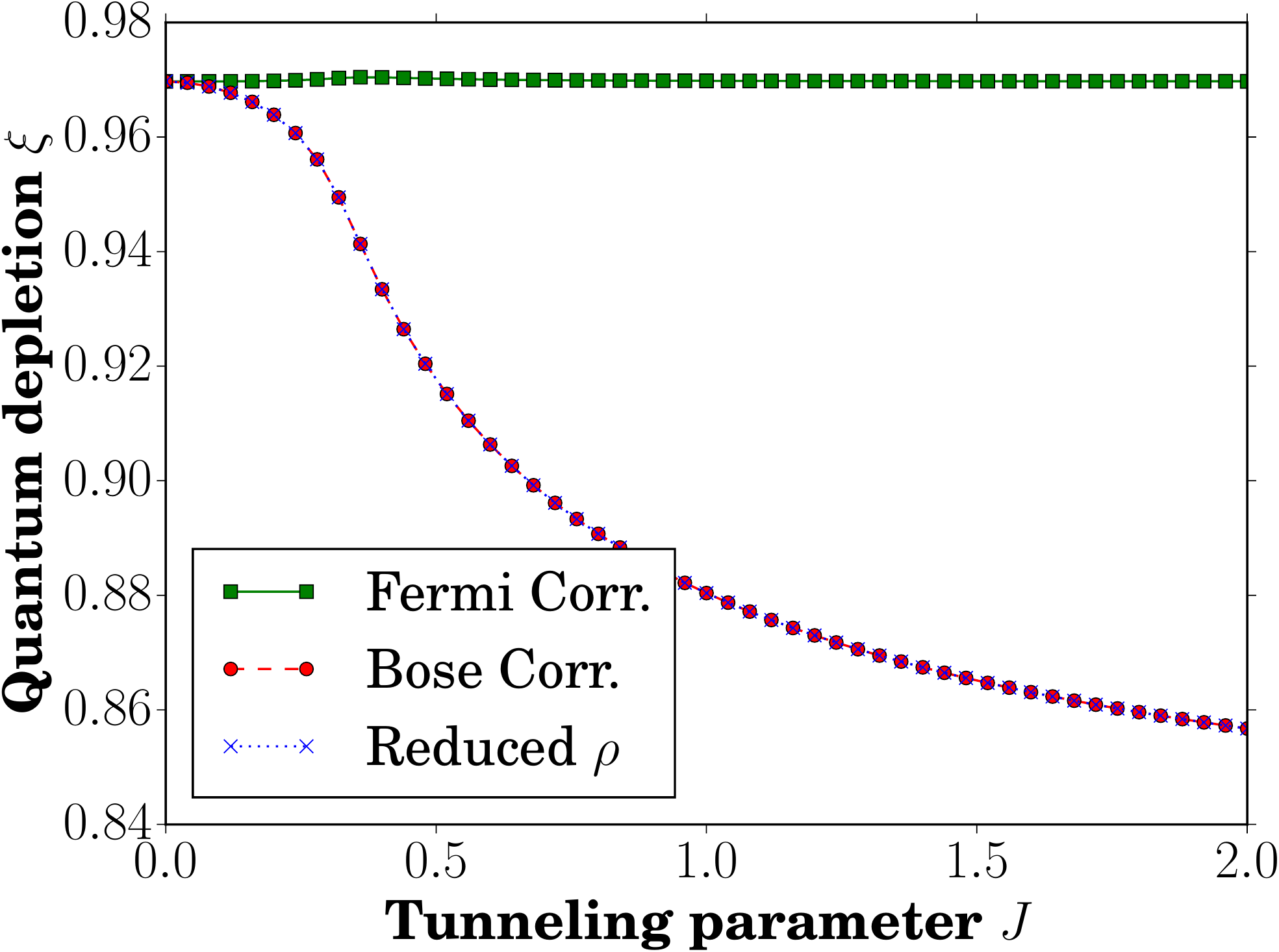}
        \put( 1,77){(b)}
      \end{overpic}
    \end{minipage}
    \caption{\emph{Spinless Fermions at half-filling.} (a) The deviation
    from the mean-filling $\delta$ (blue x's) shows the two different phases. For small
    tunneling $J < 1$ there is an alternating order of empty-occupied sites
    dominating. In contrast, the tunneling term with $J > 1$ leads to a
    superfluid-like phase. The bond entropy $S$ shows the same phase transition
    with a peak around $J=1.0$. We point out that the simulation for
    $J = 0$ fails because there are no fluctuations to feed the variational
    minimization in the ground state search. The tunneling parameter $J$
    is in units of the nearest-neighbor interaction $W$. The first point uses a
    small perturbation of $J = 10^{-4}$.
    (b) The quantum depletion $\xi$ is calculated
    from the correlation matrix $\langle c_{i}^{\dagger} c_{j} \rangle$. The
    result from the boson-type correlation without a phase term and the reduced
    density matrices are identical, corresponding to the lower curve. Both
    results do not match the correct curve when accounting for the phase
    terms of the Jordan-Wigner transformation.
                                                                                \label{fig:A2_FermiHubbard}}
  \end{center}
\end{figure}

\begin{itemize}
\item{\textbf{MPI:} Message Passing Interface (MPI) is the standard for
  distributed memory parallel high performance computing. \OSMPS{}
  supports the setup of data parallelism via MPI. Although data
  parallelism is the most basic implementation
  of a parallel algorithm, it represents the most efficient approach when
  iterating over a set of parameters. For a large fraction of problems, we
  can assume that iterations over a set of parameters are necessary as in
  the case of phase diagrams, finite-size scalings, to analyze
  Kibble-Zurek scalings, or to scan over a variety of initial conditions.
  We provide a \fort{} implementation with one master
  and $(p-1)$ workers where $p$ is the total number of cores. Furthermore, a
  \py{} implementation is available with $p$ workers where one of them is
  distributing jobs.

  As an example for the scaling of the MPI implementation, we take a look at the
  \fort{} implementation and the simulations necessary to generate the data for
  Fig.~\ref{fig:01_Ising_BE3d}. We run the simulation on one or two nodes of
  the type \mioseventeenx{} with $12$, $24$, $36$, and $48$ cores. The
  duration of the jobs can be found in Table~\ref{tab:mpiscal}. We recall
  that the set of simulations iterates over $30$ data points for the system
  size $L$ and $51$ data points for the external field $g$, i.e., $1530$
  data points in total. In the case of $48$ cores, the average workload are
  approximately $32$ data points. Since the master distributes the jobs to
  the workers one by one, each worker might handle less (more) jobs if their
  jobs take more (less) than the average time of one data point. In the
  setup of the simulation, we avoid getting stuck in long simulations at the
  end having other cores running idle, because we address the large system sizes
  first. The longest single data point takes about $4.5$ hours. This setup
  leads to a good scaling of the simulation time when increasing the number of
  cores. In fact, when increasing the number of cores from $12$ by a factor
  of $n$, the speedup is greater than $n$. This is related to the fact that
  the master distributing the jobs has less weight for more cores. The parallel
  efficiency $E_{P}$ is calculated with the cumulative CPU time of all simulations
  $T_{\Sigma}$ and the actual run time of the job $T_N$ on $N$ cores. We assume
  that the cumulative CPU time corresponds to the time of the serial job:
  \begin{eqnarray}                                                              \label{eq:pareff}
    E_{P} &=& \frac{T_{\Sigma}}{N \cdot T_{N}} \, .
  \end{eqnarray}
  The initial increase of $E_{P}$ seen in Table~\ref{tab:mpiscal} is again
  explained with the master running almost idle.

  \begin{minipage}{0.98\linewidth} 
    \begin{table}[H]
      \centering
      \begin{tabular}{@{} ccccccc @{}}
        \toprule
        Cores (Workers)       & 12 (11)   & 24 (23)   & 36 (35)   & 48 (47)   & 72 (71) & 96 (95)    \\
        \cmidrule(r){1-1}     \cmidrule(rl){2-2} \cmidrule(rl){3-3} \cmidrule(rl){4-4} \cmidrule(rl){5-5} \cmidrule(rl){6-6} \cmidrule(l){7-7}
        $\Tjob$ (hh:mm)          & 68:03     & 32:47     & 21:17     & 15:52     & 10:40    & 7:50      \\
        Efficiency $E_{P}$    & $91.7 \%$ & $95.9 \%$ & $97.3 \%$ & $97.9 \%$ & $97.1\%$ & $97.5 \%$ \\
        \bottomrule
      \end{tabular}
      \caption{\emph{MPI scaling for \OSMPS{}.} As example for the MPI
        scaling of the duration time $\Tjob$ in hours and minutes of
        all simulations, we execute the simulations for
        Fig.~\ref{fig:01_Ising_BE3d} on different numbers of cores. In order
        to evaluate the scaling the number of workers excluding the master
        job distributing the jobs has to be considered.
                                                                                \label{tab:mpiscal}}
    \end{table}
    \end{minipage}
}
\item{\textbf{Templates:} The \OSMPS{} library supports calculations with
  real and complex numbers as well as standard MPS or symmetry conserving
  MPS algorithms inside the \fort{} modules. The different data types
  lead to many redundant subroutines which we overcome
  by using templates. Subroutines are written for a generic type, e.g.
  \texttt{MPS\_TYPE}. When generating the module the necessary types are
  plugged in, e.g. \texttt{MPS}, \texttt{MPSc}, \texttt{qMPS}, and
  \texttt{qMPSc}. The corresponding call to such a subroutine is covered
  under an interface (containing all the subroutines for the different types).
  These templates reduce errors copying from type to type and keep modules
  shorter.}
\end{itemize}

\section{Convergence studies                                                   \label{app:convergence}}

\OSMPS{} can be tuned via multiple parameters modifying simulations with
regards to the convergence. Therefore, we provide some guidelines for the
convergence parameters along with examples. We divide this appendix in
one part looking at the details of the finite size statics followed by
additional studies of the time evolution methods.

\subsection{Finite size variational algorithms}

Due to the variational search we have several convergence parameters. We can
divide them into three categories:

\begin{itemize}
\item{\emph{Lanczos}: Lanczos tolerance, maximal number of Lanczos iterations}
\item{\emph{Chi/bond dimension}: maximal bond dimension, local
  tolerance, variance tolerance}
\item{\emph{Iteration}: min/max num sweeps, max outer sweeps}
\end{itemize}

Due to the number of convergence parameters, we present a simplified
picture of convergence issues with a single set of convergence parameters. We
support multiple subsequently executed sets of convergence parameters, which
allow users to refine their target state in multiple steps when increasing
the bond dimension, Lanczos iterations and number of sweeps while decreasing
the Lanczos tolerance, local tolerance, and variance tolerance. In general,
the control of convergence over the soft cutoffs defined via tolerances should
be preferred over the hard cutoff specified as the bond dimension or number
of Lanczos iterations.

\begin{figure}[t]
  \begin{center}
    \begin{minipage}{0.48\linewidth}
      \vspace{0.99cm}
      \begin{overpic}[width=1.0 \columnwidth,unit=1mm]{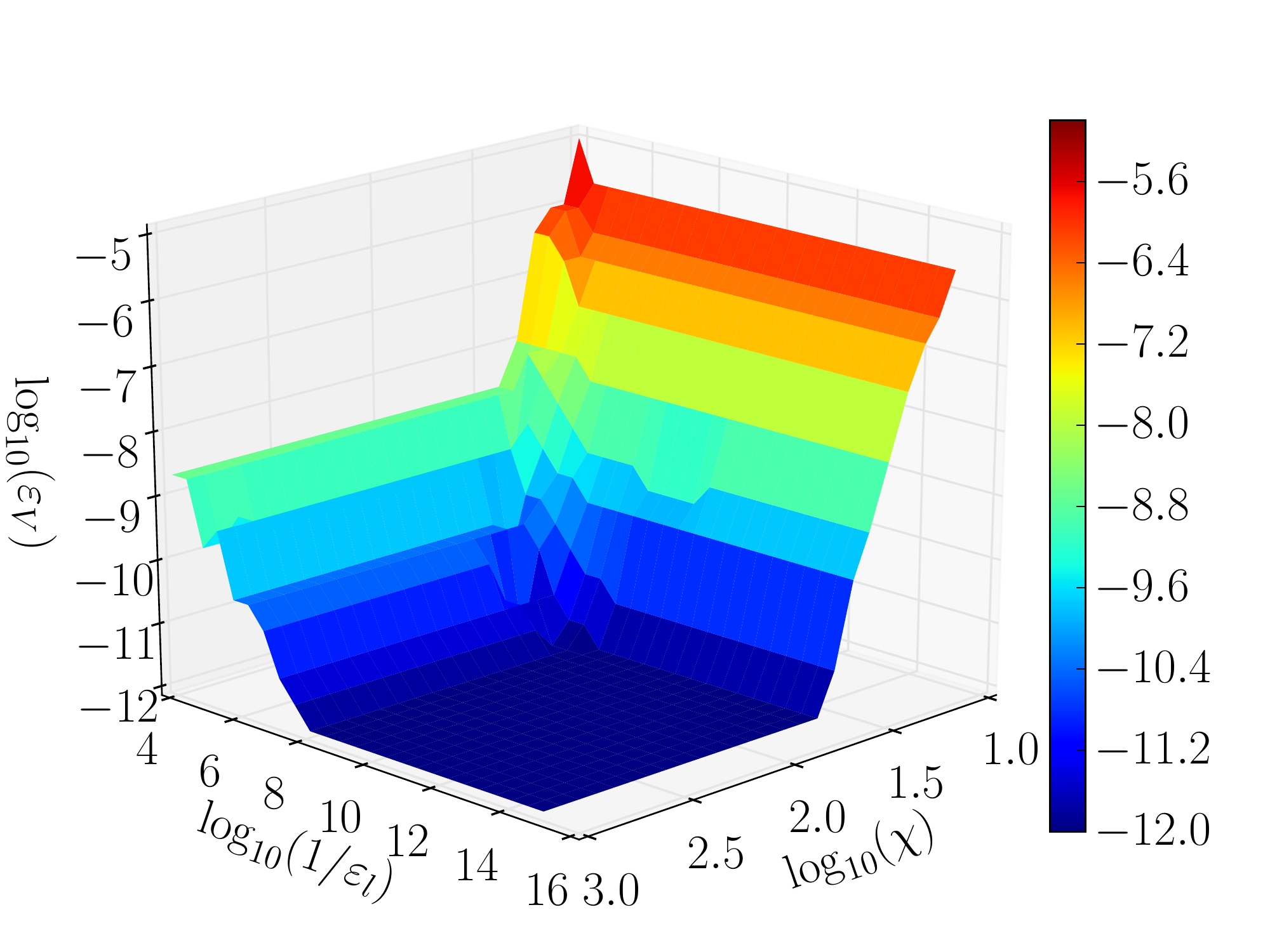}
        \put(2,67){(a)}
      \end{overpic}
    \end{minipage}\hfill
    \begin{minipage}{0.48\linewidth}
      \vspace{0.99cm}
      \begin{overpic}[width=1.0 \columnwidth,unit=1mm]{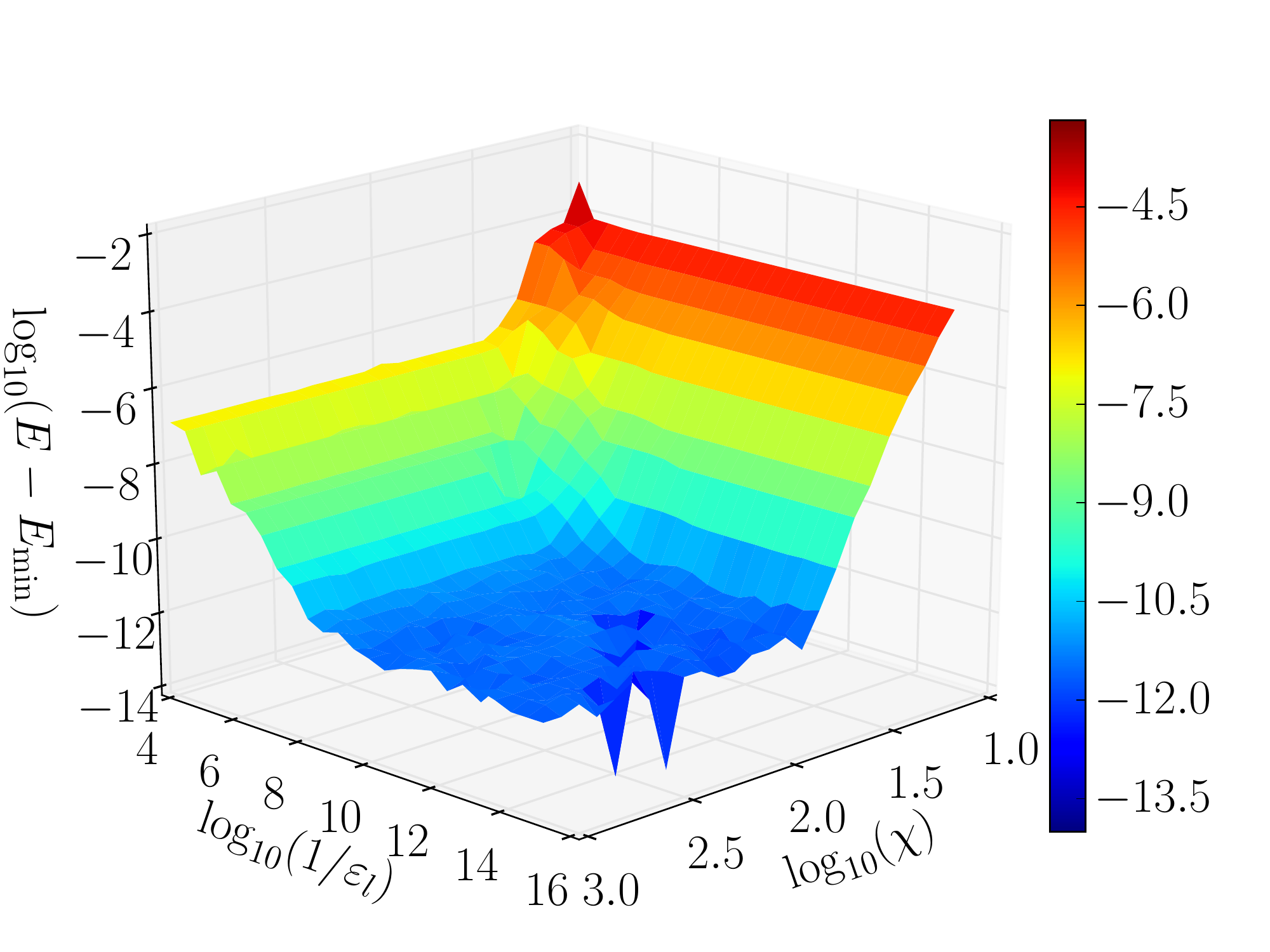}
        \put(2,67){(b)}
      \end{overpic}
    \end{minipage}
    \caption{\emph{Convergence study for the quantum Ising model.}
      (a) The boundary between converged and non-converged simulations
      as a function of the bond dimension and Lanczos tolerance. The surface
      shows the first converged simulations fulfilling the variance tolerance
      criteria. Smaller variance tolerances will lead to non-convergence
      for fixed parameters. With this plot, we can estimate the maximal
      variance tolerance that can be achieved using a given bond dimension
      and Lanczos tolerance.
      (b) The energy difference as a function of bond dimension and
      Lanczos tolerance for the first converged simulation in regard to the
      variance tolerance (see (a)) reproduces the expectation
      that the energy difference decreases as the convergence criteria
      defined over the variance decreasing tolerance is met. Same labels
      apply to color bar and $z$-axis.
                                                                                \label{fig:B1_Ising}}
  \end{center}
\end{figure}

For the convergence study presented in Table~\ref{tab:conv_mps} we concentrate
on the Lanczos tolerance $\pelanc$, the variance tolerance $\pevar$
and the bond dimension $\chi$. The maximal number of Lanczos
iterations is set to a sufficiently high value to ensure convergence ($500$,
default is $100$). The number of inner sweeps is set to exactly $2$, i.e. a
minimum and maximum of 2, except for the spinless fermions with 4. We
consider exactly one outer sweep. Since every additional outer sweep lowers
the local tolerance $\peloc$ for the cut-off of singular values, we
prevent different
$\peloc$ depending on the number of outer sweeps carried out. The warmup phase
to grow the system up to $L$ sites has a warmup tolerance $100$ times bigger
than the variance tolerance and the warmup bond dimension is half of the bond dimension.
The local tolerance is then connected with the variance tolerance as
specified in the default settings:
\begin{eqnarray}
  \peloc = \frac{\pevar}{4 L} \, .
\end{eqnarray}
For the remaining three parameters $\pelanc$, $\pevar$ and $\chi$
the variance tolerance determines if a simulation is converged according to the
criterion
\begin{eqnarray}
  \langle H^2 - \langle H \rangle^2 \rangle < \pevar L \, .
\end{eqnarray}
The actual behavior may vary for different models. We study the convergence
behavior of the Ising model, the Bose Hubbard model, and a spinless
Fermi-Hubbard model. Future possibilities for similar studies include
spinful Fermi-Hubbard or Bose-Hubbard models, XYZ models, quantum rotors,
and disordered systems. Starting with the Ising model,
we show in Fig.~\ref{fig:B1_Ising} (a) the boundary between converging and
non-converging simulations over a grid of $\pelanc$ and $\chi$. In
Fig.~\ref{fig:B1_Ising} (b) the energy difference to the smallest value can be
found.
The point is close to the quantum critical point ($g=0.98$ with $g_c = 1.0$) so
the settings also serve as an upper bound for points further away from the
critical point.

\begin{figure}[thbp]
  \begin{center}
    \begin{minipage}{0.32\linewidth}
      \vspace{0.99cm}
      \begin{overpic}[width=1.0 \columnwidth,unit=1mm]{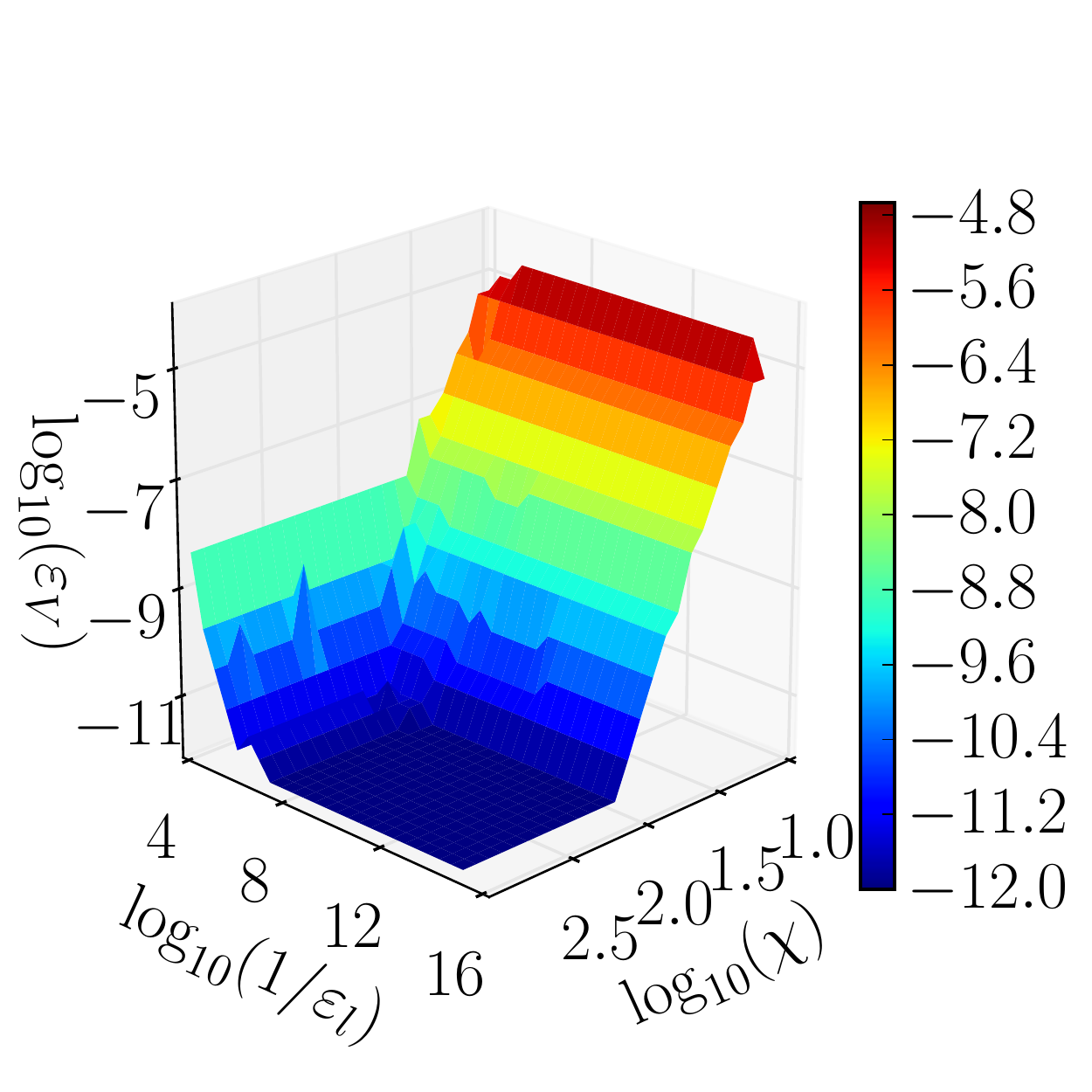}
        \put(2,87){(a)}
      \end{overpic}
    \end{minipage}\hfill
    \begin{minipage}{0.32\linewidth}
      \vspace{0.99cm}
      \begin{overpic}[width=1.0 \columnwidth,unit=1mm]{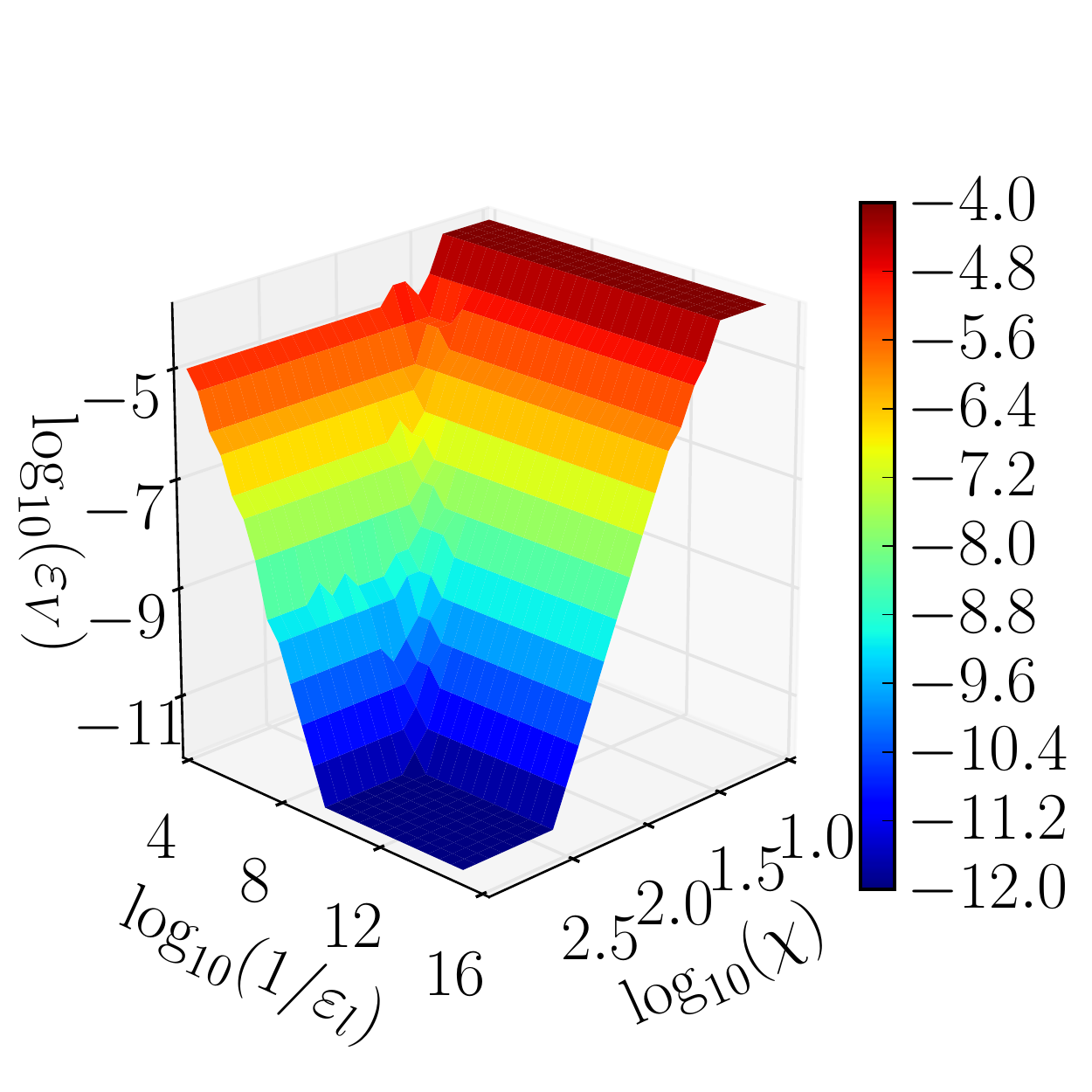}
        \put(2,87){(b)}
      \end{overpic}
    \end{minipage}\hfill
    \begin{minipage}{0.32\linewidth}
      \vspace{0.99cm}
      \begin{overpic}[width=1.0 \columnwidth,unit=1mm]{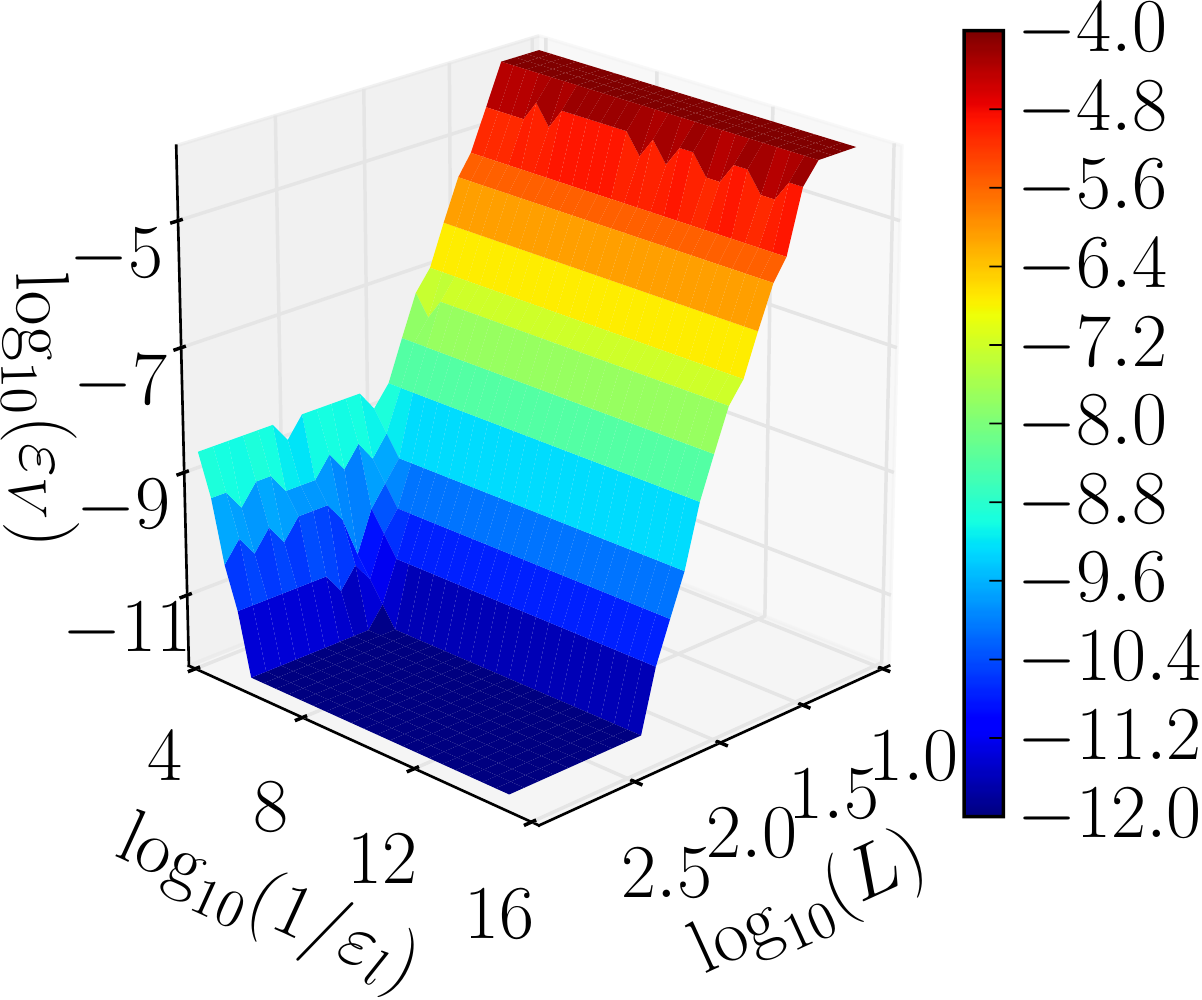}
        \put(2,87){(c)}
      \end{overpic}
    \end{minipage}
    \caption{\emph{Convergence study.} We study the variance tolerance
      achieved as a function of the bond dimension and the Lanczos tolerance
      for (a) the long-range Ising model, (b) the Bose-Hubbard model, and (c)
      a spinless Fermi-Hubbard model. (Same labels apply to color bar and
      $z$-axis.)
                                                                                \label{fig:B234_ConvSurf}}
  \end{center}
\end{figure}

In the next example we turn to the long-range quantum Ising model with the
Hamiltonian presented in Eq.~\eqref{eq:LRIsing}. We evaluate the convergence
behavior again close to the critical point. For a power-law decay with
$\alpha = 3.0$ and a system size of $L = 128$ we have a critical field of
$g_{c} \approx 1.35$ \cite{JaschkeLRQIC}. We leave the remaining parameter
fixed in comparison to the nearest neighbor quantum Ising model.
Figure~\ref{fig:B234_ConvSurf} (a) shows again the boundary between converging
and non-converging simulations based on the variance tolerance $\pevar$
iterating over the bond dimension $\chi$ and the Lanczos tolerance $\pelanc$.
We see that we need a much higher bond dimension in comparison to the
nearest-neighbor Ising model.

For the Bose Hubbard model introduced in Eq.~\eqref{eq:bosehubbard} the same
type of plot is shown in Fig.~\ref{fig:B234_ConvSurf}(b).
The algorithm includes number conservation at unit filling and the system
size is $L=32$. We choose the point with $U=0.5$, $t=0.5$ and $\mu=0$ in the
superfluid regime exhibiting long-range correlations. In contrast, the Mott
insulator has less entanglement and therefore the superfluid parameters can
serve as upper bound. In the Figs.~\ref{fig:B1_Ising} and \ref{fig:B234_ConvSurf}
we observe that the Bose-Hubbard model needs, in comparison to the Ising model,
stricter convergence parameters to arrive at an equal variance tolerance.

Finally, we consider the spinless fermions introduced in
Eq.~\eqref{eq:Hspinlessfermihubbard} and present the result on the convergence
in Fig.~\ref{fig:B234_ConvSurf} (c). As with the Ising model, we choose a
point in the region with high entanglement judged by the bond entropy evaluated in
Fig.~\ref{fig:A2_FermiHubbard}, that is a tunneling energy of $J = 1.04$. Furthermore, we take
$L = 65$, and the system is filled with $33$ fermions. In contrast to the other
simulations we increase the minimum and maximum number of inner sweeps to $4$.
Comparing the Fermi-Hubbard model to the quantum Ising model since both have
a local dimension of $d = 2$ and are nearest-neighbor, we see that the
fermionic system needs a larger bond dimension $\chi$ in comparison.

\subsection{Time evolution methods for finite size systems               \label{app:convtime}}

\begin{figure}[t!]
 \begin{center}
    \includegraphics[width=0.65 \columnwidth]{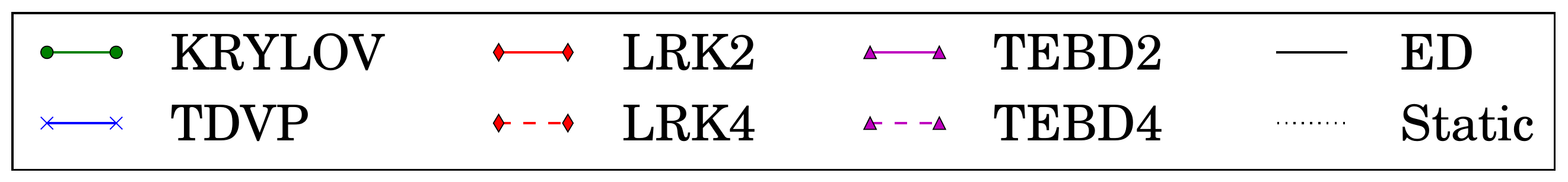}
    \begin{minipage}{0.48\linewidth}
      \begin{overpic}[width=0.9\columnwidth,unit=1mm]{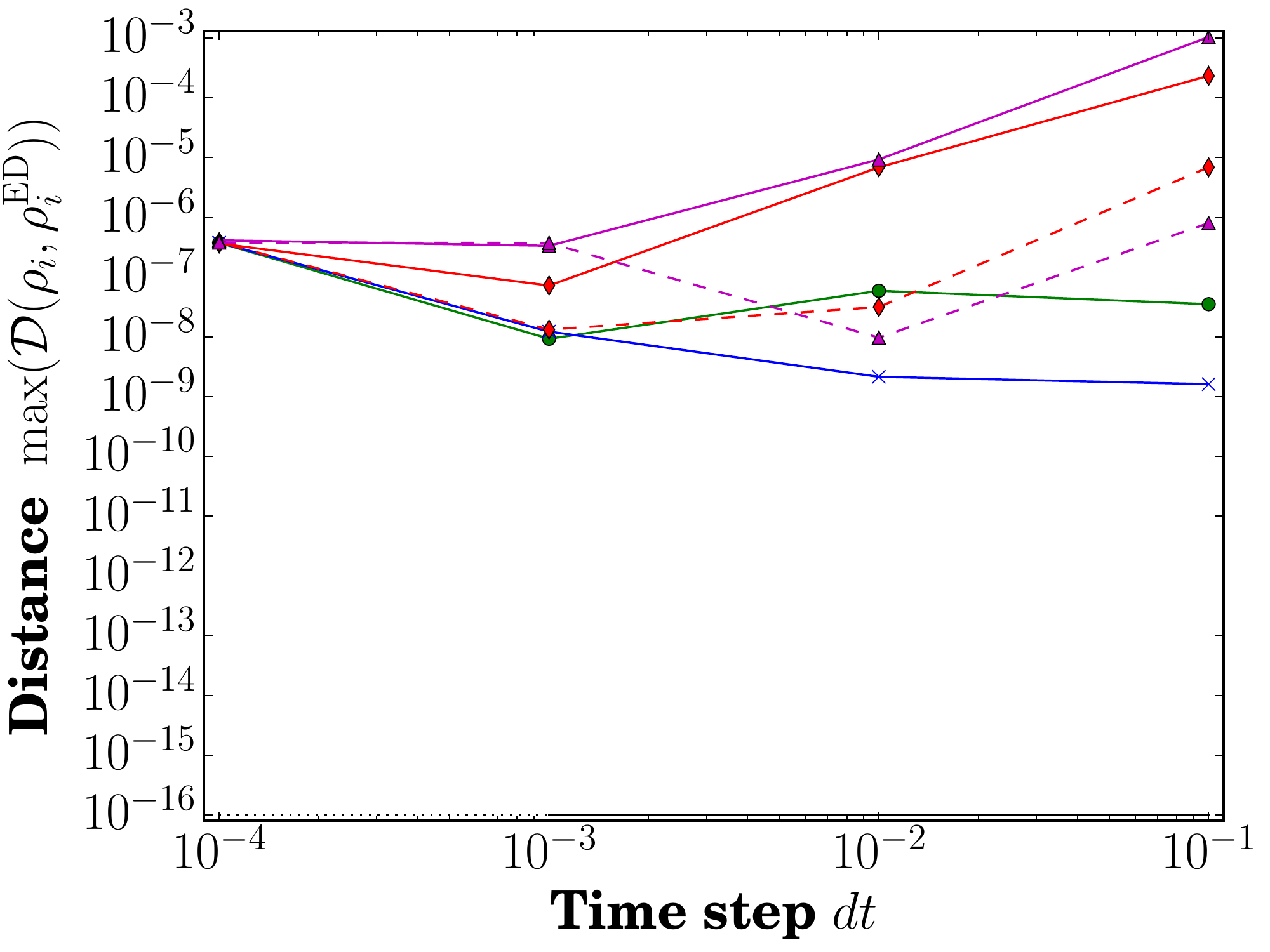}
        \put(18,67){(a)}
      \end{overpic}
    \end{minipage}\hfill
    \begin{minipage}{0.48\linewidth}
      \begin{overpic}[width=0.9 \columnwidth,unit=1mm]{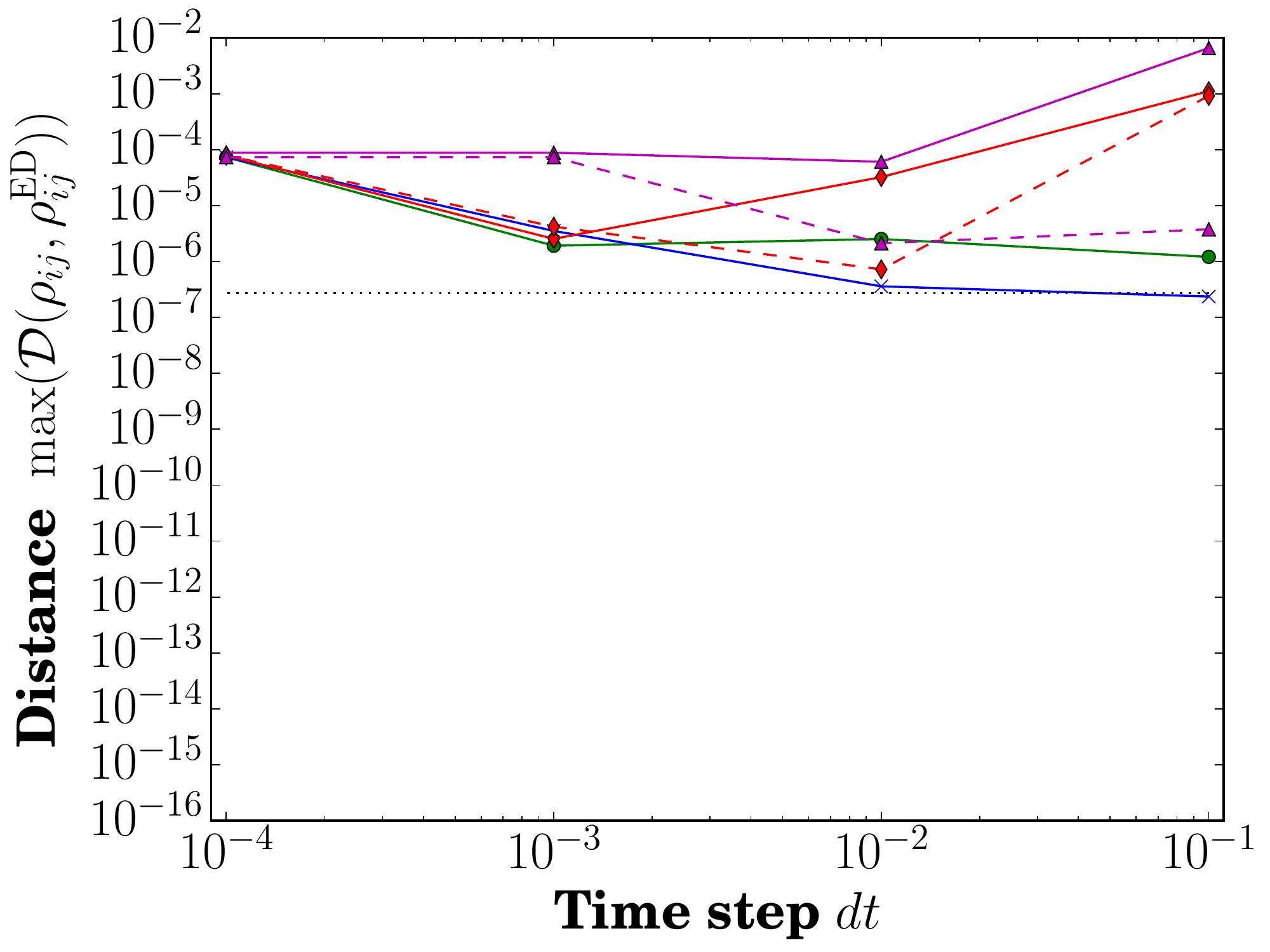}
        \put(18,67){(b)}
      \end{overpic}
    \end{minipage}

    \vspace{0.1cm}

    \begin{minipage}{0.48\linewidth}
      \begin{overpic}[width=0.9\columnwidth,unit=1mm]{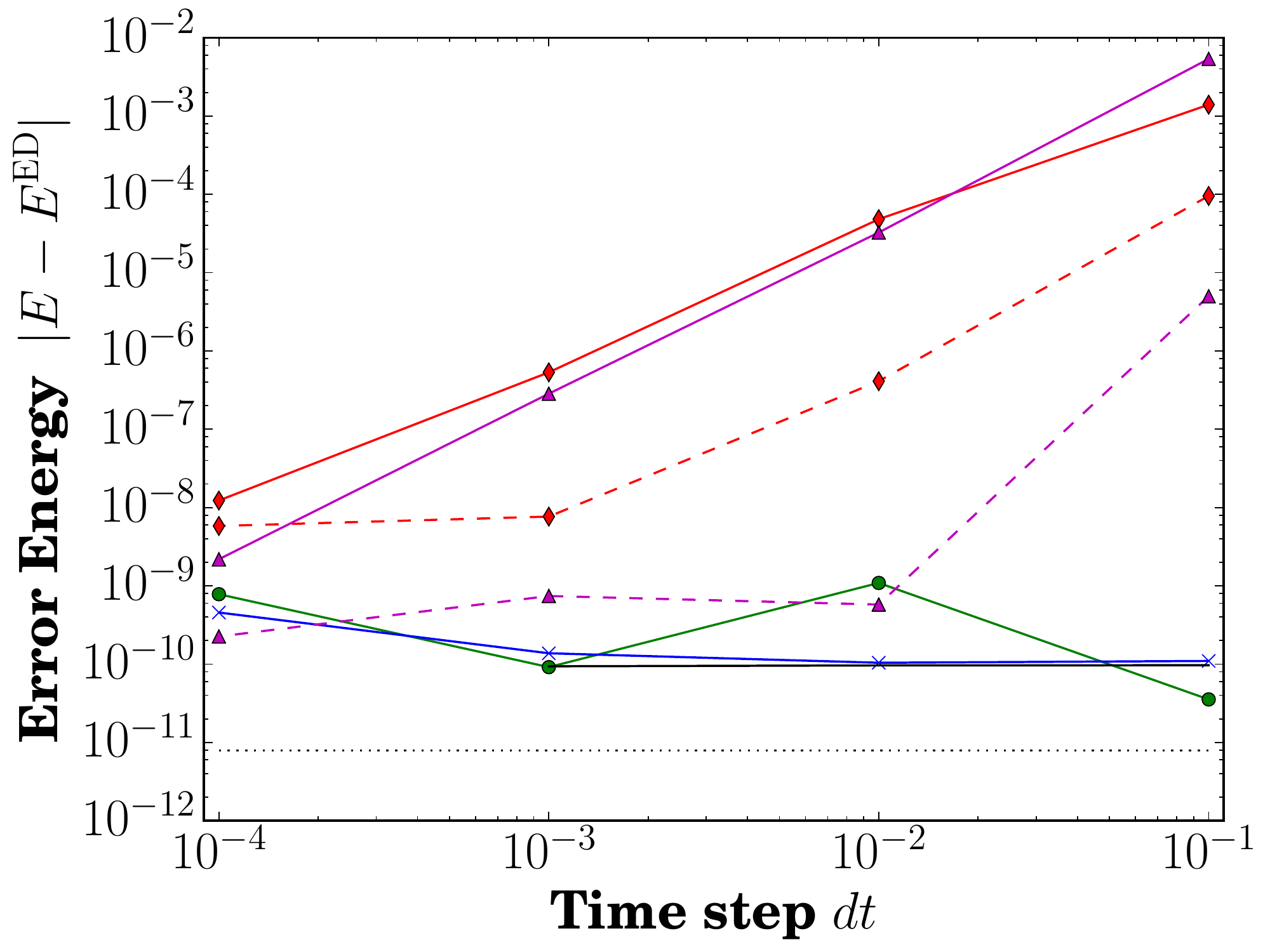}
        \put(18,67){(c)}
      \end{overpic}
    \end{minipage}\hfill
    \begin{minipage}{0.48\linewidth}
      \begin{overpic}[width=0.9 \columnwidth,unit=1mm]{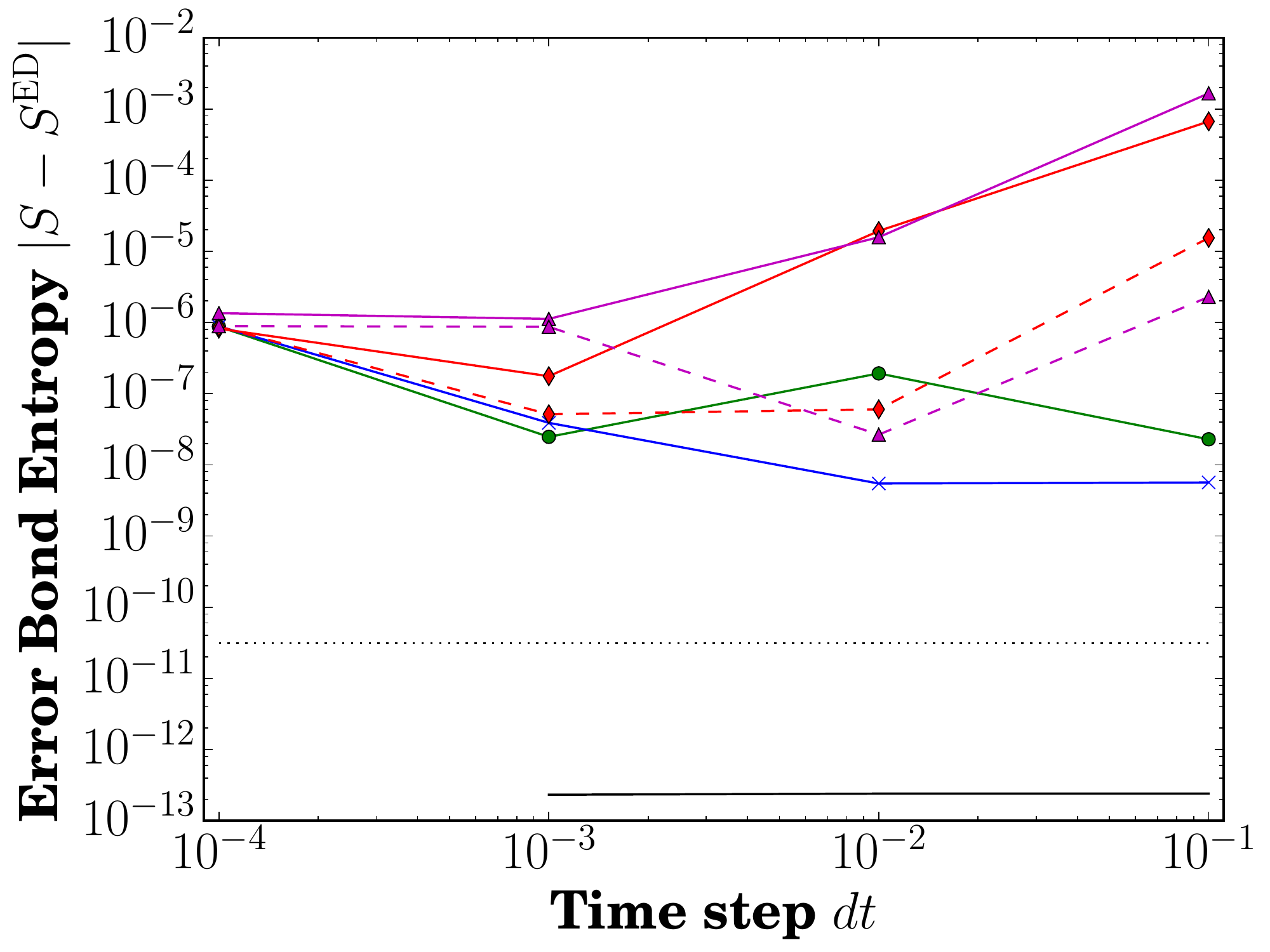}
        \put(18,67){(d)}
      \end{overpic}
    \end{minipage}

    \vspace{0.1cm}

    \begin{minipage}{0.48\linewidth}
      \begin{overpic}[width=0.9\columnwidth,unit=1mm]{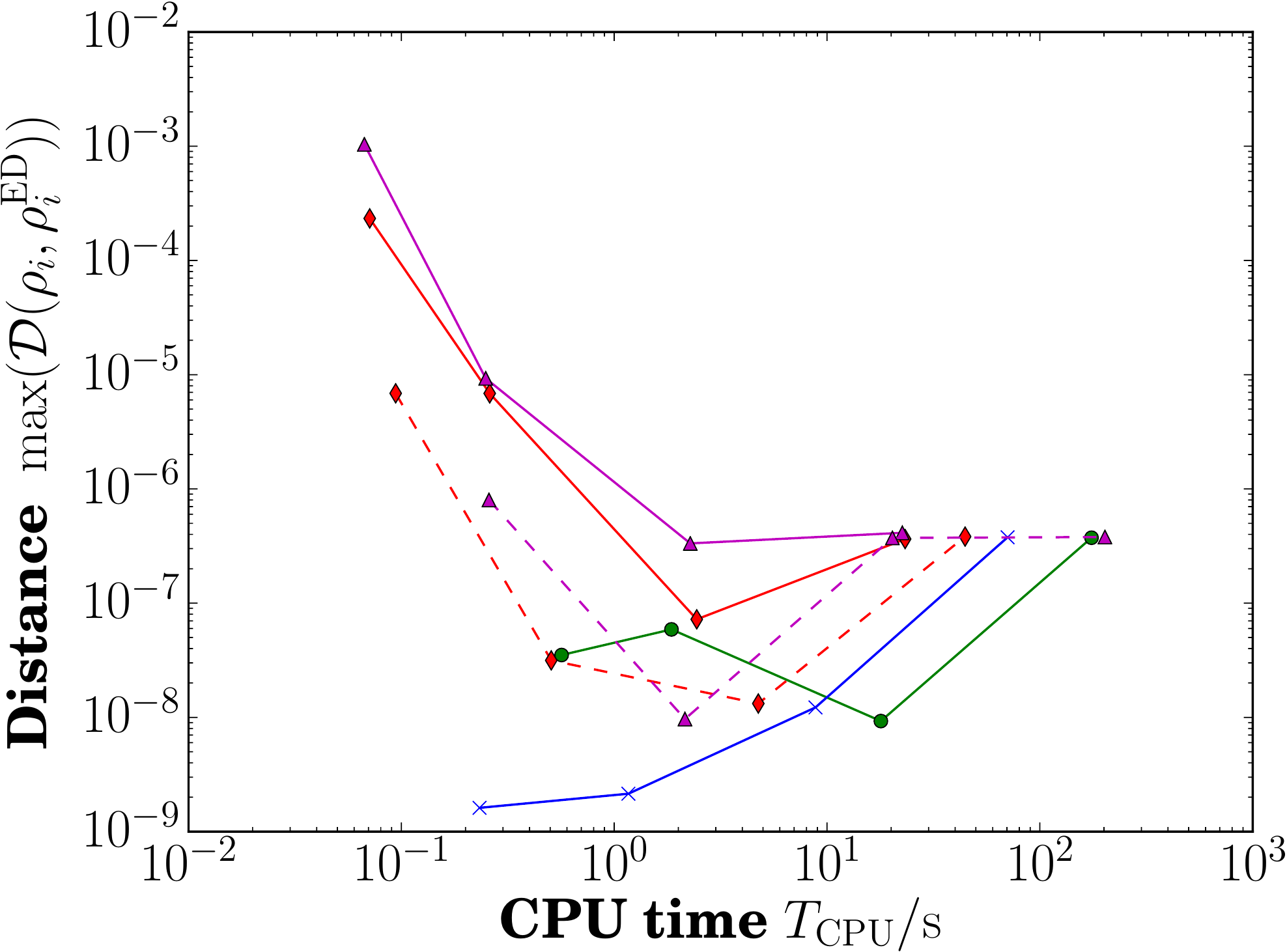}
        \put(18,66){(e)}
      \end{overpic}
    \end{minipage}\hfill
    \begin{minipage}{0.48\linewidth}
      \begin{overpic}[width=0.9 \columnwidth,unit=1mm]{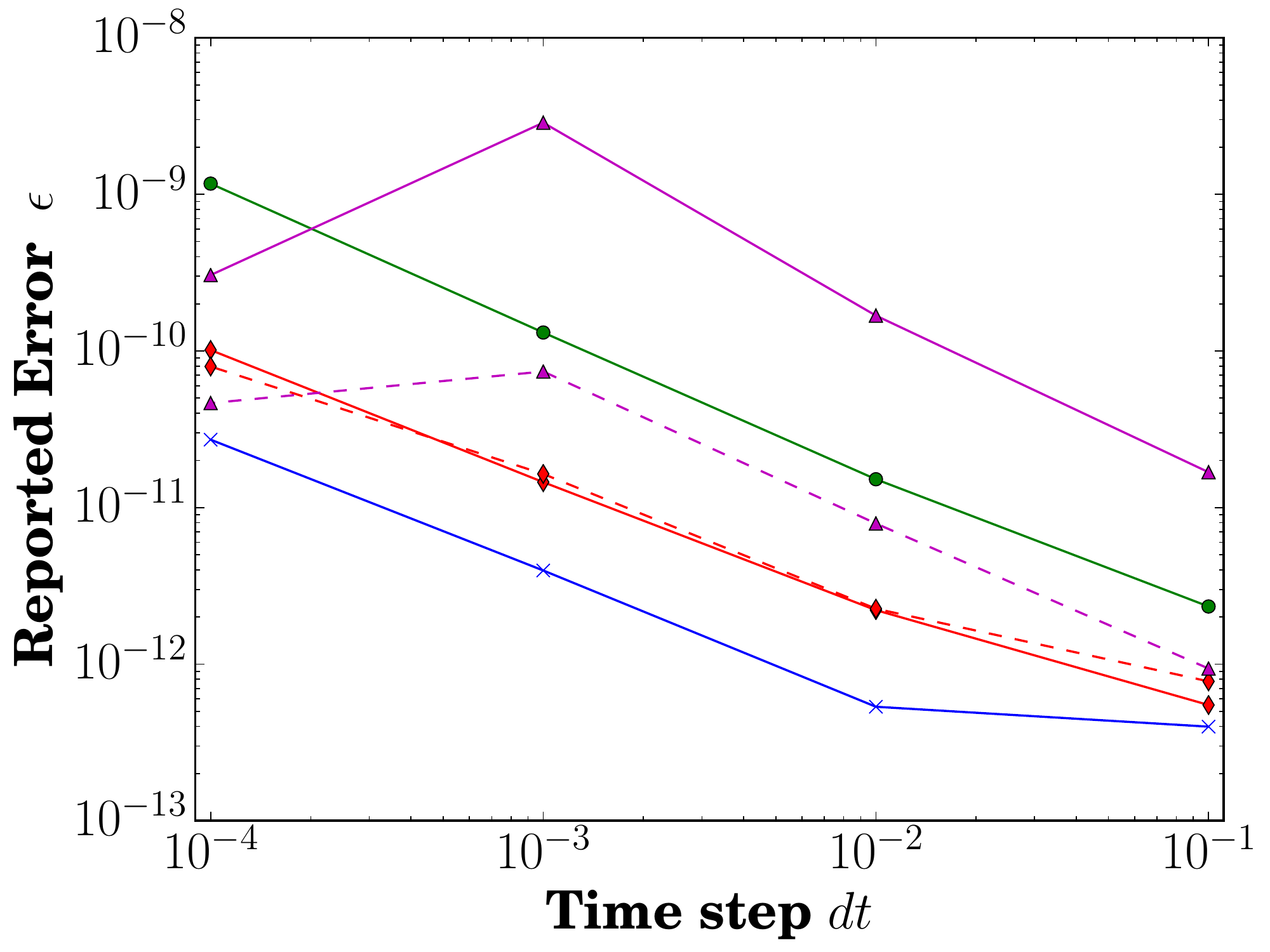}
        \put(18,67){(f)}
      \end{overpic}
   \end{minipage}
   \caption{\emph{Scaling of the error in time evolution methods} decreases as
     expected with the size of the time step for methods where the error depends
     on the time step. This is the example of a sudden quench of the Ising model
     in the paramagnetic phase evolving the ground state of an external field
     $g = 5.0$ at constant $g = 4.5$ for $L = 10$. The time step $dt$ is in
     units of $\hbar / J$.
     Curves are a guide to the eye; points represent actual data.
                                                                                \label{fig:B8_10ParaStepZ2}}
 \end{center}
\end{figure}

In this part of the appendix we expand the convergence study of the time
evolution methods with additional examples to demonstrate some key points.
Figure~\ref{fig:03_Ising} in Sec.~\ref{sec:tconv} shows aspects of a quench in the
paramagnetic phase of the Ising model for $L = 10$. Due to the time-dependence
of the Hamiltonian, we already make an error due to the evaluation of the
Hamiltonian at discrete points in time. Now, we study the convergence of the
different algorithms under the evolution of a time-independent Hamiltonian.
Therefore, we use the same ground state of the Ising model at an external
field $g = 5.0$ and evolve it under the Hamiltonian with an external field of
$g = 4.5$, which corresponds to a sudden quench. The results are shown in
Fig.~\ref{fig:B8_10ParaStepZ2}.

We start the discussion with the maximal distance of the single site
reduced density matrices. For the Krylov and TDVP algorithms, the decrease
of the time step $dt$ does not improve the result, but seems to make it
worse. Since there is no error in the method or from the time slicing of the
Hamiltonian depending on $dt$, the number of applications is the critical
variable to estimate the error. In contrast, TEBD and LRK have an error
depending on $dt$ in $\emeth$. Therefore the second order methods are worse
than the fourth order and smaller time steps improve the result if it did
not yet reach the lower bound. We recall that the lower bound is
considered to be error between the ground state results of MPS and
exact diagonalization. The two-site reduced density matrices, the
energy, and the bond entropy support this trend. Further the errors in
energy give an indication to judge on the rate of convergence for LRK and
TEBD. If we consider the data points for $dt = 0.1$ and $dt = 0.01$, the
slope for the TEBD4 curve is bigger than the one for TEBD2 indicating a
better rate of convergence. Then TEBD4 reaches the lower bound induced
by the error of the initial state or is at the level of the error of the
TDVP and Krylov method. The LRK methods have the same rate of convergence
following this argumentation.
The look at the CPU times yields then a counterintuitive result. The most
precise simulation with TDVP has the biggest time step and one of the
shortest run times. The reported error from \OSMPS{} follows the arguments
from the time-dependent Hamiltonian. Without any truncation besides the local
tolerance, more time steps add up to more error contributions.

\begin{figure}[t!]
  \begin{center}
    \begin{minipage}{0.48\linewidth}
      \begin{overpic}[width=1.0\columnwidth,unit=1mm]{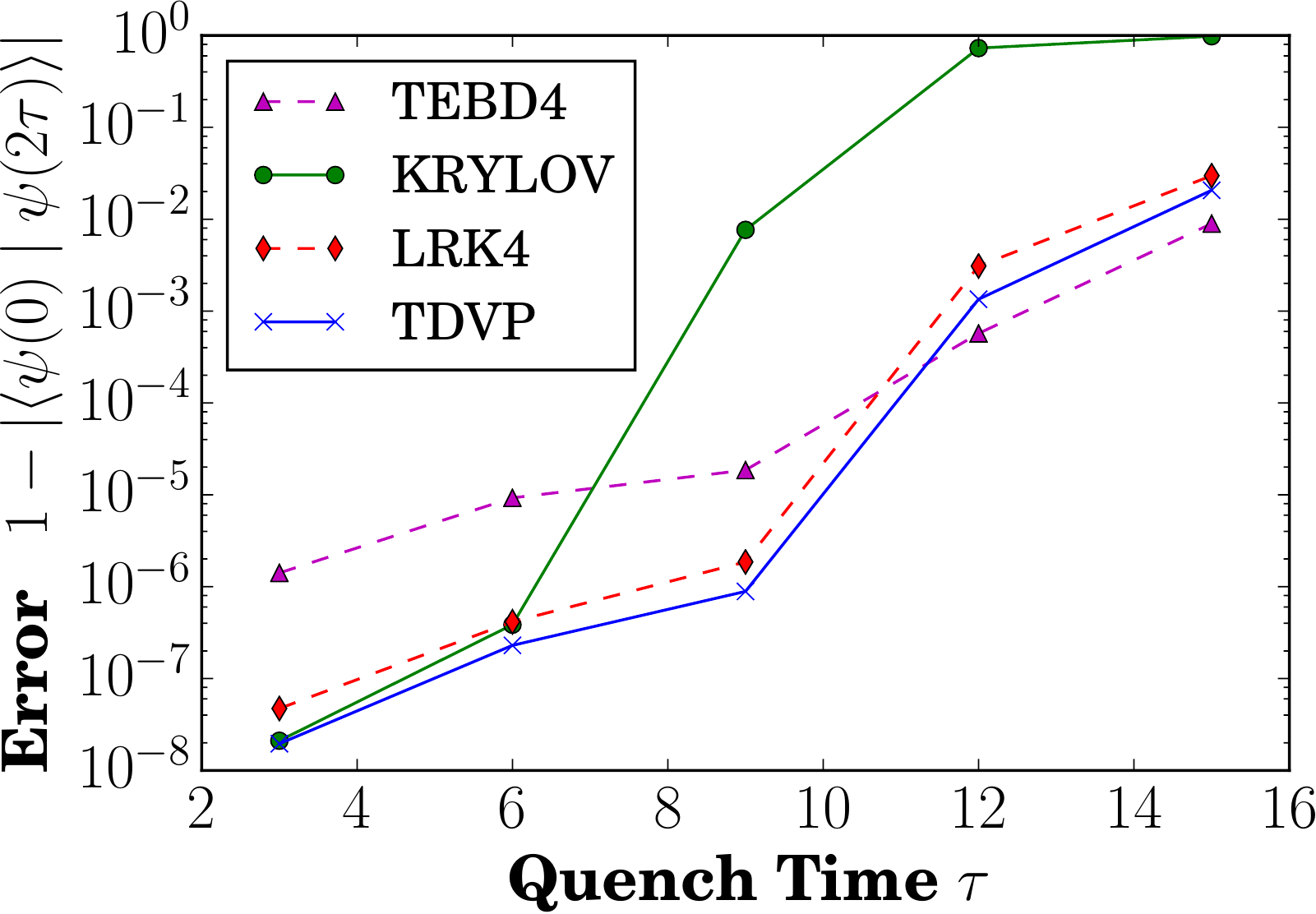}
        \put( 1,77){(a)}
      \end{overpic}
    \end{minipage}\hfill
    \begin{minipage}{0.48\linewidth}
      \begin{overpic}[width=1.0 \columnwidth,unit=1mm]{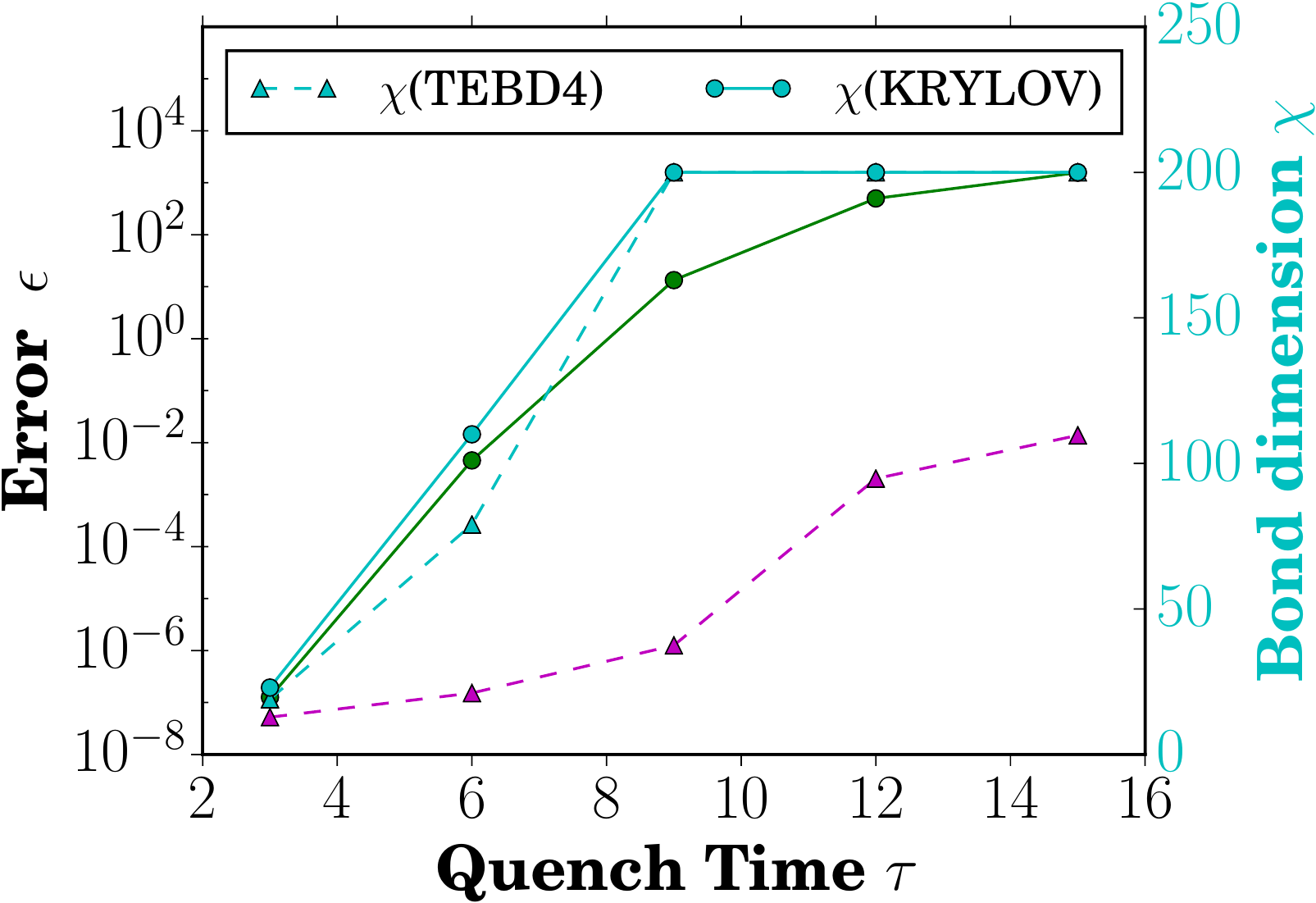}
        \put( 1,77){(b)}
      \end{overpic}
    \end{minipage}
    \caption{\emph{Error with forth-back time evolution.} Unitary time
      evolution is reversible and can help to estimate the error of time
      evolution schemes. We quench for times $\tau$ with the Hamiltonian $H$
      followed by the inverse direction, i.e., $-H$, again for $\tau$. $H$ is
      the Hamiltonian of the nearest neighbor Ising model (a)~The error defined
      over the Loschmidt echo shows that the precision of the final state
      declines for longer evolution times $\tau$. $\tau$ is in units of the
      interaction $\hbar / J$. (b)~The reason for this trend is the entanglement
      generated, which cannot be captured with $\chi_{\max} = 200$, and thus
      the cumulative error $\epsilon$ grows. $\epsilon$ includes truncation
      errors and, in the case of the Krylov method, errors from fitting states
      and $H \ket{\psi}$. (Legend for TEBD2 and KRYLOV from (a) applies for
      the errors $\epsilon$.)
                                                                                \label{fig:R3_ForthBack}}
  \end{center}
\end{figure}

A second approach to estimate the error for time evolution is the forth-back
scheme \cite{Gobert2005}. The unitary time evolution under the Hamiltonian
$H$ for a time $\tau$ followed by the evolution under $-H$ for another $\tau$
should return to the initial state. We analyze this error for the nearest
neighbor Ising model with $L = 30$ and quench times from $\tau = 3$ to
$\tau = 15$ in units of $\hbar / J$. We choose the default settings
for the time evolution, except $\chi_{\max} = 200$. The initial state is a
product state with all spins pointing in $x$-direction, except the spin on
site $16$ pointing opposite to the $x$-direction: $\ket{\psi(0)} =
\ket{\rightarrow_{1} \cdots \rightarrow_{15} \leftarrow_{16} \rightarrow_{17}
\cdots \rightarrow_{30}}$. The time step is $dt = 0.01$. We observe the error
grows with the quench time $\tau$ in Fig.~\ref{fig:R3_ForthBack} (a), which is
an effect of the growing entanglement during the time evolution. One minus the
absolute value of the Loschmidt echo corresponds to the infidelity and is
therefore a good error measure. This reasoning is supported by part
(b) of Fig.~\ref{fig:R3_ForthBack}: the maximal bond dimension is exhausted
for the larger $\tau$ and the cumulative error grows. The error includes
the truncation for the TEBD, and additional contribution from the fitting
of wave function in the Krylov method. Obviously, one can generalize this
approach to any model and study this error previous to simulating new
or different Hamiltonians and initial conditions.

\section{Scaling of computational resources                                    \label{app:scaling}}

We consider the scaling of computational resources,
especially computation time and memory, as a function of common parameters
of simulations. These include the system size $L$ and the maximal
bond dimension $\chi$. While the default parallelization is data parallelism
using the MPI interface to \OSMPS{} with a straightforward scaling
explained in Appendix~\ref{app:features}, we consider as well the openMP
(Open Multi Processing) algorithms used by the underlying libraries,
namely LAPACK and BLAS.

\begin{figure}[t!]
  \begin{center}
    \begin{minipage}{0.48\linewidth}
      \begin{overpic}[width=1.0\columnwidth,unit=1mm]{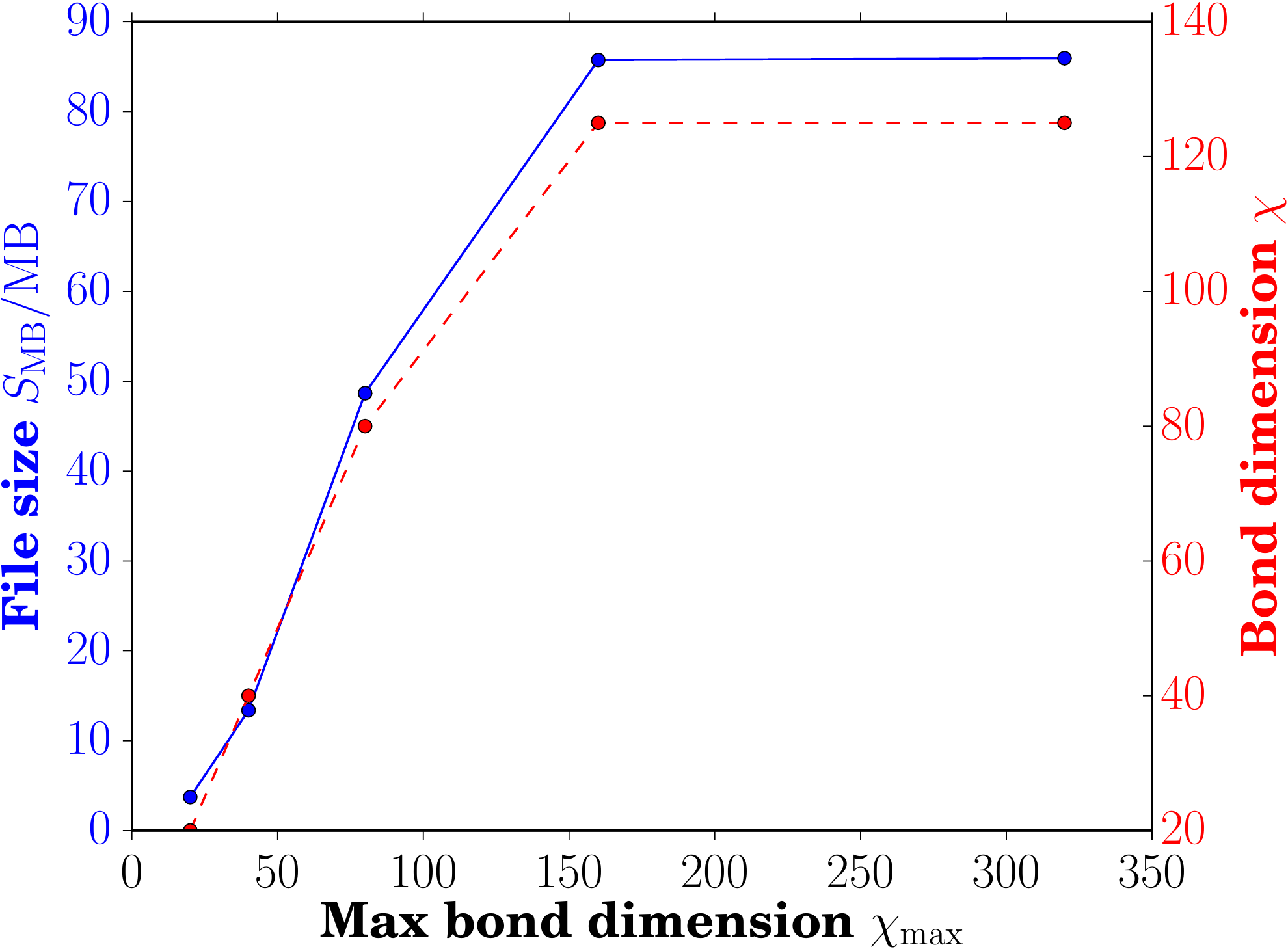}
        \put( 1,77){(a)}
      \end{overpic}
    \end{minipage}\hfill
    \begin{minipage}{0.48\linewidth}
      \begin{overpic}[width=1.0 \columnwidth,unit=1mm]{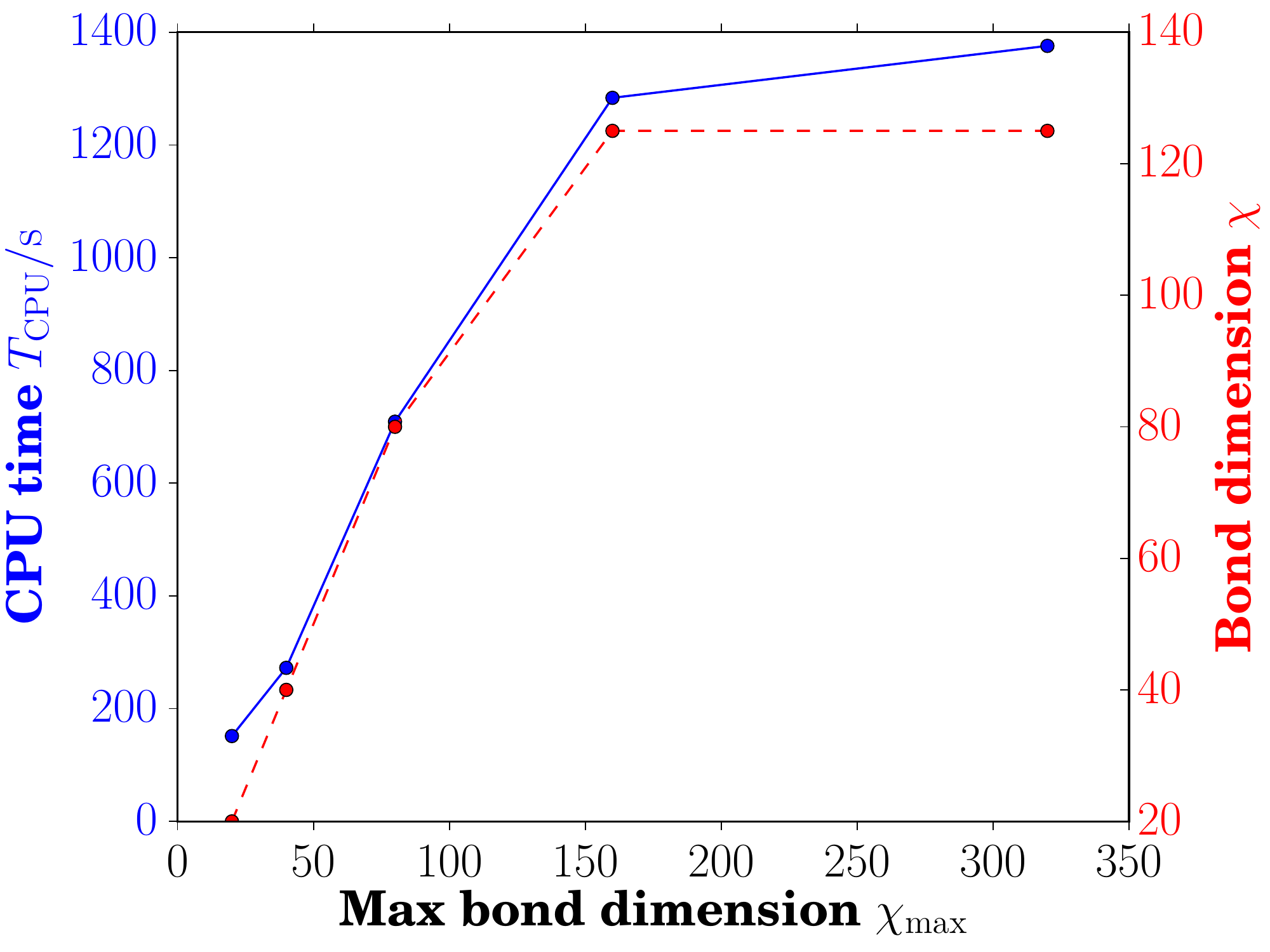}
        \put( 1,77){(b)}
      \end{overpic}
    \end{minipage}
    
    \vspace{0.5cm}

    \begin{minipage}{0.48\linewidth}
      \begin{overpic}[width=1.0\columnwidth,unit=1mm]{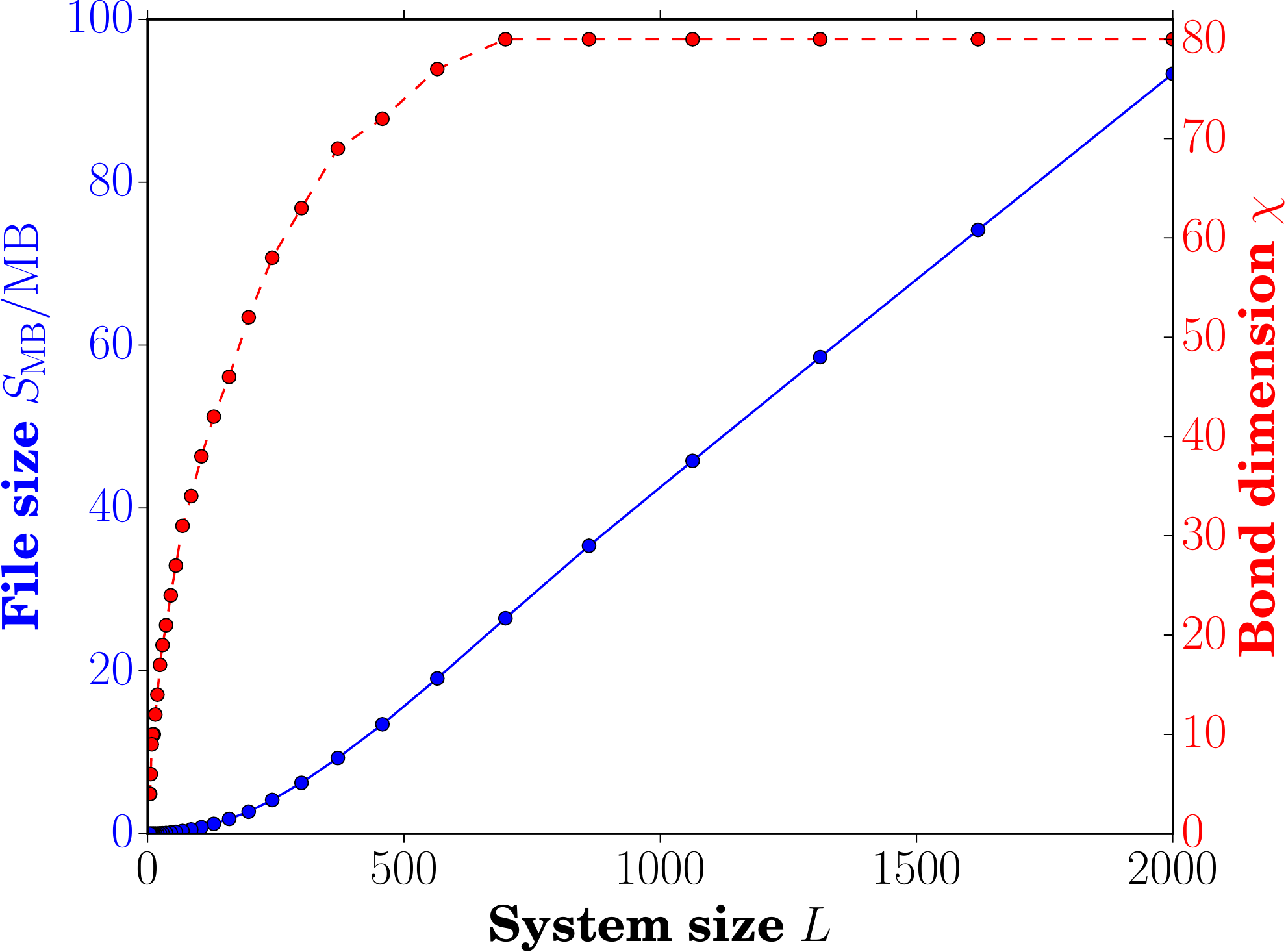}
        \put( 1,77){(c)}
      \end{overpic}
    \end{minipage}\hfill
    \begin{minipage}{0.48\linewidth}
      \begin{overpic}[width=1.0 \columnwidth,unit=1mm]{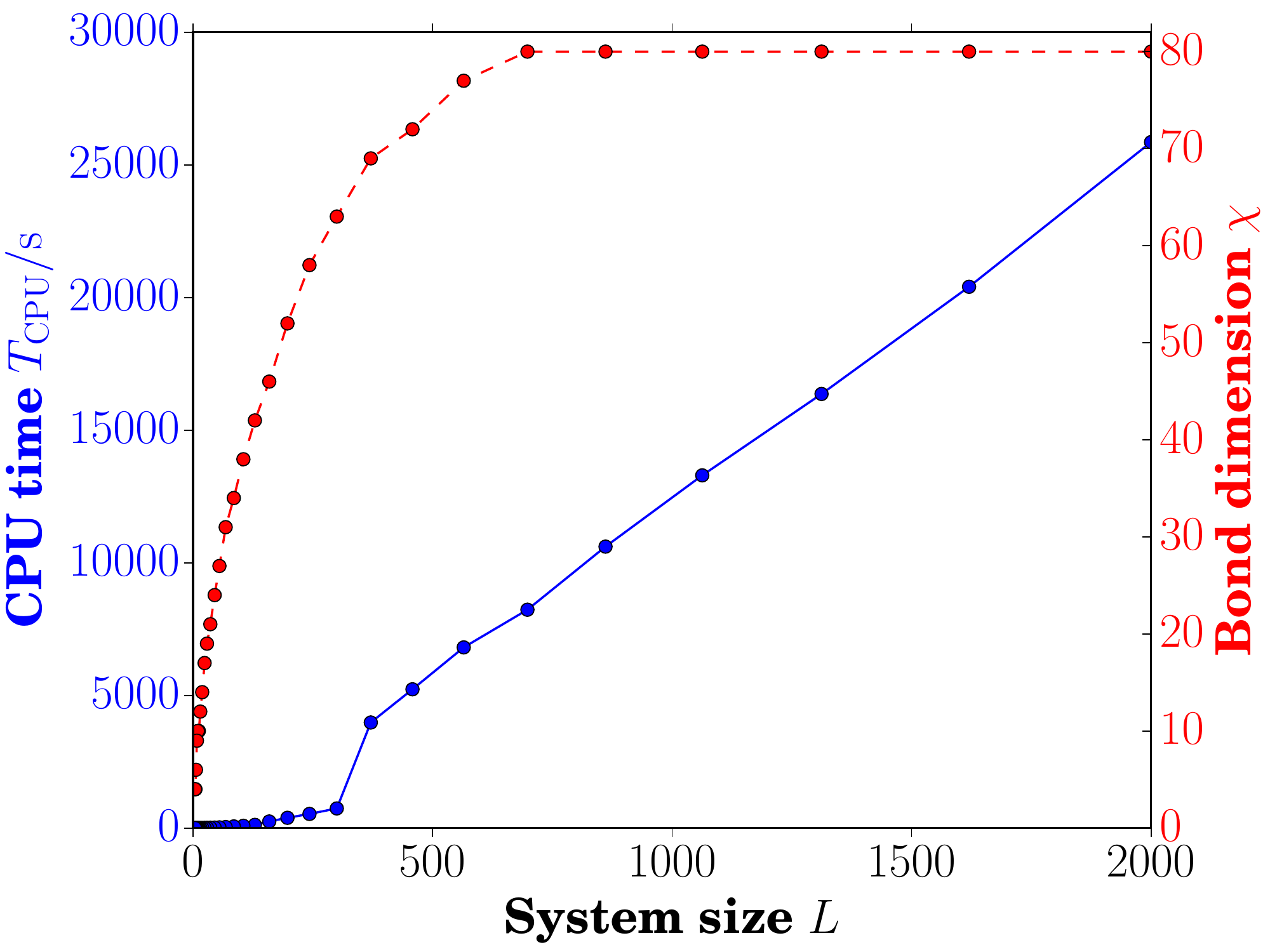}
        \put( 1,77){(d)}
      \end{overpic}
    \end{minipage}
    \caption{\emph{Scaling of computational resources.} We consider the Ising
      model and vary the bond dimension and the system size.
      (a) The file size of the ground state indicates the memory resources
      necessary for the simulation. The parameters are $L = 1063$ and $g = 1.0$
      for different maximal bond dimensions $\chi_{\max}$.
      (b) The scaling of the CPU time as a function of $\chi_{\max}$. Parameters
      equal to (a).
      (c) The file size of the ground state for $\chi_{\max} = 80$ at $g=1.0$
      for a set of different system sizes from the data of
      Fig.~\ref{fig:01_Ising_BE3d}.
      (d) The scaling of the CPU time for different system sizes. Parameters
      equal to (c).
                                                                                \label{fig:C1_ScalingChi}}
  \end{center}
\end{figure}

First, we consider the scaling with the maximal bond dimension $\chi$. We
consider in this scenario the quantum Ising model for a system size of $L=1063$,
which corresponds to one data point from Fig.~\ref{fig:01_Ising_BE3d}. The
value of the external field is the critical value for the thermodynamic limit,
i.e., $g = 1.0$. The simulations were run on a \mioseventeenx{} node and
the corresponding data is shown in Fig.~\ref{fig:C1_ScalingChi}. In order to
have a better understanding of the data we plot in each case the bond dimension.
For the memory in Fig.~\ref{fig:C1_ScalingChi}(a) we
see that the file size saturates once the bond dimension saturates. The file
contains the complete information about the state in binary format and
gives a good estimate of the memory needs for each simulation. For linear algebra
operations within LAPACK, the additional memory allocated as workspace is on
the order of the matrix size. The size of the matrices handled is bounded
by the local dimension and the maximal bond dimension leading to a maximum size
of $d \chi_{\max} \times d \chi_{\max}$. Figure~\ref{fig:C1_ScalingChi}(b)
shows the CPU time for each simulation. It naturally saturates with the bond
dimension. Before the saturation point it grows linearly with the bond
dimension actually used.

For the scaling of the resources with the system size $L$ we consider the
set of simulations generating Fig.~\ref{fig:01_Ising_BE3d} and pick the external
field $g = 1.0$. The data is for \miosixtyx{} nodes. As previously, we show
the bond dimension utilized in addition to the file size or computation time.
If the bond dimension saturates starting at $L > 500$, the file size grows
linearly with the system size $L$ as shown in
Fig.~\ref{fig:C1_ScalingChi}(c). In contrast, in the growth for $L < 500$
the file size increases faster than linear. In addition to the growing
system size, the fact that a larger system can have more entanglement leads to
the increase in file size. Figure~\ref{fig:C1_ScalingChi}(d) shows the CPU
times for the equivalent setup with a similar result as for the file size.
The scaling is linear as soon as the bond dimension actually used saturates.
Before the growth appears nonlinear with a jump.

We examine the use of openMP as the last part of the analysis
of resources. OpenMP allows for parallel computing with shared memory on one
compute node. In contrast, the tasks of MPI jobs never have access to the
same memory and have to send any data. In our type of implementation data
is sent between the master node and the workers. On current clusters this
allows one to use parallelization with up to $24$ cores. We study the
efficiency of openMP and what speedup can be gained. \OSMPS{} does not have
implementations for openMP itself, but the LAPACK and BLAS or respectively
mkl libraries can support openMP on the level of the linear algebra
operations within \OSMPS{}. We find that openMP, implemented in this form,
does not increase the speed of the simulation, as described in
Table~\ref{tab:openmpscal}. The simulation time shown is the duration of
the job and not the CPU time on a \mioseventeenx{} node.


\begin{table}[H]
  \centering
  \begin{tabular}{@{} ccccccc @{}}
    \toprule
    Cores & 1       & 2       & 4       & 8       & 16      & 24                \\
    \cmidrule(r){1-1}  \cmidrule(rl){2-2} \cmidrule(rl){3-3} \cmidrule(rl){4-4}
    \cmidrule(rl){5-5} \cmidrule(rl){6-6} \cmidrule(l){7-7}
    $\Tjob(\chi_{\max} = 20) / \mathrm{s}$
          &  150    &  154    &  154    &  155    &  154    &  168              \\
    $\Tjob(\chi_{\max} = 40) / \mathrm{s}$
          &  270    &  273    &  271    &  274    &  275    &  302              \\
    $\Tjob(\chi_{\max} = 80) / \mathrm{s}$
          &  699    &  780    &  789    &  734    &  749    &  739              \\
    $\Tjob(\chi_{\max} = 160) / \mathrm{s}$
          & 1293    & 1273    & 1232    & 1188    & 1161    & 1396              \\
    $\Tjob(\chi_{\max} = 3200) / \mathrm{s}$
          & 1368    & 1372    & 1259    & 1241    & 1237    & 1278              \\
    \bottomrule
  \end{tabular}
  \caption{\emph{OpenMP scaling for \OSMPS{}.} As an example for the scaling
    we simulate the ground state of the Ising model for $L=1063$ at an external
    field of $g = 1.0$ for different numbers of threads and different bond
    dimensions and list the duration of the simulation $\Tjob$ in seconds.
    This case corresponds to a data point of Fig~\ref{fig:01_Ising_BE3d}.
    We see that openMP does not come with any real speedup.
    The maximal bond dimension used is $124$, which affects the last
    two rows.
                                                                                \label{tab:openmpscal}}
\end{table}

\section{Error bounds for static simulations}                                  \label{app:boundstatic}

Ideally, numerical simulations yield a corresponding bound for the error of
their results. For DMRG there are calculations available discussing the
behavior of the error. Local observables in a two-dimensional Heisenberg
models are discussed in terms of the truncation error in \cite{White2007,
Schollwoeck2005,Schollwoeck2011}. The error in variational MPS methods is
mentioned in \cite{McCulloch2007,Michel2010} and relates the variance to the
squared norm of the difference between the exact and the approximate quantum
state. We present an alternate approach using the variance as well to derive
error expressions for multiple observables.

In static MPS simulations the variance of the Hamiltonian, which bounds
the error of the energy, is returned as an error estimate
for the result. Therefore, it remains for us to show how other observables or
measures are bounded by the variance of the state, where we take the ground
state as an example. The basic idea is to assume we have a final state
$| \psi \rangle$ as an MPS
\begin{eqnarray}
  | \psi \rangle
  = f | \psi_0 \rangle + \epsi | \psi_{\perp} \rangle \, ,
\end{eqnarray}
with $f = \langle \psi | \psi_0 \rangle$, $|\epsi|^2 = 1 - |f|^2$, and
$| \psi_0 \rangle$ the true ground state, while $| \psi_{\perp} \rangle$ is
orthogonal to the ground state and contains all errors. In this appendix
we keep the notation that $\ket{\psi}$ is the state from the
\OSMPS{} simulation with a variance $V_{\psi}$, $\ket{\psi_{0}}$ is the true
ground state and $\ket{\psi_{\perp}}$ contains all contributions orthogonal
to the true ground state.

\subsection{Bounding $\epsi$ with the variance delivered by open source Matrix Product States}

The first step is to bound $\epsi$ from the information gathered in
\OSMPS{}, that is the variance of $H$. Furthermore, in addition to the above
orthogonality relation $\braket{\psi_{\perp}}{\psi_0}$, any power $H^n$ is
subject to the relation $\sandwich{\psi_{\perp}}{H^n}{\psi_0} = 0$, since
$\ket{\psi_0}$ is an eigenstate of $H$. We recall that the Hamiltonian
is represented without errors except if an \texttt{InfiniteFunction} rule is
fitted through a series of \texttt{Exponential} rules. The error due to the
fitting procedure is not covered in this Appendix. Writing the variance in
terms of this decomposition we obtain
\begin{eqnarray}
  &&V_{\psi}
  = \sandwich{\psi}{H^2}{\psi} - \left(\sandwich{\psi}{H}{\psi} \right)^2     \nonumber \\
  &&= |f|^2 \sandwich{\psi_0}{H^2}{\psi_0}
     \! + \! f^{\ast} \epsi \sandwich{\psi_0}{H^2}{\psi_{\perp}}
     \! + \! f \epsi^{\ast} \sandwich{\psi_{\perp}}{H^2}{\psi_0}
     \! + \! |\epsi|^2 \sandwich{\psi_{\perp}}{H^2}{\psi_{\perp}}                      \nonumber \\
  &&  - \left(|f|^2 \sandwich{\psi_0}{H}{\psi_0}
             \! + \! f^{\ast} \epsi \sandwich{\psi_0}{H}{\psi_{\perp}}
             \! + \! f \epsi^{\ast} \sandwich{\psi_{\perp}}{H}{\psi_0}
             \! + \! |\epsi|^2 \sandwich{\psi_{\perp}}{H}{\psi_{\perp}} \right)^2      \nonumber \\
  &&= |f|^2 \sandwich{\psi_0}{H^2}{\psi_0}
      + |\epsi|^2 \sandwich{\psi_{\perp}}{H^2}{\psi_{\perp}}                      
      - |f|^4 \sandwich{\psi_0}{H}{\psi_0}^2                                      \nonumber \\
  &&          \! - \! 2 |f|^2 |\epsi|^2 \sandwich{\psi_0}{H}{\psi_0}
                \sandwich{\psi_{\perp}}{H}{\psi_{\perp}}
              \! - \! |\epsi|^4 \sandwich{\psi_{\perp}}{H}{\psi_{\perp}}^2 \, . 
\end{eqnarray}
We introduce the eigenenergy $E_{0} = \sandwich{\psi_0}{H}{\psi_0}$
of our true ground state and the energy
$E_{\perp} = \sandwich{\psi_{\perp}}{H}{\psi_{\perp}}$, which is
not an eigenenergy of the system because
$\ket{\psi_{\perp}}$ can be a linear combination of eigenstates. Next, we use
the relation $|f|^2 = 1 - |\epsi|^2$ and we add a zero in terms of
$\pm (|\epsi|^2 - |\epsi|^4) E_{\perp}^2$ to simplify the expression:
%
\begin{eqnarray}
  V_{\psi}
  &=& (1 - |\epsi|^2) \sandwich{\psi_0}{H^2}{\psi_0}
      + |\epsi|^2 \sandwich{\psi_{\perp}}{H^2}{\psi_{\perp}}      \nonumber \\
  &&  - (1 - 2 |\epsi|^2 + |\epsi|^4) E_0^2
      - 2 (|\epsi|^2 - |\epsi|^4) E_0 E_{\perp}
      - |\epsi|^4 E_{\perp}^2                                   \nonumber \\
  &&  + (|\epsi|^2 - |\epsi|^4) E_{\perp}^2
      - (|\epsi|^2 - |\epsi|^4) E_{\perp}^2       \nonumber \\
  &=& (1 - |\epsi|^2) V_0
      + |\epsi|^2 V_{\perp}
      + (|\epsi|^2 - |\epsi|^4) (E_{\perp} - E_{0})^2 \, .
\end{eqnarray}
We abbreviate the energy difference as $\Delta = E_{\perp} - E_0$, define
the variance of $\ket{\psi_{\perp}}$ as $V_{\perp}$, and the variance of
$\ket{\psi_0}$ as $V_0 = 0$. We introduce $\tilde{\epsi} = |\epsi|^2$ leading
to the following quadratic equation:
\begin{eqnarray}                                                                \label{eq:varbound_quadratic}
  \Delta^2 \tilde{\epsi}^2 - (V_{\perp} + \Delta^2) \tilde{\epsi} + V_{\psi}
  = 0 \, .
\end{eqnarray}
We remark that the variance of an eigenstate of $H$ is zero, applied to
$V_0 = 0$. The equation has two solutions returning the values for $|\epsi|^2$:
\begin{eqnarray}                                                                \label{eq:varbound_eps}
  |\epsi_1|^2 &=& \frac{1}{2} \left(1 + \frac{V_{\perp}}{\Delta^2}
                 - \sqrt{\left(1 + \frac{V_{\perp}}{\Delta^2}\right)^2
                         - 4 \frac{V_{\psi}}{\Delta^2}} \right)            \\
  |\epsi_2|^2 &=& \frac{1}{2} \left(1 + \frac{V_{\perp}}{\Delta^2}
                   + \sqrt{\left(1 + \frac{V_{\perp}}{\Delta^2}\right)^2
                           - 4 \frac{V_{\psi}}{\Delta^2}} \right) \, .          \label{eq:varbound_eps2}
\end{eqnarray}
Plugging $|\epsi|^2 = 1 - |f|^2$ into Eq.~\eqref{eq:varbound_quadratic}, we
get another quadratic equation and two solutions for $|f|^2$:
\begin{eqnarray}                                                                \label{eq:varbound_fid}
  \Delta^2 |f|^4 - (\Delta^2 - V_{\perp}) |f|^2 + (V_{\psi} - V_{\perp}) = 0 \, ,   \\
  |f_1|^2 = \frac{1}{2} \left(1 - \frac{V_{\perp}}{\Delta^2}
          + \sqrt{\left(1 + \frac{V_{\perp}}{\Delta^2} \right)^2
          - \frac{4 V_{\psi}}{\Delta^2}} \right) \, ,                           \\
  |f_2|^2 = \frac{1}{2} \left(1 - \frac{V_{\perp}}{\Delta^2}
          - \sqrt{\left(1 + \frac{V_{\perp}}{\Delta^2} \right)^2
          - \frac{4 V_{\psi}}{\Delta^2}} \right) \, .                           \label{eq:varbound_fid2}
\end{eqnarray}
According to the normalization condition we recognize that $f_1$ and
$\epsi_1$ build one solution, as well as $f_2$ and $\epsi_2$:
\begin{eqnarray}
  |f_{i}|^{2} + |\epsi_{i}|^{2} = 1 \, , \, i \in \{1, 2 \} \, .
\end{eqnarray}
Under the assumption that the major part of our state is in the ground state,
we choose the smaller $\epsi_1$. This is the first assumption in the
calculation and we proceed to bound  $\epsi^2$ using the fact that
$0 \le V_{\perp} \le V_{\psi}$ and then implementing a Taylor expansion in
$V_{\psi} / \Delta$, which is small if the variance of the ground state has
sufficiently converged targeting in the default setup a value of
$L \times 10^{-10}$ and we have an energy gap in the system. We point out
the role of the gap $\Delta$ in the following steps in detail:
\begin{eqnarray}
  |\epsi|^2 &\le& \frac{1}{2} \left(1 + \frac{V_{\psi}}{\Delta^2}
                   - \sqrt{1 - 4 \frac{V_{\psi}}{\Delta^2}} \right)             \nonumber \\
  &\approx& \frac{1}{2} \left(1 + \frac{V_{\psi}}{\Delta^2}
                   - \left(1 - 2 \frac{V_{\psi}}{\Delta^2}
                           - \mathcal{O}\left(\frac{V_{\psi}^2}{\Delta^4}
                        \right) \right) \right)                                 \nonumber \\
  &=& \frac{1}{2} \left(\frac{3 V_{\psi}}{2 \Delta^2}
                        + \mathcal{O} \left(\frac{V_{\psi}^2}{\Delta^4} \right)
                \right) \, .
\end{eqnarray}
Finally, we use the minimal gap between the ground state and the first excited
state $\Delta_{0,1} = E_{1} - E_{0}$ to approximate $\Delta$. The inverse
energy difference between the ground state and a superposition of all excited
states can be bound with the smallest gap:
\begin{eqnarray}
  |\epsi|^2 &\le& \frac{3 V_{\psi}}{4 \Delta_{0,1}^2}
                   + \mathcal{O} \left(\frac{V_{\psi}^2}{\Delta_{0,1}^4} \right)
  \quad \Longrightarrow \quad
  |\epsi| \le \frac{\sqrt{V_{\psi}}}{\Delta_{0,1}}
               + \mathcal{O} \left(\frac{V_{\psi}}{\Delta_{0,1}^{2}} \right) \, .
\end{eqnarray}
This bound shows that the variance can be a good approximation for the error as
long as the gap to the next eigenstate is finite. This gap refers to the next
accessible eigenstate in case symmetries are used. For simulations around the
critical point with a closing gap the bound becomes less precise due to the
closing gap. But for finite systems considered in this calculation, the
gap remains.

In addition, we list the implicit and explicit assumptions during the derivation
of the bound
\begin{itemize}
\item{We converged mainly to the ground state. The solutions in
  Eqs.~\eqref{eq:varbound_eps} and \eqref{eq:varbound_fid} and in the
  Eqs.~\eqref{eq:varbound_eps2} and \eqref{eq:varbound_fid2} are in general
  true for any eigenstate of $H$.}
\item{By taking the solution represented by the pair of
  Eqs.~\eqref{eq:varbound_eps} and \eqref{eq:varbound_fid} we assume that
  the main part of the solution is in the eigenstate.}
\item{The energy gap to the first state above the ground state used
  in the Taylor expansion is assumed not to be small. This approximation may
  fail around quantum critical points with a closing energy gap.}
\item{$V_{\perp} \le V_{\psi}$: the variance of the subset of states is smaller
  if the minimal energy is canceled from the set of states. This is true because
  one end of the distribution is cut.}
\end{itemize}

\subsection{Bounding observables}

The $\epsi$ derived above is only useful if bounds for other measures can
be obtained. For a local Hermitian observable $O$ with a maximal absolute
value $\mathcal{M}$ defined as
\begin{eqnarray}
  \mathcal{M} = \max_{\ket{\phi}, \ket{\phi'}} | \bra{\phi} O \ket{\phi'} | 
\end{eqnarray}
we obtain the following bound between the measurement on the true ground
state $\ket{\psi_{0}}$ and the state $\ket{\psi}$ resulting from the \OSMPS{}
simulation:
\begin{eqnarray}
 && | \bra{\psi_0} O \ket{\psi_0}
  - \bra{\psi} O \ket{\psi} |                                                   \nonumber \\
 &&= \big| (1 - |f|^2) \bra{\psi_0} O \ket{\psi_0}
            - \epsi f^{\ast} \bra{\psi_0} O \ket{\psi_{\perp}}
            - \epsi^{\ast} f \bra{\psi_{\perp}} O \ket{\psi_0}                    \nonumber \\
 &&           - |\epsi|^2 \bra{\psi_{\perp}} O \ket{\psi_{\perp}} \big|           \nonumber \\
 &&\le (1 - |f|^2) \left| \bra{\psi_0} O \ket{\psi_0} \right|
       + 2 |\epsi f| \left| \bra{\psi_0} O \ket{\psi_{\perp}} \right|
       + |\epsi|^2 \left| \bra{\psi_{\perp}} O \ket{\psi_{\perp}} \right|         \nonumber \\
 &&\le 2 |\epsi| (|f| + |\epsi|) \mathcal{M}
  \le  2 \sqrt{2} |\epsi| \mathcal{M}
  \le  3 |\epsi| \mathcal{M} \, .
\end{eqnarray}
Here in, $|f| + |\epsi|$ is always smaller than $\sqrt{2}$, which
originates in the normalization of the state and can be derived from the
maximizing the constraint problem $\mathcal{L}(f, \epsi, \gamma) =
|f| + |\epsi| + \gamma (|f|^2 + |\epsi|^2 - 1)$, where $\gamma$ is
a Lagrange multiplier for the normalization constraint. So in general
it is possible to bound values of an observable.

\subsection{Density matrices and their bounds}

In the following, we derive an expression for the complete density matrix, any
reduced density matrix of the system, and a bound for the trace distance
between the MPS result and the (reduced) density matrix of the ground state.
We start by expressing the mixed contribution of the form $\ket{\psi_0}
\bra{\psi_{\perp}}$ in a more convenient way.

\begin{lemma}[Mixed contributions $\ket{\psi_0} \bra{\psi_{\perp}}$]            \label{lem:mixedcontrib}
  A mixed contribution of the form $\e^{-\mathrm{i} \phi} \ket{\psi_0}
  \bra{\psi_{\perp}} + \e^{\mathrm{i} \phi} \ket{\psi_{\perp}}
  \bra{\psi_{0}}$ with an arbitrary phase $\phi$ can be decomposed into
  positive matrices $\sigma_{\pm}$ with one non-zero eigenvalue of value $1$
  as
  \begin{eqnarray}
    \ket{\psi_0} \bra{\psi_{\perp}} + \ket{\psi_{\perp}} \bra{\psi_{0}}
    = \sigma_{+} - \sigma_{-} \, ,
  \end{eqnarray}
  where $\sigma_{\pm}$ fulfill all characteristics of a density matrix
  (and are not the spin lowering/raising operators).
\end{lemma}
  \emph{Proof.}
  We take the pure states $\ket{\psi_{\pm}} = (\ket{\psi_{0}} \pm
  \e^{\mathrm{i} \phi} \ket{\psi_{\perp}}) / \sqrt{2}$, where $\phi$ is
  an arbitrary phase. These pure states define the density
  matrices $\sigma_{\pm}$:
  \begin{eqnarray}
   \sigma_{\pm} \!\!\!\!
   &=& \!\!\!\! \ket{\psi_{\pm}} \bra{\psi_{\pm}}                                        \nonumber \\
   &=& \!\!\!\! \frac{1}{2} \ket{\psi_0} \bra{\psi_0}
      \! + \! \frac{1}{2} \ket{\psi_{\perp}} \bra{\psi_{\perp}}
      \! \pm \! \frac{1}{2} \left( \e^{-\mathrm{i} \phi} \ket{\psi_0} \bra{\psi_{\perp}}
                            \! + \! \e^{\mathrm{i} \phi} \ket{\psi_{\perp}} \bra{\psi_0} \right) \, .
  \end{eqnarray}
  Therefore, the difference $\sigma_{+} - \sigma_{-}$ leads to the term
  we need while canceling out the contributions from $\ket{\psi_{\perp}}
  \bra{\psi_{\perp}}$ and $\ket{\psi_{0}} \bra{\psi_{0}}$:
  \begin{eqnarray}
    \sigma_{+} - \sigma_{-}
    = \e^{-\mathrm{i} \phi} \ket{\psi_0} \bra{\psi_{\perp}}
      + \e^{\mathrm{i} \phi} \ket{\psi_{\perp}} \bra{\psi_{0}} \, .
  \end{eqnarray}

\begin{lemma}[Error bound on density matrix]                                    \label{lem:boundrho}
  Knowing $\epsi$ in $\ket{\psi} = f \ket{\psi_0}
  + \epsi \ket{\psi_{\perp}}$, we can express the (reduced) density matrix
  $\rho$ representing the \OSMPS{} result as
  \begin{eqnarray}
    \rho = |f|^2 \rho_{0} + \erhop \rho_{+}
                        - \erhom \rho_{-} \, , \quad
    \erhopm < 2 |\epsi| \, ,
  \end{eqnarray}
  where $\rho_{0}$ is the density matrix of the exact ground state and
  $\rho_{\pm}$ are density matrices containing the error.
\end{lemma}
  \emph{Proof.}
  We build the density matrix $\rho$ on the complete Hilbert space as
  \begin{eqnarray}
    \rho
   &=& \ket{\psi} \bra{\psi}
    = |f|^2 \ket{\psi_0} \bra{\psi_0}
      + f \epsi^{\ast} \ket{\psi_0} \bra{\psi_{\perp}}
      + f^{\ast} \epsi \ket{\psi_{\perp}} \bra{\psi_0}
      + |\epsi|^2 \ket{\psi_{\perp}} \bra{\psi_{\perp}}                         \nonumber \\
   &=& |f|^2 \ket{\psi_0} \bra{\psi_0}
       + | f \epsi |
         \left( \e^{- \mathrm{i} \phi} \ket{\psi_0} \bra{\psi_{\perp}}
         + \e^{\mathrm{i} \phi} \ket{\psi_{\perp}} \bra{\psi_0} \right)
       + |\epsi|^2 \ket{\psi_{\perp}} \bra{\psi_{\perp}} \, ,                   \nonumber \\
  \end{eqnarray}
  where we have chosen the phase $\phi$ such that $f \epsi^{\ast} = | f \epsi |
  \e^{-\mathrm{i} \phi}$. From Lemma~\ref{lem:mixedcontrib} we obtain
  \begin{eqnarray}
    \rho
   &=& |f|^2 \rho_0 + |\epsi f| (\sigma_{+} - \sigma_{-})
       + |\epsi|^2 \rho_{\perp}
    = |f|^2 \rho_0 + |\epsi| \left((|f| \sigma_{+} + |\epsi| \rho_{\perp})
                                  - |f| \sigma_{-} \right)                      \nonumber \\
   &=& |f|^2 \rho_0 + |\epsi| \left(
      (|f| + |\epsi|) \frac{|f| \sigma_{+} + |\epsi| \rho_{\perp}}{|f| + |\epsi|}
      - |f| \sigma_{-} \right) \, .
  \end{eqnarray}
  Defining the density matrices with positive and negative sign, we obtain
  \begin{eqnarray}                                                              \label{eq:abbrevepsilon}
    \rho
   &=& |f|^2 \rho_{0} + \erhop \rho_{+} - \erhom \rho_{-} \, ,
   \quad \erhop = |\epsi| (|f| + |\epsi|)
   \le 2 |\epsi| \, , \quad
   \erhom = |\epsi f| \le |\epsi| \, ,                                              \nonumber \\
   \rho_{+} &=& \frac{|f| \sigma_{+} + |\epsi| \rho_{\perp}}{|f| + |\epsi|} \, ,
   \quad \rho_{-} = \sigma_{-} \, ,
  \end{eqnarray}
  where all matrices $\rho_{0,\pm}$ are density matrices with trace
  $1$ and fulfilling positivity.

\begin{lemma}[Bound on reduced density matrices]                                \label{lem:boundredrho}
  The bound on the reduced density matrices $\rho_{A}$ of the \OSMPS{} result is
  equal to the bound on the complete density matrix, which is
  \begin{eqnarray}
    \rho_{A}
   &=& \Ptr{B}{\rho}
    = |f|^2 \rho_{0,A} + \erhop \rho_{+,A} - \erhom \rho_{-,A} \, ,
  \end{eqnarray}
  where we consider an arbitrary bipartition of our system in parts $A$ and
  $B$, tracing out over the bipartition $B$.
\end{lemma}
  \emph{Proof.}
  In order to obtain a reduced density matrix, we define subsystem $A$ and $B$
  and take the partial trace over subsystem $B$:
  \begin{eqnarray}
    \rho_{A}
   &=& \Ptr{B}{\rho}
    = \PTr{B}{|f|^2 \rho_{0} + \erhop \rho_{+}
           - \erhom \rho_{-}} \, .
  \end{eqnarray}
  As a linear operation the previous expression can be rewritten as
  \begin{eqnarray}
    \rho_{A}
   &=& |f|^2 \Ptr{B}{\rho_{0}} + \erhop \Ptr{B}{\rho_{+}}
       - \erhom \Ptr{B}{\rho_{-}}                                               \nonumber \\
   &=& |f|^2 \rho_{0,A} + \erhop \rho_{+,A} - \erhom \rho_{-,A} \, .
  \end{eqnarray}

\subsection{Bound for the trace distance}

Now that we are able to bound the density matrices with regards to the density
matrix of the exact ground state, we can continue to prove a bound expressed
in the trace distance $\mathcal{D}$.

\begin{lemma}[Bound on the trace distance]                                      \label{lem:boundTD}
  The trace distance between the density matrix $\rho$ from \OSMPS{} and
  the density matrix of the true ground state $\rho_{0}$ can be bounded with
  \begin{eqnarray}
    \mathcal{D}(\rho, \rho_0) \le \frac{\sqrt{2 V_{\psi}}}{\Delta_{0, 1}}
    \, .
  \end{eqnarray}
\end{lemma}
  \emph{Proof.}
  We continue with a bound on the trace distance defined as
  \begin{eqnarray}
    \mathcal{D}(\rho, \rho_{0})
   &=& \frac{1}{2} \mathrm{Tr} \sqrt{(\rho - \rho_{0})^{\dagger}
                                     (\rho - \rho_{0})}
    = \frac{1}{2} \left| \rho - \rho_{0} \right| \, ,
  \end{eqnarray}
  where the simplification can be made due to the Hermitian property of
  the density matrices. We first concentrate on expressing the difference
  between the density matrices in a convenient way using the expressions for
  $\erhopm$ in Eq.~\eqref{eq:abbrevepsilon}:
  \begin{eqnarray}
    \rho - \rho_{0}
   &=& |f|^2 \rho_0 + \erhop \rho_{+} - \erhom \rho_{-} - \rho_0
    = \erhop \rho_{+} - \erhom \rho_{-} - |\epsi|^2 \rho_0                      \nonumber \\
   &=& \erhop \rho_{+}
       - | \epsi | \left(|f| \rho_{-} + |\epsi| \rho_0 \right)                  
    = \erhop (P - M) \, , \qquad                                                \nonumber \\
    P &\equiv& \rho_{+} \, , \quad
    M \equiv \frac{|f| \rho_{-} + |\epsi| \rho_{0}}{|f| + |\epsi|} \, .
  \end{eqnarray}
  The new matrices $P$ and $M$ are again defined so that they fulfill
  the requirements for a density matrix (Hermitian, positive, trace equal to 1).
  In the next step we make use of the triangular inequality of the trace norm:
  \begin{eqnarray}
    \mathcal{D}(\rho, \rho_{0})
   &=&  \frac{1}{2} \mathrm{Tr} \left| \rho - \rho_{0} \right|
    =   \frac{|\epsi| (|f| + |\epsi|)}{2} \mathrm{Tr} \left| P - M \right|      \nonumber \\
   &\le& \frac{|\epsi| (|f| + |\epsi|)}{2} \left( \mathrm{Tr} |P|
        + \mathrm{Tr} |M| \right)                                               
    =  |\epsi| (|f| + |\epsi|)                                                  \nonumber \\
   &\le& \sqrt{2} |\epsi|
    \le \frac{\sqrt{2 V_{\psi}}}{\Delta_{0, 1}} \, .
  \end{eqnarray}

\subsection{Bound on the bond entropy}

In order to get the bound on the bond entropy, we use Fannes' inequality
\cite{NielsenChuang} stating that the difference of the entropy $S$ of 
two density matrices of dimension $D \times D$ is bounded by the trace
distance $\mathcal{D}$ between those two density matrices:
\begin{eqnarray}
  | S(\rho) - S(\sigma)|
  \le \mathcal{D} \log_{a}(D) - \mathcal{D} \log_a(\mathcal{D}) \, , \quad
  \forall \mathcal{D} \le \frac{1}{e} \approx 0.36 \, ,
\end{eqnarray}
with the logarithm for the von Neumann entropy base $a$ and $D$ the
dimension of the Hilbert space for the density matrices $\rho$ and $\sigma$.
The inequality was derived for logarithm base two, $a = 2$, but holds for any
basis. Therefore, the bound for the bond entropy is
\begin{eqnarray}
  \eentr \equiv | S(\rho_0) - S(\rho)|
  \le \frac{\sqrt{2 V_{\psi}}}{\Delta_{0, 1}} \log(D)
      - \frac{\sqrt{2 V_{\psi}}}{\Delta_{0, 1}}
        \log \left(\frac{\sqrt{2 V_{\psi}}}{\Delta_{0, 1}}\right) \, .
\end{eqnarray}

\section{Bounding measurement with the trace distance                          \label{app:boundtracedist}}

We have used the trace distance of reduced density matrices to judge the
convergence of our results. We now show that the trace distance bounds from
above any observable defined on the reduced density matrix. As a preliminary
result we need that the absolute value of the eigenvalues of a Hermitian matrix
correspond to the singular values of the matrix. Therefore, we use the fact
that the matrix can be decomposed via an eigenvalue decomposition
$A = U \Lambda U^{\dagger}$ and a singular value decomposition
$A = \tilde{U} \lambda \tilde{V}^{\dagger}$. Both decompositions
are unique up to a permutation of the orthonormal basis vectors and basis
transformations within degenerate sets of eigenvalues and singular values,
respectively. Calculating $AA^{\dagger}$, which is by definition
positive semi-definite even for a general $A$, with both expressions leads
to
\begin{eqnarray}                                                                \label{eq:conneigsvd}
  A A^{\dagger}
  = U \Lambda^2 U^{\dagger}
  = \tilde{U} \lambda^2 \tilde{U}^{\dagger}\, ,
\end{eqnarray}
which shows that the absolute values of the eigenvalues $\Lambda$ are equal
to the singular values $\lambda$.
In the case where the observable is positive semi-definite and
the the ordering of the eigenvalues and singular values is
equal, e.g. descending, the decompositions are equal, except for rotations
within degenerate subsets. In contrast, for observables with at least one
negative eigenvalue and sorting in descending order by the absolute value, the
decompositions are not equal as negative signs are contained in the
unitary matrices of the singular value decomposition.
We use the fact that the absolute values of the eigenvalues of a Hermitian
matrix are the singular values in the following when estimating the difference
between the measurement of an observable $Q$ for two different density matrices
$\rho$ and $\sigma$ defined as
\begin{eqnarray}
  \epsilon(\rho, \sigma, Q)
  &=& \left| \mathrm{Tr}(Q \rho) - \mathrm{Tr}(Q \sigma) \right|
   =  \left| \mathrm{Tr}\left( Q (\rho - \sigma) \right) \right| \, .
\end{eqnarray}
In order to bound this with the trace distance of $\rho$ and $\sigma$ we
use the von Neumann trace inequality \cite{Mirsky1975} on the matrices $Q$ and
$(\rho - \sigma)$ leading to
\begin{eqnarray}                                                                \label{eq:vNtraceinequal}
  \epsilon(\rho, \sigma, Q)
 &\le& \sum_{i}^{D} \kappa_i | \Lambda_i | \, ,
\end{eqnarray}
where $\kappa_i$ are the singular values of $Q$ sorted in descending order with
$\kappa_{i} > \kappa_{i+1}$. The $\Lambda_i$ are the eigenvalues
of $(\rho - \sigma)$ sorted in descending order with $|\Lambda_i| > |\Lambda_{i+1}|$.
$D$ is the dimension of the density matrix and the corresponding Hilbert
space. We derived the relation Eq.~\eqref{eq:conneigsvd} for this purpose.
We approximate Eq.~\eqref{eq:vNtraceinequal} by choosing the maximal singular
values $\kappa_{\max}$ from all $\kappa_{i}$:
\begin{eqnarray}
  \epsilon(\rho, \sigma, Q)
  &\le& \kappa_{\max} \sum_{i} | \Lambda_{i} |
   =    \kappa_{\max} \mathrm{Tr} \left| \Lambda_{i} \right|
   =    \kappa_{\max} 2 \mathcal{D}(\rho, \sigma) \, .
\end{eqnarray}
For further convenience we estimate $\kappa_{\max}$ as a function of the
observable $Q$. In the case of a Hermitian observable $Q = Q^{\dagger}$
we use again the connection between eigenvalues and singular values from
Eq.~\eqref{eq:conneigsvd}. The eigenvalues of a square matrix can be estimated
with \gershgorin{} circles \cite{Gerschgorin1931,Qi1984,Golub_VanLoan_96}.
Each \gershgorin{} circle has its origin in the diagonal element of the matrix;
the radius is the sum of the absolute values of the non-diagonal of the
corresponding row (or column). In this case we are only interested in the
absolute values and the expression for $\kappa_{\max}$ in terms of the
\gershgorin{} circles simplifies to
\begin{eqnarray}
  \kappa_{\max} \le \max_{i} \left( \sum_{j} \left| Q_{i,j} \right| \right) \, .
\end{eqnarray}
In the case of a non-Hermitian $Q$, i.e., the correlation measurement
$\langle b_{i} b_{j}^{\dagger} \rangle$, we can obtain an estimate using the fact that the
singular values are connected to the eigenvalues of $Q Q^{\dagger}$. If $Q$ has
the singular value decomposition $Q = U S V^{\dagger}$, then $Q Q^{\dagger}$ has the
eigenvalue decomposition $Q Q^{\dagger} = U S^2 U^{\dagger}$. Therefore,
the square root of the non-negative eigenvalues of $Q Q^{\dagger}$ are the singular
values of $Q$. For the non-Hermitian $Q$ we can estimate $\kappa_{\max}$ with
the \gershgorin{} circles over $Q Q^{\dagger}$ as
\begin{eqnarray}
  \kappa_{\max}
  \le \max_{i} \sqrt{\left( \sum_{j} \left| \left[Q Q^{\dagger} \right]_{i,j}
                     \right| \right)} \, .
\end{eqnarray}
In conclusion, these bounds show that the trace distance is a meaningful
quantity to bound the error on any other observable.

\section{Details of the Krylov method                                          \label{app:krylov}}

For the real-time TEBD algorithm we used the Krylov approximation to the
propagated state for a local propagator on two sites. In this appendix, we clarify
why sums throughout the algorithm can be done locally without involving the
other parts of the subsystem. As an example we take the first step of the
Gram-Schmidt orthogonalization procedure which can be generalized to sums.
We denote the two local sites $i$ and $j = i + 1$ acted on with $\ket{C}$ and
the other parts of the system to left and right with $\ket{L_i}$ and
$\ket{R_j}$. Moreover, the orthogonality center is in $\ket{C}$ which could
be either site $i$ or $j$. Using the Schmidt decomposition within the MPS,
$\ket{\psi}$ can be written as
\begin{eqnarray}                                                                \label{eq:app:krylov:decomp}
  \ket{v_1} &=& \ket{\psi} = \sum_{i, j} \ket{L_i} \ket{C_{i,j}} \ket{R_j}
\end{eqnarray}
where the indices $i$ and $j$ run to the corresponding bond dimension at the
splitting. In order to find the second basis vector $\ket{v_2}$ we calculate
\begin{eqnarray}
  \ket{v_2} \propto
  H_C \ket{v_1} - \bra{v_1} H_{C}^{\dagger} \ket{v_1} \ket{v_1} \, ,
\end{eqnarray}
with $H_C$ the local Hamiltonian acting on the two sites of $\ket{C}$. Using
Eq.~\eqref{eq:app:krylov:decomp} to expand the second basis vector, we obtain
the local summation:
\begin{eqnarray}                                                                \label{eq:app:krylov:decompmps}
  \ket{v_2}  &\propto& 
  H_{C}   \sum_{i,j} \ket{L_i} \ket{C_{i,j}} \ket{R_j}                          \nonumber \\
  &&-  \left[\sum_{i,i',j,j'}   \bra{L_{i'}} \bra{C_{i',j'}} \bra{R_{j'}} H_C^{\dagger}
                           \ket{L_{i}} \ket{C_{i,j}} \ket{R_{j}} \right]
    \sum_{i,j} \left(\ket{L_{i}} \ket{C_{i,j}} \ket{R_j} \right) \, .           \nonumber \\
\end{eqnarray}
The expression in brackets in Eq.~\eqref{eq:app:krylov:decompmps} is
simplified due to the canonical form of the MPS. We obtain Kronecker deltas
for $\langle L_{i'} | L_{i} \rangle = \delta_{i,i'}$ and $\langle R_{j'} |
R_{j} \rangle = \delta_{j,j'}$. We point out that the expression
$H_{C} \ket{L_i} \ket{C_{i,j}} \ket{R_j}$ is a short hand notation for
$(\1_{L} \otimes H_{C} \otimes \1_{R}) \ket{L_i} \otimes \ket{C_{i,j}}
\otimes \ket{R_j} = \1_{L} \ket{L_i} \otimes H_{C} \ket{C_{i,j}} \otimes
\1_{R} \ket{R_{j}}$. These steps lead us to:
\begin{eqnarray}
  \ket{v_2}
  &=& \sum_{i,j} \ket{L_{i}} \left(H_C \ket{C_{i,j}} \right) \ket{R_{j}}
  - \sum_{i,j} \left(\ket{L_{i}} \ket{C_{i,j}} \ket{R_j} \right)
    \sum_{i,j} \bra{C_{i,j}} H_C \ket{C_{i,j}}                                  \nonumber \\
  &=& \sum_{i,j} \ket{L_{i}} \left(H_C \ket{C_{i,j}} \right) \ket{R_{j}}
  - \sum_{i,j} h_{1,1} \ket{L_{i}} \ket{C_{i,j}} \ket{R_j} \, ,
\end{eqnarray}
where we introduced the overlap $h_{1,1} \equiv \sum_{i,j} \bra{C_{i,j}} H_C
\ket{C_{i,j}}$ representing a scaling of the MPS. For readers familiar
with the Krylov method, we point out that $h_{1,1}$ is the corresponding entry
in the upper Hessenberg matrix used to build the exponential in the Krylov
subspace. In order to scale the
MPS with a scalar the orthogonality center is modified. The identity
operator is acting $\ket{L_i}$ and $\ket{R_j}$ leaving them unchanged.
%
%
Knowing that the orthogonalization procedure does not change any $\ket{L_i}$
and $\ket{R_j}$, a similar arguments apply to show that we can sum locally
over the scaled Krylov basis $\ket{v_{k}}$.

We consider now the scaling of TEBD-Krylov in comparison to a usual
TEBD taking the matrix exponential. In the usual TEBD implementation
we can consider four steps: \emph{1)} Build two-site tensor, \emph{2)}
calculate local propagator with Hamiltonian, \emph{3)} contract propagator
to two site tensor, and \emph{4)} split two site tensor. Without considering
speedup due to symmetries, a basic computational complexity analysis yields
\begin{eqnarray}
  \mathcal{O}_{\mathrm{TEBD}}
 &=& \mathcal{O}(\chi^3 d^2) + \mathcal{O}(d^6) + \mathcal{O}(\chi^2 d^4)
     + \mathcal{O}(\chi^3 d^3) \, ,
\end{eqnarray}
where we assume cubic scaling for the matrix exponential and the splitting.
In case of the Krylov-TEBD (KTEBD) we have the following steps: \emph{1)}
contract the single site tensors to a two site tensor, \emph{2)} build $n$
Krylov vectors, \emph{3)} $m'$ calculations of overlap to previous vectors
and $m''$ subtraction for orthogonalization ($m = m' + m''$), \emph{4)}
taking the matrix exponential in the Krylov basis, \emph{5)} adding the
weighted Krylov vectors to the solution, and \emph{6)} finally splitting
the two site tensor. All the scalings are chronological, resulting in
\begin{eqnarray}
  \mathcal{O}_{\mathrm{KTEBD}}
 &=&   \mathcal{O}(\chi^3 d^2)
     + \mathcal{O}(n \chi^2 d^4)
     + \mathcal{O}(m \chi^2 d^2)                                                \nonumber \\
 &&  + \mathcal{O}(n^3)
     + \mathcal{O}(n \chi^2 d^2)
     + \mathcal{O}(\chi^3 d^3) \, .
\end{eqnarray}
We again assumed cubic scaling for matrix exponential and splitting, which
is an upper bound in the case of the matrix exponential being tridiagonal and
symmetric. The overall scaling seems to be dominated by the splitting with
$\mathcal{O}(\chi^3 d^3)$, but we calculate the difference to see which
algorithm is favorable is which cases:
\begin{eqnarray}
  \mathcal{O}_{\mathrm{KTEBD}} - \mathcal{O}_{\mathrm{TEBD}}
 &=&   \mathcal{O}(n^3) - \mathcal{O}(d^6)
     + \mathcal{O}((n - 1) \chi^2 d^4)                                          \nonumber \\
 &&   + \mathcal{O}((m + n) \chi^2 d^2) \, ,
\end{eqnarray}
where TEBD is faster when the expression is greater than zero. If we assume
in favor of KTEBD $d = \chi$ and $n = 2$, the second and the third term
cancel each other. Thus, the remaining terms show that TEBD is always faster
than KTEBD according to this scaling. These equations do not include quantum
numbers. A careful study is contemplated to proof or disproof the scaling
equations once both methods are implemented as pointed out in the future
developments in Sec.~\ref{sec:future}.

\section{Auxiliary calculations}                                               \label{app:aux}

The overlap between the truncated state $\ket{\psi'}$ and the untruncated
normalized state $\ket{\psi}$ is defined over the singular values
$\lambda_{i}'$ of the truncated state and the singular values $\lambda_{i}$ of
the untruncated state. In the following $\chi$ is the number of singular values
in $\ket{\psi'}$ and $\chi_{\max}$ is the untruncated number of singular values.
The singular values for the truncated state are calculated via a normalization
\begin{eqnarray}
  \lambda_{i}' &=& \frac{\lambda_{i}}{\sqrt{\sum_{j=1}^{\chi} \lambda_{j}^2}} \, .
\end{eqnarray}
This expression leads us directly to the overlap between the two states
\begin{eqnarray}                                                                \label{eq:app:osch}
  \braket{\psi'}{\psi}
 &=& \sum_{i=1}^{\chi_{\max}} \lambda_{i}' \lambda_{i}
  =  \sum_{i=1}^{\chi} \lambda_{i}' \lambda_{i}
  =  \frac{1}{\sqrt{\sum_{i=1}^{\chi} \lambda_{i}^2}}
     \sum_{i=1}^{\chi} \lambda_{i}^{2}
  =  \sqrt{\sum_{i=1}^{\chi} \lambda_{i}^2} \, .
\end{eqnarray}
For the error we calculate $1 - \braket{\psi'}{\psi}$ and abbreviate with
$x \equiv \braket{\psi'}{\psi}$:
\begin{eqnarray}                                                                \label{eq:app:J3}
  1 - \braket{\psi'}{\psi}
 &=& 1 - x                                                                      \nonumber \\
 &=& \frac{1}{2} \left(1 - 2x + x^2 \right) + \frac{1}{2} \left(1 - x^2 \right)
  =  \frac{1}{2} \left(1 - x \right)^{2} + \frac{1}{2} \left(1 - x^2 \right)    \nonumber \\
 &\le& (1 - x^2) \,
\end{eqnarray}
where the inequality is based on $x \le 1$. By definition via
Eq.~\eqref{eq:app:osch}, $x$ is real and bounded as $0 \le x \le 1$. This is
based on the normalization constraint and knowing that the singular values
$\lambda_i$ are positive. The following inequality is only valid for $x \le 1$:
%
%
Equation~\eqref{eq:app:J3} is further simplified to the sum over the truncated
singular values using the fact that the original state with $\chi_{\max}$
singular values was normalized:
\begin{eqnarray}
  1 - \braket{\psi'}{\psi}
  \le 1 - \sum_{i=1}^{\chi} \lambda_{i}^{2}
  = 1 - \left(1 - \sum_{i=\chi+1}^{\chi_{\max}} \lambda_{i}^{2} \right)
  = \sum_{i=\chi+1}^{\chi_{\max}} \lambda_{i}^{2} \, .
\end{eqnarray}

\section{Supplemental Material                                                \label{app:suppl}}

We provide further, supplemental, material in our forum
\cite{OSMPSPaperSupplMaterial} and our package with version 2.1. This
material includes:

\begin{itemize}
\item{Building and installing the open source Matrix Product States library:
  instructions for minimalists who do not want to read the full manual.}
\item{User support and contributions to the code.}
\item{Code for selected examples to reproduce plots in this paper, i.e.,
  the nearest-neighbor Ising model, the long-range Ising model, and the
  dynamics of the Bose-Hubbard model.}
\end{itemize}



\newpage\thispagestyle{empty}
\section*{Building and installing the open source Matrix Product States library \label{app:build}}

In order to install \OSMPS{}, one needs to provide at a minimum the following
packages covering the \fort{} and \py{} installation and linear algebra
tools:

\begin{itemize}
\item{\py{} (at least 2.6 or 2.7 recommended, 3.x tested successfully)}
\item{numpy and scipy}
\item{Fortran2003 compiler, e.g. \emph{gfortran} for Linux.}
\item{BLAS and LAPACK}
\item{ARPACK}
\end{itemize}

More packages might be convenient e.g. for plotting with \py{}, but are not
mandatory. The remaining part of the installation is covered by \py{}. We
distinguish between a global and a local installation where the local
installation can only be accessed in the same directory or through providing
the path. For global installation only one must execute on the command
line
\begin{eqnarray}
  \texttt{sudo python setup.py install} \nonumber
\end{eqnarray}
The \fort{} library is compiled then with the command
\begin{eqnarray}
  \texttt{python BuildOSMPS.py --os=unixmpi} \nonumber
\end{eqnarray}
with prefix \texttt{sudo} for the global installation in order to provide
the admin privileges. Some default settings are available via command
line options as shown for compiling with unix and MPI. Similarly, a local
installation can be specified with the option \pyline{--local='./'}. More
specific settings can be made inside the \py{} script, which builds the
makefile for the \fort{} libraries.

\newpage\thispagestyle{empty}
\section*{User support and contributing to the code                            \label{app:project}}

Being an open source project we welcome anyone willing to help improve
\OSMPS{} or share her knowledge with the library. On the project website
\url{http://sourceforge.net/projects/openmps/} on SourceForge we maintain a
forum for discussions about \OSMPS{}. These discussions include questions about
the installation and use of the algorithms for new users, suggestions for
future implementations, and help requests for implementing new features on your
own.

The current version of \OSMPS{} is organized via the \emph{svn} version
control system via SourceForge. It is possible without a user account to
fetch the newest version of the library via \emph{svn}. In order to contribute
to \OSMPS{}, a user account on \emph{SourceForge} is necessary so that we can
add you to the developers team.

\newpage
\section*{Files to reproduce plots in this work                                \label{app:files}}

We provide the files for the quantum Ising model, the long-range quantum
Ising model and the dynamics of the Bose-Hubbard model as examples how
\py{} scripts for \OSMPS{} are designed from the definition of the
model to post-processing.

\pycode{./code/01_Ising.py}
       {Example of nearest neighbor Ising model}
       {py:01_Ising}

\pycode{./code/02_LRIsing.py}
       {Example of the long-range Ising model}
       {py:02_LRIsing}

\pycode{./code/05_RotSaddle.py}
       {Example of dynamics in the Bose-Hubbard model}
       {py:05_RotSaddle}

\end{document}